%%%%%%%%%%%%%%%% Version  12/15/05    %%%%%%%%%%%%%%
\newif\iflanl
\openin 1 lanlmac
\ifeof 1 \lanlfalse \else \lanltrue \fi
\closein 1
\iflanl
    \input lanlmac
\else
    \message{[lanlmac not found - use harvmac instead}
    \input harvmac
    \fi

\input epsf
\input amssym
\input xy
\input tables
\xyoption{all}

%\draftmode

\noblackbox

%%%%%%%%%%%%%%%%%%%%%%%%%%%%%%%%%%%%%%%%%%%%%%
% macros
%%%%%%%%%%%%%%%%%%%%%%%%%%%%%%%%%%%%%%%%%%%%%%
\newif\ifhypertex
\ifx\hyperdef\UnDeFiNeD
    \hypertexfalse
    \message{[HYPERTEX MODE OFF}
    \def\hyperref#1#2#3#4{#4}
    \def\hyperdef#1#2#3#4{#4}
    \def\hypernoname{}
    \def\e@tf@ur#1{}

\else
    \hypertextrue
    \message{[HYPERTEX MODE ON}
%hypertex links to xxx.lanl.gov:
%  \def\hth/#1#2#3#4#5#6#7{\special{html:<a
%   href="http://xxx.lanl.gov/abs/hep-th/#1#2#3#4#5#6#7">}
%  {\tt hep-th/#1#2#3#4#5#6#7}\special{html:</a>}}

\fi
%%%%%%%%%%%%%%%%%%%%%%% %%%%%%%%%%%%%%%%%%%%%%%
\newif\ifdraft

\noblackbox
%
%\catcode`\@=11
\newif\iffrontpage
%%%%%%%%%%%%%%%%%%% %%%%%%%%%%%%%%%%%%%%%%%%%%%%%%%%%%%%%%%%%%%%%%
%%%%% sizes, offsets etc
%%%%%%%%%%%%%%%%%%% %%%%%%%%%%%%%%%%%%%%%%%%%%%%%%%%%%%%%%%%%%%%%%
\ifx\answ\bigans

\magnification=1200\baselineskip=15pt plus 2pt minus 1pt
%
%%%%% unreduced mode: %%%%
%\voffset=0.35truein\hoffset=0.250truein
%\advance\hoffset by-0.075truein
\advance\voffset by.6truecm
\hsize=6.15truein\vsize=600.truept\hsbody=\hsize\hstitle=\hsize
\else\let\lr=L

\magnification=1000\baselineskip=15pt plus 2pt minus 1pt
%
%%%%% reduced mode: %%%%%%%
\hoffset=-0.75truein\voffset=-.0truein
%?\hoffset=-.25truein\voffset=-.0truein
\vsize=6.5truein
\hstitle=8.truein\hsbody=4.75truein
\fullhsize=10truein\hsize=\hsbody
\fi
\parskip=4pt plus 15pt minus 1pt
%
%%%%%%%%%%%%%%%%%%% %%%%%%%%%%%%%%%%%%%%%%%%%%%%%%%%%%%%%%%%%%%%%%
%%%%% figures
%%%%%%%%%%%%%%%%%%% %%%%%%%%%%%%%%%%%%%%%%%%%%%%%%%%%%%%%%%%%%%%%%
\newif\iffigureexists
\newif\ifepsfloaded
\def\epsfcheck{
\ifdraft% to speed up
\input epsf\epsfloadedtrue
\else
  \openin 1 epsf
  \ifeof 1 \epsfloadedfalse \else \epsfloadedtrue \fi
  \closein 1
  \ifepsfloaded
    \input epsf
  \else
\immediate\write20{NO EPSF FILE --- FIGURES WILL BE IGNORED}
  \fi
\fi
\def\epsfcheck{}}
\def\checkex#1{
\ifdraft
\figureexistsfalse\immediate%
\write20{Draftmode: figure #1 not included}
\figureexiststrue
\else\relax
    \ifepsfloaded \openin 1 #1
        \ifeof 1
           \figureexistsfalse
  \immediate\write20{FIGURE FILE #1 NOT FOUND}
        \else \figureexiststrue
        \fi \closein 1
    \else \figureexistsfalse
    \fi
\fi}
\def\missbox#1#2{$\vcenter{\hrule
\hbox{\vrule height#1\kern1.truein
\raise.5truein\hbox{#2} \kern1.truein \vrule} \hrule}$}
\def\lfig#1{%  this is to call the figure in the text
\let\labelflag=#1%
\def\numb@rone{#1}%
\ifx\labelflag\UnDeFiNeD%
{\xdef#1{\the\figno}%
\writedef{#1\leftbracket{\the\figno}}%
\global\advance\figno by1%
}\fi{\hyperref{}{figure}{{\numb@rone}}{Fig.~{\numb@rone}}}}
\def\figinsert#1#2#3#4{%  this inserts the figure
\epsfcheck\checkex{#4}%
\def\figsize{#3}%
\let\flag=#1\ifx\flag\UnDeFiNeD
{\xdef#1{\the\figno}%
\writedef{#1\leftbracket{\the\figno}}%
\global\advance\figno by1%
}\fi
\goodbreak\midinsert%
\iffigureexists
\centerline{\epsfysize\figsize\epsfbox{#4}}%
\else%
\vskip.05truein
  \ifepsfloaded
  \ifdraft
  \centerline{\missbox\figsize{Draftmode: #4 not included}}%
  \else
  \centerline{\missbox\figsize{#4 not found}}
  \fi
  \else
  \centerline{\missbox\figsize{epsf.tex not found}}
  \fi
\vskip.05truein
\fi%
{\smallskip%
\leftskip 4pc \rightskip 4pc%
\noindent\ninepoint\sl \baselineskip=11pt%
{\bf{\hyperdef\hypernoname{figure}{{#1}}{Fig.~{#1}}}:~}#2%
\smallskip}\bigskip\endinsert%
}

\def\boxit#1{\vbox{\hrule\hbox{\vrule\kern8pt
\vbox{\hbox{\kern8pt}\hbox{\vbox{#1}}\hbox{\kern8pt}}
\kern8pt\vrule}\hrule}}
\def\mathboxit#1{\vbox{\hrule\hbox{\vrule\kern8pt\vbox{\kern8pt
\hbox{$\displaystyle #1$}\kern8pt}\kern8pt\vrule}\hrule}}

%
%
%%%%%%%%%%%%%%%%%%% %%%%%%%%%%%%%%%%%%%%%%%%%%%%%%%%%%%%%%%%%%%%%%
%%%%% tables 
%%%%%%%%%%%%%%%%%%% %%%%%%%%%%%%%%%%%%%%%%%%%%%%%%%%%%%%%%%%%%%%%%
\newcount\tabno
\tabno=1
\def\ltab#1{%  this is to call the table in the text
\let\labelflag=#1%
\def\numb@rone{#1}%
\ifx\labelflag\UnDeFiNeD%
{\xdef#1{\the\tabno}%
\writedef{#1\leftbracket{\the\tabno}}%
\global\advance\tabno by1%
}\fi{\hyperref{}{table}{{\numb@rone}}{Table~{\numb@rone}}}}
\def\tabinsert#1#2#3{%  this inserts the figure
\let\flag=#1\ifx\flag\UnDeFiNeD
{\xdef#1{\the\tabno}%
\writedef{#1\leftbracket{\the\tabno}}%
\global\advance\tabno by1%
}\fi
\vbox{\bigskip
#3
\smallskip%
\leftskip 4pc \rightskip 4pc%
\noindent\ninepoint\sl \baselineskip=11pt%
{\bf{\hyperdef\hypernoname{table}{{#1}}{Table~{#1}}}.~}#2%
\smallskip}
\bigskip
}

%%%%%%%%%%%%%%%%%%%%%%%%%%%%%%%%%%

%%%%%%%%%%%%%%%%%%%%%%%%%%%%% %%%%%%%%%%%%%%%%%%%%%%%%%%%%%%%%%
%%%%%%  macros for titlepage etc
%%%%%%%%%%%%%%%%% %%%%%%%%%%%%%%%%%%%%%%%%%%%%%%%%%%%%%%%%%%%%%%
%
\def\abstract#1{
\vskip .5in\vfil\centerline
{\bf Abstract}\penalty1000
{{\smallskip\ifx\answ\bigans\leftskip 1pc \rightskip 1pc 
\else\leftskip 1pc \rightskip 1pc\fi
\noindent \abstractfont  \baselineskip=12pt
{#1} \smallskip}}
\penalty-1000}
\def\endpage{\tenpoint\supereject\global\hsize=\hsbody%
\frontpagefalse\footline={\hss\tenrm\folio\hss}}

%\catcode`\@=12

%%%%%%%%%%%%%%%%%%%%%%%%%%%%%%%%%%%%%%%%%%%%%%
% Definitions
%%%%%%%%%%%%%%%%%%%%%%%%%%%%%%%%%%%%%%%%%%%%%%

\newif\ifnref

\def\doubref#1#2{\refs{{#1},{#2}}}

\nreffalse

\def\coeff#1#2{\relax{\textstyle {#1 \over #2}}\displaystyle}
\def\bar#1{\overline{#1}}
\def\sqr#1#2{{\vcenter{\vbox{\hrule height.#2pt
\hbox{\vrule width.#2pt height#1pt \kern#1pt \vrule width.#2pt}
\hrule height.#2pt}}}}

\def\al{\alpha}
\def\be{\beta}
\def\ga{\gamma}

\def\del{\partial}

   \def\cD{{\cal
D}} \def\cE{{\cal E}}  \def\cF{{\cal F}}    \def\cL{{\cal L}} \def\cM{{\cal
M}}  \def\cO{{\cal O}} \def\cP{{\cal P}} \def\cQ{{\cal
Q}} \def\cR{{\cal R}}    \def\cV{{\cal V}} \def\cV{{\cal V}}   
\def\ZZ{\Bbb{Z}}
\def\IC{\Bbb{C}}

\def\IP{\Bbb{P}}

\def\oneone{\rlap 1\mkern4mu{\rm l}}
\font\fourteenmi=cmmi10 scaled\magstep 1
\def\Chi{{\hbox{$\textfont1=\fourteenmi \chi$}}}

%%%%%%%%%%%%%%%%%%%%%%%
\def\thrbythr{{3\times 3}}
\def\twobytwo{{2\times 2}}
\def\fobyfo{{4\times 4}}
\def\condense{{\mathop{\succ}}}
%%%%%%%%%%%%%%%%%%%%%%%
\font\cmss=cmss10
\font\cmsss=cmss10 at 7pt

\def\inbar{\,\vrule height1.5ex width.4pt depth0pt}
\def\IC{{\relax\hbox{$\inbar\kern-.3em{\rm C}$}}}
\def\IZ{{\relax\ifmmode\mathchoice
        {\hbox{\cmss Z\kern-.4em Z}}{\hbox{\cmss Z\kern-.4em Z}}
        {\lower.9pt\hbox{\cmsss Z\kern-.4em Z}}
        {\lower1.2pt\hbox{\cmsss Z\kern-.4em Z}}\else{\cmss Z\kern-.4em Z}\fi}}
\def\Diag#1{\hbox{Diag}\left(#1\right)}
\def\id{\oneone}                        % Identity operator
\def\xym#1{\vcenter{\xymatrix{#1}}}        

\def\Transpose#1{#1^{\rm T}}
\def\Hom{\hbox{Hom}}

%%%%%%%%%%%%%%%%%%%%%%%
\def\frac#1#2{{{#1}\over{#2}}}
\def\dalpha{\hat\alpha}
%%%%%%%%%%%%%%%%%%%%%%%
%\def\comment#1{\noindent  \raggedright {\tt [#1]} \raggedright }
\def\ie{{\it i.e.}}
\def\cf{{\it c.f.}}

%%%%%%%%%%%%%%%%%%%%%%%%%%%%%%%%%%%%%%%%%%%%
% References
%%%%%%%%%%%%%%%%%%%%%%%%%%%%%%%%%%%%%%%%%%%%%%
%

%\OrlovA
\lref\OrlovA{
  D.\ Orlov,
  ``Triangulated categories of singularities and D-branes in Landau-Ginzburg models,''
  [arXiv:math.AG/0302304].
}

%\OrlovB
\lref\OrlovB{
  D.\ Orlov,
  ``Derived categories of coherent sheaves and triangulated categories of singularities,''
  [arXiv:math.AG/0503632].
}

 %\PolishchukDB
\lref\PolishchukDB{
  A.~Polishchuk and E.~Zaslow,
``Categorical mirror symmetry: The Elliptic curve,''
  Adv.\ Theor.\ Math.\ Phys.\  {\bf 2}, 443 (1998)
  [arXiv:math.ag/9801119].
  %%CITATION = MATH-AG 9801119;%%
}
\lref\PolishchukAppell{
  A.~Polishchuk,
``M. P. Appell's function and vector bundles of rank 2 on elliptic curve,''
  [arXiv:math.ag/9810084].
  %%CITATION = MATH-AG 9810084;%%
}
\lref\Laza{R. Laza, G. Pfister and D. Popescu, 
``Maximal Cohen-Macaulay modules over the cone of an elliptic curve,'' 
Journal of Algebra {\bf 253} (2002), no. 2, 209--236.}
\lref\Atiyah{M.F. Atiyah, 
``Vector bundles over an elliptic curve,'' 
Proc. London Math. Soc. {\bf VII (3)} (1957) 414-- 452.}

%\BrunnerMT
\lref\BrunnerMT{
  I.~Brunner, M.~Herbst, W.~Lerche and J.~Walcher,
 ``Matrix factorizations and mirror symmetry: The cubic curve,''
[arXiv:hep-th/0408243].
  %%CITATION = HEP-TH 0408243;%%
}

%\HerbstZM
\lref\HerbstZM{
  M.~Herbst, C.~I.~Lazaroiu and W.~Lerche,
``D-brane effective action and tachyon condensation in topological  minimal
models,''
  JHEP {\bf 0503}, 078 (2005)
  [arXiv:hep-th/0405138].
  %%CITATION = HEP-TH 0405138;%%
}

%\BrunnerDC
\lref\BrunnerDC{
  I.~Brunner, M.~Herbst, W.~Lerche and B.~Scheuner,
``Landau-Ginzburg realization of open string TFT,''
  [arXiv:hep-th/0305133].
  %%CITATION = HEP-TH 0305133;%%
}

%\BrunnerFV
\lref\BrunnerFV{
  I.~Brunner and M.~R.~Gaberdiel,
``Matrix factorisations and permutation branes,''
  JHEP {\bf 0507}, 012 (2005)
  [arXiv:hep-th/0503207].
  %%CITATION = HEP-TH 0503207;%%
}

%\EngerJK
\lref\EngerJK{
  H.~Enger, A.~Recknagel and D.~Roggenkamp,
``Permutation branes and linear matrix factorisations,''
  [arXiv:hep-th/0508053].
  %%CITATION = HEP-TH 0508053;%%
}

%\WalcherTX
\lref\WalcherTX{
  J.~Walcher,
``Stability of Landau-Ginzburg branes,''
  J.\ Math.\ Phys.\  {\bf 46}, 082305 (2005)
  [arXiv:hep-th/0412274].
  %%CITATION = HEP-TH 0412274;%%
}

%\HoriZD
\lref\HoriZD{
  K.~Hori and J.~Walcher,
``D-branes from matrix factorizations,''
  Comptes Rendus Physique {\bf 5}, 1061 (2004)
  [arXiv:hep-th/0409204].
  %%CITATION = HEP-TH 0409204;%%
}

%\HoriJA
\lref\HoriJA{
  K.~Hori and J.~Walcher,
``F-term equations near Gepner points,''
  JHEP {\bf 0501}, 008 (2005)
  [arXiv:hep-th/0404196].
  %%CITATION = HEP-TH 0404196;%%
}

%\HoriBX
\lref\HoriBX{
  K.~Hori,
``Boundary RG flows of N = 2 minimal models,''
  [arXiv:hep-th/0401139].
  %%CITATION = HEP-TH 0401139;%%
}

%\KapustinRC
\lref\KapustinRC{
  A.~Kapustin and Y.~Li,
``D-branes in topological minimal models: The Landau-Ginzburg approach,''
  JHEP {\bf 0407}, 045 (2004)
  [arXiv:hep-th/0306001].
  %%CITATION = HEP-TH 0306001;%%
}

%\KapustinGA
\lref\KapustinGA{
  A.~Kapustin and Y.~Li,
``Topological correlators in Landau-Ginzburg models with boundaries,''
  Adv.\ Theor.\ Math.\ Phys.\  {\bf 7}, 727 (2004)
  [arXiv:hep-th/0305136].
  %%CITATION = HEP-TH 0305136;%%
}

%\KapustinBI
\lref\KapustinBI{
  A.~Kapustin and Y.~Li,
``D-branes in Landau-Ginzburg models and algebraic geometry,''
  JHEP {\bf 0312}, 005 (2003)
  [arXiv:hep-th/0210296].
  %%CITATION = HEP-TH 0210296;%%
}

%\GreeneUT
\lref\GreeneUT{
  B.~R.~Greene, C.~Vafa and N.~P.~Warner,
``Calabi-Yau Manifolds And Renormalization Group Flows,''
  Nucl.\ Phys.\ B {\bf 324}, 371 (1989).
  %%CITATION = NUPHA,B324,371;%%
}

%\WarnerAY
\lref\WarnerAY{
  N.~P.~Warner,
``Supersymmetry in boundary integrable models,''
  Nucl.\ Phys.\ B {\bf 450}, 663 (1995)
  [arXiv:hep-th/9506064].
  %%CITATION = HEP-TH 9506064;%%
}

%\AshokZB
\lref\AshokZB{
  S.~K.~Ashok, E.~Dell'Aquila and D.~E.~Diaconescu,
  ``Fractional branes in Landau-Ginzburg orbifolds,''
  Adv.\ Theor.\ Math.\ Phys.\  {\bf 8}, 461 (2004)
  [arXiv:hep-th/0401135].
  %%CITATION = HEP-TH 0401135;%%
}

%\DellAquilaJG
\lref\DellAquilaJG{
  E.~Dell'Aquila,
``D-branes in Toroidal Orbifolds and Mirror Symmetry,''
 [arXiv:hep-th/0512051].
  %%CITATION = HEP-TH 0512051;%%
}

%\EngerJK
\lref\EngerJK{
  H.~Enger, A.~Recknagel and D.~Roggenkamp,
  ``Permutation branes and linear matrix factorisations,''
 [arXiv:hep-th/0508053].
  %%CITATION = HEP-TH 0508053;%%
}

%\BrunnerPQ
\lref\BrunnerPQ{
  I.~Brunner and M.~R.~Gaberdiel,
``The matrix factorisations of the D-model,''
  J.\ Phys.\ A {\bf 38}, 7901 (2005)
  [arXiv:hep-th/0506208].
  %%CITATION = HEP-TH 0506208;%%
}

%\BrunnerFV
\lref\BrunnerFV{
  I.~Brunner and M.~R.~Gaberdiel,
 ``Matrix factorisations and permutation branes,''
  JHEP {\bf 0507}, 012 (2005)
  [arXiv:hep-th/0503207].
  %%CITATION = HEP-TH 0503207;%%
}

%\WalcherTX
\lref\WalcherTX{
  J.~Walcher,
``Stability of Landau-Ginzburg branes,''
  J.\ Math.\ Phys.\  {\bf 46}, 082305 (2005)
  [arXiv:hep-th/0412274].
  %%CITATION = HEP-TH 0412274;%%
}

%\HoriZD
\lref\HoriZD{
  K.~Hori and J.~Walcher,
``D-branes from matrix factorizations,''
  Comptes Rendus Physique {\bf 5}, 1061 (2004)
  [arXiv:hep-th/0409204].
  %%CITATION = HEP-TH 0409204;%%
}

%\HerbstAX
\lref\HerbstAX{
  M.~Herbst and C.~I.~Lazaroiu,
``Localization and traces in open-closed topological Landau-Ginzburg
models,''
  JHEP {\bf 0505}, 044 (2005)
  [arXiv:hep-th/0404184].
  %%CITATION = HEP-TH 0404184;%%
}

%\AshokXQ
\lref\AshokXQ{
  S.~K.~Ashok, E.~Dell'Aquila, D.~E.~Diaconescu and B.~Florea,
``Obstructed D-branes in Landau-Ginzburg orbifolds,''
  Adv.\ Theor.\ Math.\ Phys.\  {\bf 8}, 427 (2004)
  [arXiv:hep-th/0404167].
  %%CITATION = HEP-TH 0404167;%%
}

%\LazaroiuZI
\lref\LazaroiuZI{
  C.~I.~Lazaroiu,
``On the boundary coupling of topological Landau-Ginzburg models,''
  JHEP {\bf 0505}, 037 (2005)
  [arXiv:hep-th/0312286].
  %%CITATION = HEP-TH 0312286;%%
}

%\DouglasGI
\lref\DouglasGI{
  M.~R.~Douglas,
  ``D-branes, categories and N = 1 supersymmetry,''
  J.\ Math.\ Phys.\  {\bf 42}, 2818 (2001)
  [arXiv:hep-th/0011017].
  %%CITATION = HEP-TH 0011017;%%
}

\lref\kontsevich{
  M.~Kontsevich,
  ``Homological Algebra of Mirror Symmetry,''
  [arXiv:alg-geom/9411018].
  %%CITATION = ALG-GEOM 9411018;%%
}

\lref\kontsevichU{
  M.~Kontsevich,
 unpublished.  
}

\lref\HeHo{
 M.\ Herbst and K.\ Hori, to appear.
}

%\AspinwallJR
\lref\AspinwallJR{
  P.~S.~Aspinwall,
``D-branes on Calabi-Yau manifolds,''
  [arXiv:hep-th/0403166].
  %%CITATION = HEP-TH 0403166;%%
}

%\DouglasQW
\lref\DouglasQW{
  M.~R.~Douglas, B.~Fiol and C.~R\"omelsberger,
  ``The spectrum of BPS branes on a noncompact Calabi-Yau,''
  JHEP {\bf 0509}, 057 (2005)
  [arXiv:hep-th/0003263].
  %%CITATION = HEP-TH 0003263;%%
}

%\GovindarajanJS
\lref\GovindarajanJS{
  S.~Govindarajan, T.~Jayaraman and T.~Sarkar,
  ``Worldsheet approaches to D-branes on supersymmetric cycles,''
  Nucl.\ Phys.\ B {\bf 580}, 519 (2000)
  [arXiv:hep-th/9907131].
  %%CITATION = HEP-TH 9907131;%%
}

%\GovindarajanKR
\lref\GovindarajanKR{
  S.~Govindarajan and T.~Jayaraman,
  ``Boundary fermions, coherent sheaves and D-branes on Calabi-Yau
  manifolds,''
  Nucl.\ Phys.\ B {\bf 618}, 50 (2001)
  [arXiv:hep-th/0104126].
  %%CITATION = HEP-TH 0104126;%%
}

%\EzhuthachanJR
\lref\EzhuthachanJR{
  B.~Ezhuthachan, S.~Govindarajan and T.~Jayaraman,
  ``A quantum McKay correspondence for fractional 2p-branes on LG orbifolds,''
  JHEP {\bf 0508}, 050 (2005)
  [arXiv:hep-th/0504164].
  %%CITATION = HEP-TH 0504164;%%
}

%\HellermanBU
\lref\HellermanBU{
  S.~Hellerman, S.~Kachru, A.~E.~Lawrence and J.~McGreevy,
  ``Linear sigma models for open strings,''
  JHEP {\bf 0207}, 002 (2002)
  [arXiv:hep-th/0109069].
  %%CITATION = HEP-TH 0109069;%%
}

%\BerensteinFI
\lref\BerensteinFI{
  D.~Berenstein and M.~R.~Douglas,
  ``Seiberg duality for quiver gauge theories,''
  [arXiv:hep-th/0207027].
  %%CITATION = HEP-TH 0207027;%%
}

%\VafaQF
\lref\VafaQF{
  C.~Vafa,
``Brane/anti-brane systems and U(N$|$M) supergroup,''
 [arXiv:hep-th/0101218].
  %%CITATION = HEP-TH 0101218;%%
}

%\Baciu
\lref\Baciu{
  C.~Baciu,
``Maximal Cohen-Macaulay Modules over the Affine Cone of the Simple Node,''
  [arXiv:math.AC/0511405].
  %%CITATION = MATH-AC 0511405;%%
}

%\RecknagelSB
\lref\RecknagelSB{
  A.~Recknagel and V.~Schomerus,
  ``D-branes in Gepner models,''
  Nucl.\ Phys.\ B {\bf 531}, 185 (1998)
  [arXiv:hep-th/9712186].
  %%CITATION = HEP-TH 9712186;%%
}

%\BrunnerJQ
\lref\BrunnerJQ{
  I.~Brunner, M.~R.~Douglas, A.~E.~Lawrence and C.~R\"omelsberger,
  ``D-branes on the quintic,''
  JHEP {\bf 0008}, 015 (2000)
  [arXiv:hep-th/9906200].
  %%CITATION = HEP-TH 9906200;%%
}

%\LazaroiuJM
\lref\LazaroiuJM{
  C.~I.~Lazaroiu,
  ``Generalized complexes and string field theory,''
  JHEP {\bf 0106}, 052 (2001)
  [arXiv:hep-th/0102122].
  %%CITATION = HEP-TH 0102122;%%
}

%\LazaroiuMD
\lref\LazaroiuMD{
  C.~I.~Lazaroiu,
 ``D-brane categories,''
  Int.\ J.\ Mod.\ Phys.\ A {\bf 18}, 5299 (2003)
  [arXiv:hep-th/0305095].
  %%CITATION = HEP-TH 0305095;%%
}

%\AspinwallPU
\lref\AspinwallPU{
  P.~S.~Aspinwall and A.~E.~Lawrence,
  ``Derived categories and zero-brane stability,''
  JHEP {\bf 0108}, 004 (2001)
  [arXiv:hep-th/0104147].
  %%CITATION = HEP-TH 0104147;%%
}

%\SenSM
\lref\SenSM{
  A.~Sen,
  ``Tachyon condensation on the brane antibrane system,''
  JHEP {\bf 9808}, 012 (1998)
  [arXiv:hep-th/9805170].
  %%CITATION = HEP-TH 9805170;%%
}

%\SenYI
\lref\SenYI{
  A.~Sen,
  ``Dyon - monopole bound states, selfdual harmonic forms on the multi -
  monopole moduli space, and SL(2,Z) invariance in string theory,''
  Phys.\ Lett.\ B {\bf 329}, 217 (1994)
  [arXiv:hep-th/9402032].
  %%CITATION = HEP-TH 9402032;%%
}

\lref\mirbook{
K.~Hori, S.~Katz, A.~Klemm, R.~Pandharipande, R.~Thomas, C.~Vafa, R.~Vakil, and
  E.~Zaslow, ``Mirror {S}ymmetry,''
American Mathematical Society, 2003.
}
%
%%%%%%%%%%%%% End References %%%%%%%%%%%%%

%%%%%%%%%%%%%%%   Title Page  %%%%%%%%%%%%%
\Title{
\vbox{
\hbox{\tt hep-th/0512208}\vskip -.15cm
\hbox{\tt CERN PH-TH/2005-259}
}}
{\vbox{
%\vskip -4.5cm
\ifx\answ\bigans
\vskip -5cm
\else
\vskip -4.5cm
\fi
\centerline{\hbox{Tachyon Condensation on the Elliptic Curve}}
}}
\vskip -.3cm
\centerline{Suresh Govindarajan\footnote{${}^\dagger$}{On leave from the 
{\it Department of Physics, Indian Institute of Technology Madras, 
Chennai, 600036, India.}},
 Hans Jockers, Wolfgang Lerche   and 
Nicholas P.\ Warner\footnote{*}{On leave from the {\it Department of Physics and 
Astronomy,   University of Southern California, Los Angeles, CA 90089-0484, USA. }}}
\medskip
\centerline{{ \it Department of Physics, Theory Division}}
\centerline{{\it CERN, Geneva, Switzerland}}
%\medskip
%\bigskip

\ifx\answ\bigans
\vskip -1.0cm
\else
\vskip -.2cm
\fi

%%%%%%%%%%%%%%%%%%%%%%%%%%%%%%%%%%%%%%%%%
%:Abstract
\abstract{
We use the framework of matrix factorizations to study topological
$B$-type $D$-branes on the cubic curve. Specifically, we elucidate
how the brane $RR$ charges are encoded in the matrix factors, by
analyzing their structure in terms of sections of vector bundles
in conjunction with equivariant $R$-symmetry.  One particular
advantage of matrix factorizations is that explicit moduli dependence
is built in, thus giving us full control over the open-string moduli
space. It allows one to study phenomena like discontinuous jumps
of the cohomology over the moduli space, as well as formation of
bound states at threshold.  One interesting aspect is that certain
gauge symmetries inherent to the matrix formulation lead to a
non-trivial global structure of the moduli space.  We also investigate
topological tachyon condensation, which enables us to construct,
in a systematic fashion, higher-dimensional matrix factorizations
out of smaller ones; this amounts to obtaining branes with higher
$RR$ charges as composites of ones with minimal charges.  As an
application, we explicitly construct all rank two matrix factorizations.
}
%%%%%%%%%%%%%%%%%%%%%%%%%%%%%%%%%%%%%%%%%

\noindent

%\vskip .3in
\Date{\sl {December, 2005}}

%\vfill\eject
\endpage

%%%%%%%%%%%%%%%%%%%%%%%%%%%%%%%%%%
%:intro
\newsec{Introduction}
\subsec{Motivation and overview}
%%%%%%%%%%%%%%%%%%%%%%%%%%%%%%%%%%

$D$-branes play a very important r\^ole in our  understanding
non-perturbative properties of string and field theories, as well
as in building semi-realistic models. However, the naive geometrical
notion of a $D$-brane, in which it is thought of as wrapping some
$p$-dimensional cycle of a Calabi-Yau manifold, is a classical
concept that is valid only in certain limiting situations, such as
the large radius limit. When distances are small or curvatures
large, quantum corrections tend to blur notions of classical geometry,
such as   the dimension of a wrapped submanifold. Moreover, branes
can become unstable and decay in ways that are not visible classically.

Therefore one needs to adopt a more suitable language for describing
general $D$-brane configurations. For topological $B$-type $D$-branes,
the proper mathematical framework is a certain enhanced, bounded
derived category of coherent sheaves
\refs{\DouglasGI,\LazaroiuJM,\AspinwallPU} (and via homological
mirror symmetry  \doubref\kontsevich\mirbook, this maps to the Fukaya category of
$A$-type branes wrapping special Lagrangian cycles). This framework
retains more data  than the more familiar characterization just in
terms of $K$-theory (\ie, $RR$ charges) and thus provides a much
sharper description of $D$-branes.   That is, the category
also  contains the information about the brane locations, and other
possible (bundle or sheaf) moduli.  For instance, a configuration
consisting of an anti-$D0$-brane located at some point $\zeta_1$
of the compactification manifold, plus a $D0$-brane located at some
other point $\zeta_2$, is trivial from the $K$-theory point of view,
but is a non-trivial object in the categorical description as long
as $\zeta_1\not=\zeta_2$. Obviously,
this extra information is crucial for understanding questions such
as whether, in a given $D$-configuration, deformations are obstructed
or not (\ie, what is the effective superpotential and the moduli
space of its flat directions). Moreover, the language of categories
is tailor-made for addressing questions about stability and bound
state formation, which can be described more physically by tachyon
condensation. Excellent reviews of these matters may be found in
refs.~\doubref\LazaroiuMD\AspinwallJR.

Often physicists associate derived categories with just an abstract
collection of objects (the $D$-branes) and maps (open strings)
between them, and wonder what concrete physical benefit such a
picture might provide. Indeed, by merely tracing arrows around a
quiver diagram, all one obtains is a list of possible terms
in the effective superpotential and these terms are merely added up
with unit coefficients. However, there is more to these maps than just
being pointers between objects: in general they depend on the various
parameters like brane-location moduli, and thus encode valuable extra
information beyond mere combinatorics. Thus, superpotential terms
derived from quiver diagrams will, in general, have pre-factors depending
on the various moduli of the geometry, a fact that is often neglected in
the physics literature.

The abstract notions of objects and morphisms (maps) can, in fact,
be easily translated into a language more familiar to physicists
via the following two logical steps. First, as has been proven
recently \OrlovB\ in quite some generality (see also \HeHo), the relevant category of
topological $B$-type $D$-branes is isomorphic to a certain category
of matrix factorizations \refs{\kontsevichU,\OrlovA}, which encodes
the specific $D$-geometry in question.  Second, such matrix
factorizations have a direct interpretation \refs{\KapustinBI,\BrunnerDC}
in terms of two-dimensional topological (twisted $N=2$ supersymmetric)
boundary Landau-Ginzburg theory \refs{\WarnerAY,\GovindarajanJS}.
Specifically, the maps alluded to above feature (partly) as boundary
superpotentials.  Thus, via this chain of arguments, boundary
Landau-Ginzburg theory provides a very explicit realization of the
topological field theory of $B$-type $D$-branes. A sample computation
was presented in ref. \BrunnerMT\ demonstrating how it can be used,
in conjunction with mirror symmetry, to explicitly determine
moduli-dependent, instanton-corrected contributions to superpotentials
on intersecting branes.  

The purpose of the present paper is to use
the language of matrix factorizations to develop,  from a physicist's
point of view, a better understanding of tachyon condensation and
the process of composite formation of $B$-type $D$-branes.
 Specifically, we will analyze, in some detail, $D$-branes
on the cubic curve, $\Sigma$, which is the simplest situation with
both bulk and boundary (brane) moduli and can be studied fairly
explicitly.
 
Mathematically, the classification of bundles on the
elliptic curve is a completely solved problem \Atiyah: An
indecomposable bundle is uniquely determined by rank and
first Chern class of the bundle $\cE$, plus a continuous parameter
$\zeta$: $(r(\cE),c_1(\cE),\zeta)\equiv(N_2,N_0,\zeta)$.  
In physical terms, this corresponds to the number of $D2$ and
$D0$-branes plus, essentially, the location of the $D0$-brane on $\Sigma$. 
The mirror map to the Fukaya category of $A$-type branes is 
understood as well \PolishchukDB.  Matrix factorizations describing bundles
on $\Sigma$ have been described in ref.~\Laza, but only for fixed
moduli and not via tachyon condensation.\foot{While writing up this paper, 
we received a paper \Baciu\ that also deals with matrix factorizations pertaining 
to the elliptic curve, and which has some overlap with our work.} We will make use
of these mathematical results to construct and analyze matrix
factorizations explicitly depending on moduli, extending prior work
\doubref\HoriJA\BrunnerMT\ in a systematic fashion. Specifically
we will show how the bundle data $(r(\cE),c_1(\cE),\zeta)$ are 
explicitly encoded by certain properties of the matrices. This
leads to an algorithm that allows one to recover the brane data encoded
in a given matrix factorization.

As an application of this we study (topological) tachyon
condensation \SenSM\ of pairs of branes. This problem has two parts: first,
determining the open-string spectrum by solving the relevant
cohomology problem, and second, identifying the bundle data
$(N_2,N_0,\zeta)$ of the matrix factorization that results from
perturbing the direct product of matrix factorizations with an open
string cohomology element.  While we will encounter a few minor
subtleties (such as discontinuous jumps of the cohomology upon
varying moduli, and the formation of composites at threshold), the
physical results are entirely as expected: brane composites can be
formed according to the vector addition of brane charges, provided
one properly chooses the perturbing tachyonic operators and 
appropriately tunes the moduli.

We will show explicitly how all configurations with rank
$r=N_2\leq2$ can be built out of a minimal generating set of
two-dimensional factorizations, which correspond to $D$-branes whose
$RR$~charges generate the full charge lattice. It is pretty clear
that by iteratively applying the same logic, brane composites
corresponding to arbitrary points $(N_2,N_0)$ on the charge lattice
can be generated. Of course, this does not come as a surprise,
but this isn't the point of the present paper -- the
point is to understand how rather abstract mathematical concepts
can be realized in a concrete physical framework, namely boundary
Landau-Ginzburg theory and how to understand condensation
within that framework. We expect that the insight we gain will
be useful for attacking more complicated geometries, like branes
on threefolds.

The plan of the paper is as follows: In the remainder of this
section, we will review the description of $B$-branes by matrix
factorizations in very simple terms; this is aimed at non-experts.
Moreover, we will outline the main points of tachyon condensation
in such models and present a few examples. Section 2 is then devoted
to a general discussion of how the bundle data of a given brane
configuration are encoded in the corresponding matrix factorization.
An important r\^ole is played by the holomorphic sections of bundles on the
elliptic curve, which are given by Riemann theta functions and Appell
functions, for rank one line bundles and rank two vector bundles,
respectively.
 
In Section 3 we will reconsider the known $\twobytwo$ and $\thrbythr$
dimensional factorizations, and discuss in detail their structure
in terms of transition functions of the relevant bundles. Section
4 deals with the open-string moduli space of the $\thrbythr$
factorization, which has a non-trivial global structure due to gauge
symmetries inherent to the factorization; also, we will find how
its moduli space can be compactified by adding an exceptional
$\fobyfo$ factorization at the boundary, which appears to describe
a pure, rigid anti-$D2$-brane.

In Section 5 we address how to properly formulate tachyon
condensation in term of equivariant $R$-symmetry; this is necessary
for disentangling the various different branes that are described
by a given matrix factorization, and for subsequently identifying
the various tachyon channels between pairs of them. Section 6
provides some more tools for obtaining new matrix factorizations
from known ones; in particular we introduce certain bound states
at threshold, which resolve a certain singularity in the multi-brane
moduli space. Finally, in Section 7 we apply the techniques developed
in the preceding sections, and show how the factorizations
describing certain rank two vector bundles can be systematically generated
by condensation of lower rank branes. We have also relegated some more
technical material to the appendices.

%%%%%%%%%%%%%%%%%%%%%%%%%%%%%%%%%%
%:recap
\subsec{Recapitulation: $B$-type branes on the elliptic curve $\Sigma$}
%%%%%%%%%%%%%%%%%%%%%%%%%%%%%%%%%%

We will consider $B$-type branes on the
cubic curve, $\Sigma$, defined by the following hypersurface in 
$\IP^2$: 
\eqn\cubic{ W(x) ~\equiv~ 
\big( {x_1}^3 + {x_2}^3 + {x_3}^3 \big) ~-~ 
3\, a\, x_1 x_2 x_3 ~=~0  \,.}
Here, $a$ is the complex structure modulus that is related to the
standard modulus of the torus via:
\eqn\ataureln{\Big({3\, a\, (a^3 +8) \over  a^3 - 1} \Big)^3~=~  j(\tau) \,,}
where $j(\tau)  = q^{-1} + 744 + \dots$ is the familiar modular invariant
function in terms of $q=e^{2\pi i\tau}$, and $\tau$ is the flat coordinate 
of the complex structure moduli 
space, which coincides with the K\"ahler modulus of the mirror curve, $\hat\Sigma$. 
The coordinates on the cubic can be uniformized in terms of theta functions 
by writing $x_\ell = \mu_\ell(\xi)$, where
\eqn\mudefn{
\mu_\ell(\xi) ~\equiv~ \mu_\ell(\xi|\tau)  ~=~   \omega^{(\ell -1)}  \, 
\Theta \bigg[{ {{1 \over 3} (1-\ell) -{1 \over 2} \atop -{1 \over 2}} }\, \bigg |\,3\,\xi ,3
\,\tau \bigg]\ ,}
with $\omega \equiv e^{2 \pi i /3}$. For further details we refer
the reader to Appendix A. Here, $\xi$ is an arbitrary point on the 
Jacobian of $\Sigma$, which coincides with $\Sigma$ itself. One should also note
that $\mu_{\ell}(\xi)$ is not a function on $\Sigma$, but
is actually a section of the line bundle, $\cL^3$, 
with first Chern number $c_1 =3$. This will lead
to a natural ambiguity in determining the bundle data associated with a
given matrix factorization.

\phantom{\KapustinGA,\KapustinRC,\LazaroiuZI,\AshokZB,\HoriBX,\AshokXQ,\HerbstAX,\HoriJA,\HerbstZM,\BrunnerMT,\HoriZD,\WalcherTX,\BrunnerFV,\BrunnerPQ,\EngerJK,\DellAquilaJG}
As has been discussed by now in many papers
\refs{\KapustinBI,\BrunnerDC,\KapustinGA{--}\DellAquilaJG}, topological
$B$-type $D$-branes can be described by a two-dimensional, twisted
$N=2$ supersymmetric boundary Landau-Ginzburg model based on matrix
factorizations of the form
\eqn\matfac{J(x) \, E(x) ~=~ E(x) \, J(x) ~=~W(x) \, 
\oneone_{n\times n} \,.} 
Here $J(x)$ and $E(x)$ are $n \times n$ polynomial matrices\foot{We do 
not require $n$ to be the dimension of a Clifford algebra, which would be needed
if we introduced boundary fermions \WarnerAY. The implementation of
more general boundary couplings not involving fermions, was discussed in
refs. \refs{\LazaroiuZI,\HerbstAX}.}
with values in $\IC[x_1,x_2,x_3]$, whose precise form depends on the 
$D$-brane configuration in question. It will be important later to make use of the
 fact that $\matfac$ is invariant under gauge transformations of the
 form:
\eqn\gauge{\eqalign{J(x) &\ \rightarrow \ U_L(x)\,J(x)\,U_R(x) \ ,\cr
                    E(x) &\ \rightarrow \ U_R^{-1}(x)\,E(x)\,U_L^{-1}(x)\ ,}}
for polynomial matrices $U_{L,R}(x)$ that are invertible over $\IC[x_1,x_2,x_3]$.  
In particular, this means 
we can do arbitrary row and column reduction operations on
$J$ (respectively, $E$) so long as one does the corresponding
inverse operations on $E$ (respectively, $J$). For further details of 
this procedure we refer the reader to Appendix B. 

Mathematically, matrix factorizations form a category 
\doubref\kontsevichU\OrlovA\ with objects $P$ of the form
\eqn\DbP{P\equiv\left[\xym{P_1 \ar@/^/[d]^{E_P} \cr P_0 \ar@/^/[u]^{J_P}}\right]
 \ ,}
where $P_0$ and $P_1$ are certain projective modules over
$\IC[x_1\dots,x_m]$.  This ``composite'' form of objects, $P$, has
a simple physical interpretation.  Recall that the usual bulk
Landau-Ginzburg action is not conformally invariant, but represents
an action that will flow, in the infra-red, to the conformal field
theory of interest, and the superpotential will remain unrenormalized
along the flow.  By the same token, in the boundary theory, the
``constituents'' $P_{0,1}$ correspond to $D$-branes and anti-$D$-branes
in $\IC[x_1\dots,x_m]$ and the maps $J$ and $E$ correspond to tachyon
profiles that trigger the condensing of $P_{0,1}$ into $P$  
\doubref\KapustinBI\LazaroiuZI. 
Thus we construct a particular $D$-brane, $P$, by setting up
an action that will generate it via an infra-red boundary flow (\ie,
tachyon condensation).   In this formulation, $J$ has the interpretation
as a boundary superpotential while $E$ appears in a modified chirality
condition of fermionic boundary superfields
\refs{\GovindarajanKR,\HellermanBU}.

Anti-$D$-branes are associated with objects
commonly denoted by $P[1]$, and look like $P$ in \DbP\ except that
 $J_P\to-E_P$ and  $E_P\to-J_P$; we will often use the
 notation~$\bar P$ for them. 

The trivial object in the category, denoted by $V$, is described
by the simplest factorization, $P_{1\times 1}$, for which $J =
\oneone_{1\times 1}$ and $E= W \oneone_{1\times 1}$, or vice-versa.
It corresponds to the situation where $P_0$ and $P_1$ have completely
annihilated to leave no net branes at all.  Factorizations of
different dimension are thus considered to be equivalent if they differ
by appending or removing such trivial matrix blocks.\foot{In the derived 
category of matrix factorizations these objects are called 
perfect complexes and are divided out \OrlovB, because they do not have any
non-trivial morphisms with any other object in the category. This means
that there are no open-string states associated to these perfect complexes, and
hence such configurations are isomorphic to the open-string vacuum \AspinwallJR.}

On the cubic curve, $\Sigma$, the simplest non-trivial factorizations are two- 
and three-dimensional
\doubref\HoriJA\BrunnerMT.  The first ones are given by:%
\goodbreak
\eqn\Mtwo{
P_\twobytwo:\ \ \cases{ J_\twobytwo&=\  $\pmatrix{ Q_1 & -Q_2 \cr 
                                 L_2 & \hphantom{-}L_1 }  $\cr
                        E_\twobytwo&=\ $\pmatrix{\hphantom{-}L_1 & Q_2\cr 
                                -L_2 & Q_1 } $\ ,}
}
where the linear entries read 
\eqn\Mtwolin{\eqalign{L_1 \, &=\, \alpha_3 x_1 - \alpha_2 x_3 \ , \cr
                      L_2 \, &=\,-\alpha_3 x_2 + \alpha_1 x_3 \ ,}}
and the quadratic entries are\foot{This particular
matrix factorization is valid for $\alpha_3\ne 0$ because then $L_1$
and $L_2$ are linear independent.  For the singular limit $\alpha_3=0$
one needs to go to a different coordinate patch \BrunnerMT. For
$\alpha_1=0$ the matrix $E$ also becomes singular. However, this
can be fixed by apply a gauge transformation of the form $Q_1\rightarrow
Q_1+{\alpha_2^2 \over \alpha_1\alpha_2\alpha_3^2} x_1 L_2$,
$Q_2\rightarrow Q_2-{\alpha_2^2 \over \alpha_1\alpha_2\alpha_3^2}
x_1 L_1 $. Similarly one proceeds for $\alpha_2=0$.} 
\eqn\Mtwoquad{\eqalign{Q_1 \, &=\,{1\over \alpha_1\alpha_2\alpha_3}
                                  \left( \alpha_1\alpha_2 x_1^2+\alpha_2^2 x_1 x_2-\alpha_1^2 x_2^2
                                         -\alpha_1\alpha_3 x_3^2\right) \ , \cr
                       Q_2 \, &=\,{1\over \alpha_1\alpha_2\alpha_3}
                                  \left( \alpha_2^2 x_1^2 - \alpha_1^2 x_1 x_2 -\alpha_1 \alpha_2 x_2^2
                                   +\alpha_3^2 x_1 x_3\right) \ . }}
There is also the three-dimensional factorization given by:
\eqn\JEthree{
P_\thrbythr: \cases{J_ \thrbythr\!\!\!&= 
$\left(\matrix{\alpha_1 \, x_1  & \alpha_2 \, x_3   & 
\alpha_3 \, x_2   \cr 
\alpha_3 \, x_3   & \alpha_1 \, x_2  & \alpha_2 \, x_1  \cr
\alpha_2 \, x_2   & \alpha_3 \, x_1  & \alpha_1 \, x_3}\right)$
\crcr
E_\thrbythr\!\!\!&=
$\left(\matrix{ 
{1 \over \alpha_1}x_1^2 -  {\alpha_1 \over \alpha_2\alpha_3} x_2 \, x_3 &
{1 \over \alpha_3}x_3^2 -  {\alpha_3 \over \alpha_1\alpha_2} x_1\, x_2 &
{1 \over \alpha_2}x_2^2 - {\alpha_2 \over \alpha_1\alpha_3} x_1\, x_3 \cr
{1 \over \alpha_2}x_3^2 - {\alpha_2 \over \alpha_1\alpha_3} x_1\, x_2 &
{1 \over \alpha_1}x_2^2 - {\alpha_1 \over \alpha_2\alpha_3} x_1\, x_3 &
{1 \over \alpha_3}x_1^2 - {\alpha_3 \over \alpha_1\alpha_2} x_2\, x_3 \cr
{1 \over \alpha_3}x_2^2 - {\alpha_3 \over \alpha_1\alpha_2} x_1\, x_3 &
{1 \over \alpha_2}x_1^2 - {\alpha_2 \over \alpha_1\alpha_3} x_2\, x_3 &
{1 \over  \alpha_1}x_3^2 - {\alpha_1 \over \alpha_2\alpha_3} x_1\, x_2}\right)
$. }
}
Both \Mtwo\ and \JEthree\ represent valid matrix factorizations satisfying \matfac\ 
precisely when the parameters, just like the coordinates $x_\ell$, 
satisfy the cubic equation:
\eqn\Cubicalpha{ 
\big( {\alpha_1}^3 + {\alpha_2}^3 + {\alpha_3}^3 \big) ~-~ 
3\, a\, \alpha_1 \alpha_2 \alpha_3 ~=~0
 \ .}
One may therefore uniformize these parameters by taking, once again:
\eqn\alphauni{\alpha_\ell ~=~ \mu_\ell(\zeta) \,,}
for some parameter, $\zeta$. The physical interpretation is that
the factorization parameters, $\alpha_\ell$, are moduli of the
$D$-branes, and $\zeta$ is the associated flat coordinate that
labels a point on the curve corresponding to the
location of the $D0$-brane component of the $(D2,D0)$ brane
configuration.

As mentioned above, (indecomposable) $B$-type $D$-branes are
labeled by their $RR$~charges and location:
$(r(\cE),c_1(\cE),\zeta)\equiv(N_2,N_0,\zeta)$.  In discussing
the corresponding holomorphic vector bundles we will adopt a
common notation, $\cE(r,c_1)$, that suppresses the parameter,
$\zeta$.  We will also use the notation, $\cL^n$, to denote the
$n^{\rm th}$ power of the degree-one line bundle, $\cL$.

The question naturally arises as to the precise map between
these bundle data and the structure of the matrices $J$ and $E$ that 
define a given factorization, $P$.  This will be discussed in detail in
Sections 2 and 3 below.    We  recall here that the two
factorizations under discussion have been shown to 
correspond each to a triplet of branes with the following charges:
\eqn\branecharges{
\eqalign{
S\equiv{P_\twobytwo}:\qquad(\,r,\,c_1\,)^{LG}(S_a)\ &=\ \Big\{(1,0),(0,1),(-1,-1) \Big\}
\cr
L\equiv{P_\thrbythr}:\qquad(\,r,\,c_1\,)^{LG}(L_a)\ &=\ \Big\{(2,1),(-1,1),(-1,-2) \Big\}.
}}
As indicated above, we will denote the two factorizations by $S$
and $L$, each comprising three branes denoted by $S_a$
and $L_a$.\foot{In BCFT language, the $L_a$ are the ``Recknagel-Schomerus'' 
branes \doubref\RecknagelSB\BrunnerJQ\ with smallest charges, which
correspond to holomorphic vector bundles inherited from the ambient
$\IP^2$ and as such do not generate the full $RR$ charge lattice.
The $S_a$ correspond to the recently-discovered ``permutation
branes'' \refs{\AshokZB,\BrunnerFV,\EngerJK}\ that form an integral
basis of the charge lattice on the curve.}  These charge assignments,
at least for the $\twobytwo$ factorization, were originally computed
rather indirectly \refs{\AshokZB,\BrunnerMT}: The intersection
matrices were computed using the matrix formulation and then the
results were compared with the intersection matrices computed from
the conformal field theory and from the geometry.  Part of our
purpose here is to give a far more direct algorithm for computing
these geometric data from the matrices.

Under mirror symmetry, K\"ahler and complex structure
moduli exchange and the $B$-type branes map into $A$-type
$D1$-branes, which are labeled by the winding numbers $p$ and $q$:
$(r,\,c_1)_B=(p,q)_A$. Moreover, the brane modulus, $\zeta$, maps
into a complex modulus comprising both position and Wilson line
moduli of the $D1$-brane. Thus, the lattice of brane charges can
be drawn on the covering space of $\Sigma$, as shown in \lfig\branes.
%%%%%%%%%%%%%%%%%%%%%%%%%%%%
\figinsert\branes{The $\twobytwo$ and $\thrbythr$ factorizations
correspond, via mirror symmetry, to $D1$-branes that stretch
along the ``short'' ($S$) and ``long''
($L$) diagonals on the elliptic curve, respectively. 
Here we show these $D1$ branes (suppressing
the anti-branes with opposite charges, which correspond
to factorizations where $J$ and $E$ are exchanged).
This figure also provides a useful and simple graphical 
representation of the $RR$~charge lattice for the 
$B$-branes. }{2.in}{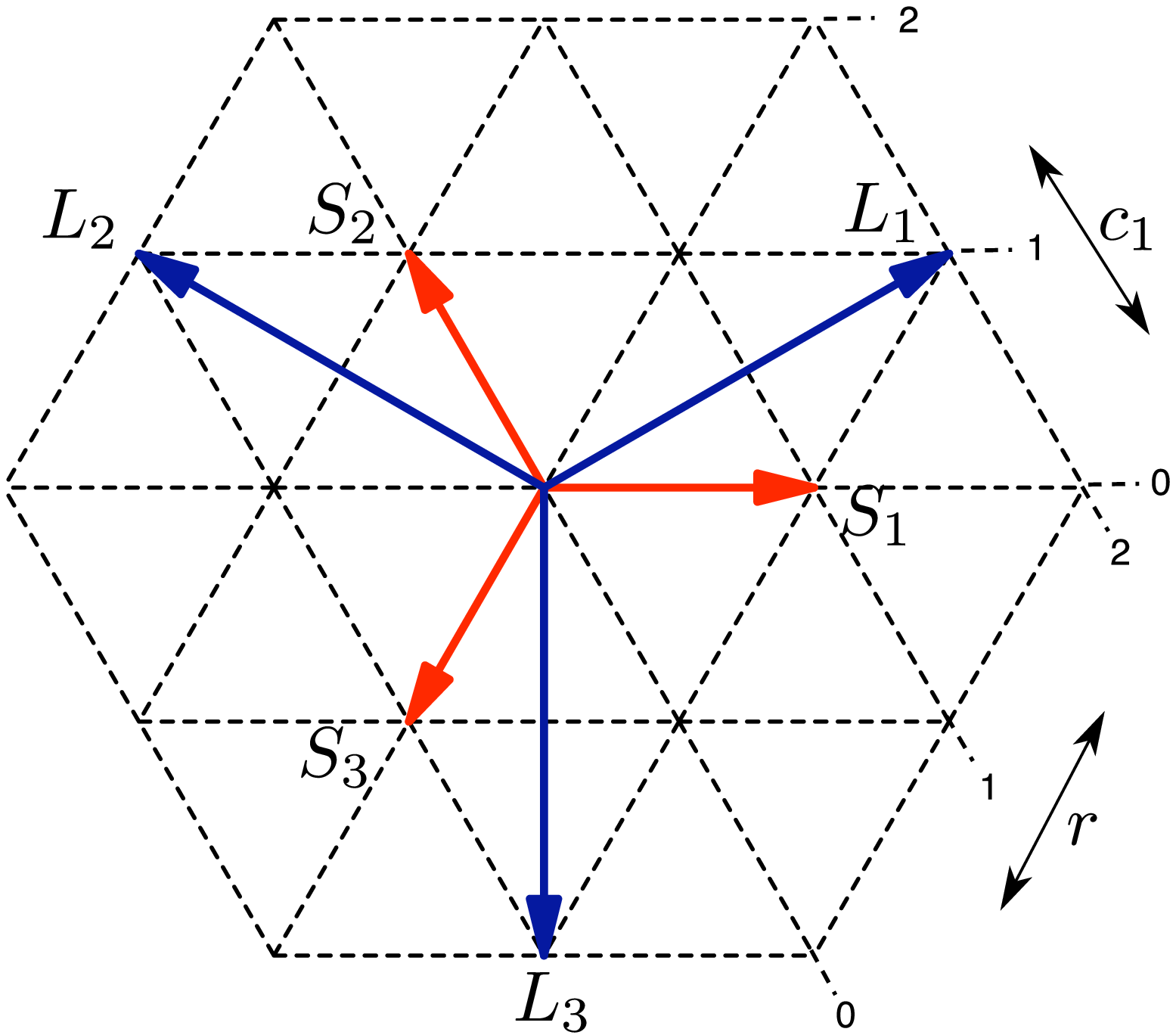}
%%%%%%%%%%%%%%%%%%%%%%%%%%%%

We have listed in \branecharges\ the charges corresponding to the
``Gepner-point'' in the K\"ahler moduli space, which is natural
from the Landau-Ginzburg perspective. However, recall that $RR$~charges
are ambiguous due to monodromy, or flow of gradings
\doubref\DouglasGI\DouglasQW, in the K\"ahler moduli. Matrix
factorizations, which pertain to the topological $B$-model and thus
depend only on the complex structure moduli, are insensitive to
variations of the K\"ahler moduli
 and therefore cannot distinguish bundles differing
by such monodromies.  For example, looping around the large radius
limit induces a monodromy that amounts to tensoring  with powers
of the line bundle $\cL^{3}$, which means that the first Chern
number of a rank-$r$ bundle will jump by   $\pm 3r$.

On the other hand, performing a
``partial monodromy'' by moving from the Gepner point to the large
radius limit, the charges \branecharges\ flow according to tensoring
with the line bundle\foot{This can be understood from a linear sigma
model point of view \HeHo. 
Note that this is also closely related to Seiberg dualities \BerensteinFI.}~$\cL^{-2}$, \ie,
$(r,\,c_1)\rightarrow(r,\,c_1-2r)$, and this results in the following
list of charges \refs{\DouglasQW,\EzhuthachanJR}:
\eqn\shiftedcharges{
\eqalign{
{\twobytwo}:\qquad(\,r,\,c_1\,)^{LR}(S_a)\ &=\ \Big\{(1,-2),(0,1),(-1,1)\Big\}
\cr
{\thrbythr}:\qquad(\,r,\,c_1\,)^{LR}(L_a)\ &=\ \Big\{(2,-3),(-1,3),(-1,0)\Big\}
\,.}}
In the following, we will adopt this labeling convention because
it refers to the large radius limit, which is semi-classical from
the point of view of the sigma-model and coincides with the labeling in the
mathematics literature.

%%%%%%%%%%%%%%%%%%%%%%%%%%%%%%%%%%
%:open cohom
%%%%%%%%%%%%%%%%%%%%%%%%%%%%%%%%%%

In order to discuss topological tachyon condensation, we need to
determine the relevant part of the open-string spectrum. Since, in
the twisted theory, the tachyons become fermionic operators  \VafaQF\
(coupling to bosonic deformation parameters), we will consider only
fermionic operators here. There are two classes of such operators.
First, there are boundary preserving operators, represented by
$2n_P\times2n_P$ dimensional, block off-diagonal matrices of the
form: $\Omega_P\equiv\Psi_{(P,P)}=\big({0\ \ \delta J_P\atop\delta
E_P\ 0}\big)$, which are tied to a single brane and describe moduli
corresponding to infinitesimal deformations of the brane (such as
position shifts).  Most of the physics literature on open-string
TFT deals with this class only. Mathematically these operators
correspond to endomorphisms of the object~$P$.

The other class consists of boundary changing operators,
$\Psi_{(P,Q)}$, which correspond to open strings stretching between
pairs of branes $P$ and $Q$ and thus are localized at their
intersection.\foot{Non-intersecting branes have a trivial topological
open-string spectrum between them because the open strings are
massive and so are not part of the cohomology.} See \lfig\intersecting.
These are the topological version of tachyons, and indeed they
typically have $R$-charges $q<1$, which means that they are
relevant operators inducing a non-trivial boundary RG flow.
 They can be written in terms of $2n_P\times2n_Q$ dimensional, block
 off-diagonal matrices of the form:
$\Psi_{(P,Q)}=\big({0\ \ \psi_0\atop\psi_1\ 0}\big)$.  The r\^ole
of $\psi_0$ and $\psi_1$ as maps between the composite objects $P$
and $Q$ can be visualized by the following diagram:

\eqn\FermCoh{\xym{P_1 \ar@/^/[d]^{E_P} \ar[rrd] && Q_1 \ar@/^/[d]^{E_Q}  \cr
                  P_0 \ar@/^/[u]^{J_P} \ar[rru] && Q_0 \ar@/^/[u]^{J_Q} }
                  \ . 
                  \setbox0=\hbox{$\scriptstyle \psi_1$}
                  \setbox1=\hbox{\kern-\wd0 $\scriptstyle\psi_0$}
                  \hbox{\kern-19ex \raise 13pt \box0 \lower 10pt \box1}
                  \hbox{\kern+17ex}}
There is a similar structure for the bosonic operators, $\Phi$.

%%%%%%%%%%%%%%%%%%%%%%%%%%%%
\vskip5mm
\figinsert\intersecting{
This shows schematically where boundary preserving ($\Omega$) and 
boundary changing ($\Psi,\,\Phi$) operators are located on intersecting 
$D$-branes and on the boundary of the world-sheet,~$Z$. }{1.2in}{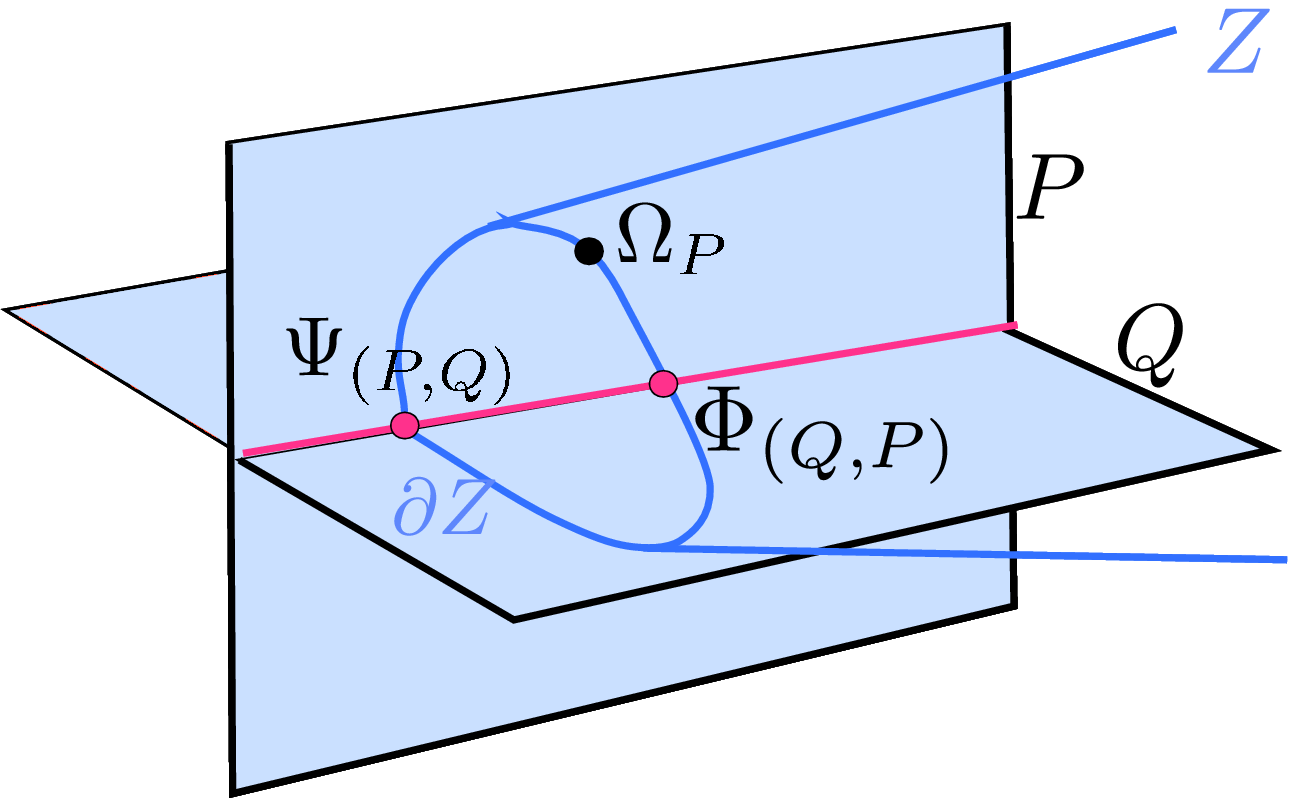} 
%%%%%%%%%%%%%%%%%%%%%%%%%%%%

The physical operators then correspond to the non-trivial cohomology 
elements of the BRST operator,
which can be written in the form: $\cQ=\big({0\ \ J\atop E\ 0}\big)$. 
That is, we require them to be closed:
\eqn\PsiPhys{\eqalign{0\,&=\,E_Q\,\psi_0+\psi_1\,J_P \ , \cr
                      0\,&=\,J_Q\,\psi_1+\psi_0\,E_P \ , }}
modulo exactness
\eqn\PsiExact{\eqalign{\psi_0^{\rm ex}\,&=\,J_Q\,\varphi_0-\varphi_1\,J_P \ , \cr
                       \psi_1^{\rm ex}\,&=\,E_Q\,\varphi_1-\varphi_0\,E_P \ , }}
for any choice of matrices $\varphi_0$ and $\varphi_1$.

The open-string spectrum pertaining to the $\twobytwo$ and $\thrbythr$
matrix factorizations can be represented by the quiver diagram shown
in \lfig\quiver. The number of arrows indicates the number of
inequivalent fermionic cohomology elements, and coincides with the
number of intersection points of the mirror $A$ $D1$-branes on the
curve.  For each arrow there is implicitly another one running in
the opposite direction, which corresponds to the Serre dual, bosonic
operator and which we do not show.\foot{Fermionic operators $\Psi_{(P,Q)}$
correspond to ${\rm Ext}(P,Q)$ while bosonic operators $\Phi_{(Q,P)}$ 
correspond to  ${\rm Hom}(Q,P)$. Serre duality implies ${\rm Hom}(Q,P)\sim
{\rm Ext}(P,Q)$.}   Moreover, closed loops
denote the boundary preserving deformations, $\Omega_P$.  
%%%%%%%%%%%%%%%%%%%%%%%%%%
\figinsert\quiver{The quiver diagram displaying the fermionic open
string states related to the short- and long-diagonal branes, $S_a$
and $L_a$. The arrows depict the multiplicities. For simplicity, we do
not show the anti-branes $\bar S_a$, $\bar L_a$.}{2.5in}{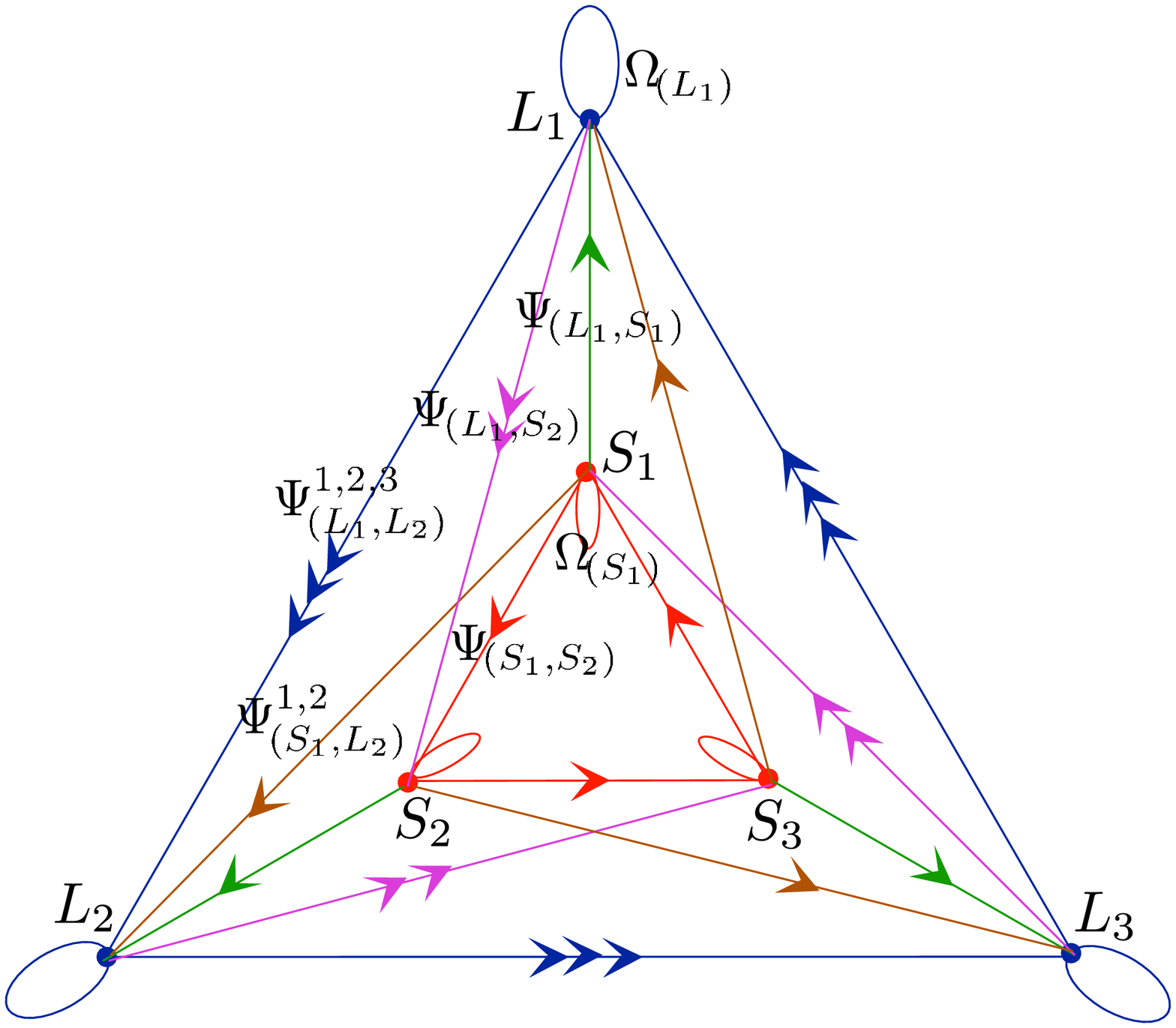} 
%%%%%%%%%%%%%%%%%%%%%%%%%%

\goodbreak
%%%%%%%%%%%%%%%%%%%%%%%%%%%%%%%%%%
%:tc examples
\subsec{Examples of tachyon condensation}
%%%%%%%%%%%%%%%%%%%%%%%%%%%%%%%%%%

To illustrate one of the basic techniques we use in this paper, we sketch
the two simplest possible examples of tachyon condensation; a more extensive 
analysis will be presented  in the subsequent sections. To recapitulate the basic point: 
One wants to find, and turn on, a  suitable boundary changing operator $\Psi_{(P,Q)}$, 
whose effect is to condense two sets of branes, $P$ and $Q$, to form some 
composite, $R$. This process can be visualized by collapsing the two-brane 
system shown in eq.~\FermCoh\ into a single object as follows:
\eqn\tachcondiagram{
\xym{P_1 \ar@/^/[d]^{E_P} \ar[rrd] && Q_1 \ar@/^/[d]^{E_Q}  \cr
                  P_0 \ar@/^/[u]^{J_P} \ar[rru] && Q_0 \ar@/^/[u]^{J_Q} }
                  \  
                  \setbox0=\hbox{$\scriptstyle \psi_1$}
                  \setbox1=\hbox{\kern-\wd0 $\scriptstyle\psi_0$}
                  \hbox{\kern-19ex \raise 13pt \box0 \lower 10pt \box1}
                  \hbox{\kern+17ex}
\ \Longrightarrow\ 
\left[\xym{R_1 \ar@/^/[d]^{E_R} \cr R_0\ar@/^/[u]^{J_R}}
\right]\ \equiv\ R\ ,
}
where the composite maps, $J_R$ and $E_R$, can be thought of as 
$(n_P+n_Q)\times(n_P+n_Q)$   block matrices of the form:
\eqn\Rmaps{
J_R\,=\,\pmatrix{J_Q & \psi_0 \cr 0 & J_P} \ , \qquad
                 E_R\,=\,\pmatrix{E_Q & \psi_1\cr 0 & E_P } \ . 
}
If $\Psi$ is a non-trivial cohomology element, these matrices satisfy
the factorization condition \matfac\ and represent a new $B$-type
of $D$-brane.\foot{Note that in principle the maps need not be upper
triangular. Factorization then becomes a highly non-trivial condition,
which in general is satisfied only on a sub-locus of the combined
open/closed string moduli space; for examples, see ref.~\HerbstZM.} In the following,
we will often use the following shorthand notation to denote the
process of tachyon condensation: 
\eqn\suxx{P\,\condense_{\Psi}\,Q\,\Longrightarrow\,R\ .
}
Note that this construction is well-known in the
mathematical literature and goes by the name of the ``cone construction''.
That is, one writes a sequence of maps in the form of a ``distinguished
triangle'':
\eqn\cone{\xymatrix{ P[1]  \ar@{->}[r]^{\Psi[1]}  &   Q  
\ar@{->}[r]   & R \ar@{->}[r]  &P}\ ,}
where the composite $R$ coincides with what is called the ``mapping cone'' 
(for details see,  for example, ref.~\AspinwallJR, and for subtleties
concerning off-shell versus on-shell physics, see 
refs.~\doubref\LazaroiuJM\LazaroiuMD).

In practice, one would like to find a simple way to determine exactly what  this 
new $D$-brane is.  From the point of view of matrix factorization, the obvious,
but rather impractical method is to try to reduce the matrices to some
standard set of canonical forms.  The most efficient way of achieving 
this end is, in fact, to determine the underlying bundle data for the
new brane, $R$.  Of course, as far as the $K$-theoretic data are concerned,  
ranks and first Chern classes of bundles are additive
under condensation and thus can be trivially determined. 
It is, however, not so obvious how to determine data beyond $RR$~charges,
that is, the extent to which $R$ is decomposable, and how the parameters
of $P$ and $Q$ combine into the parameter(s) of $R$.   Here we illustrate
this issue with two examples. 

The simplest possible example is combining a pair of $2\times
2$-matrix factorizations.  There are two ways to achieve this,
namely either combining a pair of branes, or a brane with an
anti-brane.  We start with the second possibility and take the first
$D$-brane to be given by the $2\times 2$-matrix factorization
$S(\alpha)=(J(\alpha),E(\alpha))$ in eq.~\Mtwo, while the anti-D-brane
is represented by the $2\times 2$-matrix factorization $\bar
S(\beta)=(-E(\beta),-J(\beta))$. The outcome of the condensation
depends on which precise members, $S_a$ and $\bar S_b$, of the two
factorizations $S$ and $\bar S$ we choose to condense, and this is
tied to which specific tachyonic operator in the cohomology between
the factorizations we choose to switch on.

From the vector addition of $RR$ charges 
shown in \lfig\condensation\ we expect that there should be two types
of tachyonic perturbations, which either lead to complete annihilation,
or to a composite corresponding to a $3\times 3$ factorization, $L$ or 
 $\bar L$.  The simplest possibility is of course the complete
annihilation of the brane/anti-brane pair, which we will discuss
momentarily; the other, more involved situation will be analyzed in Section~5.

%%%%%%%%%%%%%%%%%%%%%%%%%%
\figinsert\condensation{In addition to the two simple condensation processes 
discussed in this section, we show here how the rank two composites
can be generated by condensing $S$ and $L$ branes or anti-branes,
$\bar S$ and $\bar L$, respectively, by switching on suitable tachyons. 
The anti-branes are denoted by dashed arrows.  Details
will be discussed in Section 7.}{2.in}{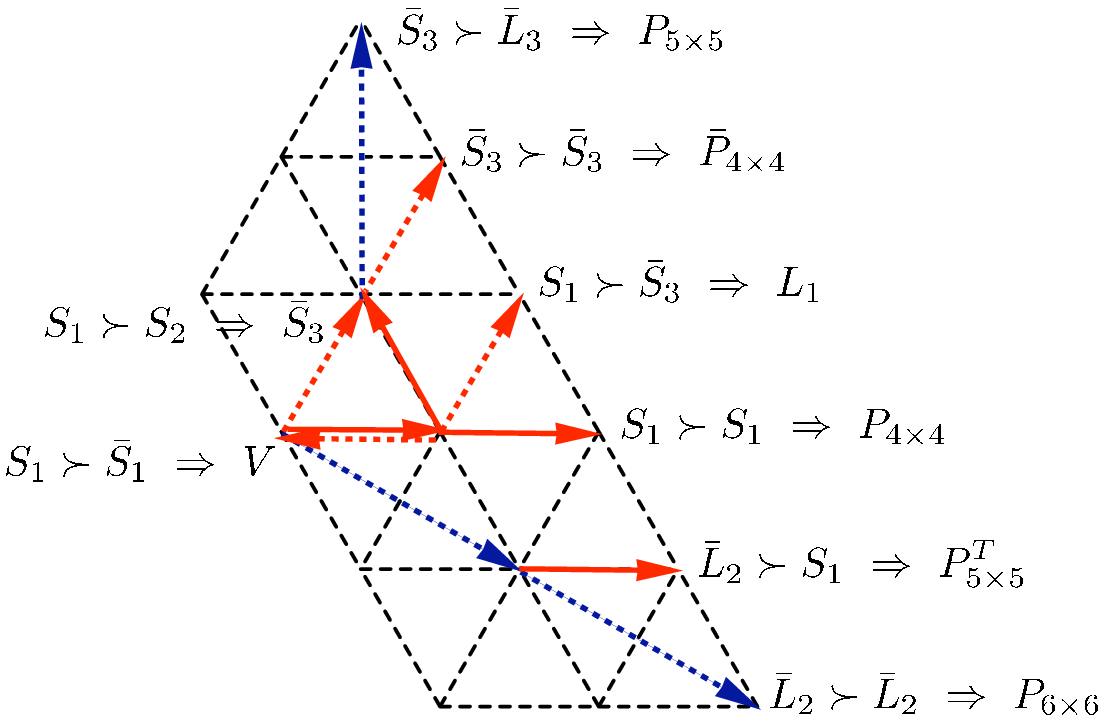}
%%%%%%%%%%%%%%%%%%%%%%%%%%

For generic values of the moduli, $\alpha_i$ and $\beta_i$, 
of the branes $S_1(\al)$ and $\bar S_1(\be)$, the
relevant cohomology of fermionic open-string operators (determined
by the physical state condition \PsiPhys{} modulo \PsiExact{}) turns
out to be empty.  This reflects the fact that, in the $A$-model mirror picture,
the anti-parallel $D1$-brane/anti-brane pair does not intersect,
so that there is no operator that can be used to form a tachyon condensate.
However, upon tuning $\alpha_i=\beta_i$ it is easy to check that
the cohomology jumps and now contains the fermionic operator
$\Psi_{(S_1,\bar S_1)}\sim(\psi_0,\psi_1)$ given by
\eqn\TachB{\psi_0 \,=\, \id_{2\times 2} \ , \qquad \psi_1 \,=\, \id_{2\times 2} \ . }
This tuning of the open-string moduli corresponds to the situation
where the anti-parallel brane and anti-brane move on top of each
other and where the Wilson line moduli of the branes are tuned to
match.  In fact the cohomology then also contains an extra bosonic
operator, so that the intersection index $\Chi_{S_1,\bar
S_1}=\Tr_{S_1,\bar S_1}(-1)^F={\rm dim} ({\rm Hom}(S_1,\bar S_1))
- {\rm dim}({\rm Ext}(S_1,\bar S_1))=0$ does not change.\foot{
The appearance of such pairs is not a special
feature of the $2\times 2$ factorizations, but instead this operator, as well as
its bosonic partner, appear for any anti-parallel brane/anti-brane
pair as long as the open-string moduli of both branes coincide.}

To see that the result of the condensation induced by \TachB\ is
indeed the expected trivial ground state, one can make use of the gauge
transformations \gauge, and most particularly, of the constant
entries of the $\psi_{0,1}$, to make elementary row and column reductions
of $J_R$ and $E_R$ to show that they are gauge equivalent to the trivial
factorization (given by matrices with either $1$ or $W$ on the
diagonal).  See Appendix B for more discussion of gauge transformations
and  row and column reduction.

The next-to-simplest possibility is to condense two of the same
type of matrix factorization, but with different values of the moduli:
$S(\alpha)$ condensed with $S(\beta)$.   We find that there are two
distinct choices for the tachyon, and the choice of the tachyon
again relates to which particular branes, $S_a(\alpha)$ and
$S_b(\beta)$, in the families described by the factorizations,
$S(\alpha)$ and $S(\beta)$, participate in the condensation.  In section~4
we will analyze  the interrelation between the tachyonic spectrum
and the choice of brane pairs in complete detail, using equivariant $R$-symmetry.
For the present we will choose  the tachyon that leads to the
condensation of $S_1$ and $S_2$ to form a composite anti-brane,
$\bar S_3$ ({\it c.f.}~\lfig\condensation).

This tachyon, represented by the boundary changing fermionic
operator~$\Psi_{(S_1,S_2)}(\al,\be)$, is found to be
\foot{These expressions have also been obtained by J.\ Walcher
who participated in early stages of this project.}
\eqn\TachC{\psi_0(\alpha,\beta) \,=\, \pmatrix{H_1(\alpha,\beta) & H_2(\alpha,\beta) \cr
                                  G_1(\alpha,\beta) & G_2(\alpha,\beta) }, \ \
           \psi_1(\alpha,\beta) \,=\, \pmatrix{-G_2(\beta,\alpha) & \hphantom{-}H_2(\beta,\alpha) \cr
                      \hphantom{-}G_1(\beta,\alpha) & -H_1(\beta,\alpha) }.  }
The constant entries are given by\foot{These vanish for $\al_i=\be_i$.
However, at $\al_i=\be_i$ there exists a cohomology
element of different form, for which the following
arguments hold analogously.}
\eqn\TachCconst{
\eqalign{G_1(\alpha,\beta)\,&=\,\alpha_1\beta_3^2-\alpha_3\beta_1\beta_3 \ ,\cr
         G_2(\alpha,\beta)\,&=\,-\alpha_2\beta_3^2+\alpha_3 \beta_2 \beta_3  \ , }}
while the linear entries are:
\eqn\TachClin{\eqalign{
   H_1(\alpha,\beta)\,&=\,
         \left(\alpha_2-{\alpha_1\beta_2\over\beta_1}\right)x_1 
         -\alpha_1\left({\alpha_1\over\alpha_2}-{\beta_1\over\beta_2}\right)x_2 
         -\alpha_3\left({\alpha_3\over\alpha_2}-{\beta_3\over\beta_2}\right)x_3  \ , \cr
   H_2(\alpha,\beta)\,&=\,
         \left({\alpha_1^2\over \alpha_2} + {\alpha_2\beta_2\over \beta_1}- 
                        {\alpha_3\beta_1^2\over\beta_2\beta_3}-
                        {\alpha_3\beta_2^2\over\beta_1\beta_3} \right) x_1
         +\left(\alpha_1-{\alpha_2\beta_1 \over \beta_2} \right) x_2  \ .  }}

The fermionic operator~\TachC\ can be used to construct a condensate~$R$
via the cone construction mentioned above, and one obtains the following
$4\times 4$-matrix factorization:
\eqn\MtwotwoC{J_R(\alpha,\beta)\,=\,\pmatrix{J(\beta) & \psi_0(\alpha,\beta) \cr 
0 & J(\alpha) } \ , \qquad
  E_R(\alpha,\beta)\,=\,\pmatrix{E(\beta) & \psi_1(\alpha,\beta) \cr 0 & E(\alpha) } \ . }
Since the tachyon operator, $\Psi_{(S_1,S_2)}$, contains the constant
entries~\TachCconst, the matrix-factorization can again be
simplified  by the process of row and column elimination. After a 
few straightforward steps of algebra, the
matrix factorization \MtwotwoC\ can be cast into its gauge-equivalent
form:
\eqn\MtwotwoCgauge{J_R(\ga)=\pmatrix{0 & -L_1(\ga) & 0 & -Q_2(\ga) \cr
                                    0 & 0 & 1 & 0 \cr W & 0 & 0 & 0 \cr
                                    0 & \hphantom{-}L_2(\ga) & 0 & -Q_1(\ga)}, \ \ 
                   E_R(\ga)=\pmatrix{0 & 0 & 1 & 0\cr -Q_1(\ga) & 0 & 0 &\hphantom{-}Q_2(\ga) \cr
                                    0 & W & 0 & 0\cr -L_2(\ga) & 0 & 0 &-L_1(\ga) } , }
where
\eqn\Defzeta{\eqalign{\ga_1\,&=\,\alpha_2^2\beta_1\beta_2-\alpha_1\alpha_3\beta_3^2 \ , \cr
                      \ga_2\,&=\,\alpha_3^2\beta_2\beta_3-\alpha_1\alpha_2\beta_1^2 \ , \cr
                      \ga_3\,&=\,\alpha_1^2\beta_1\beta_3-\alpha_2\alpha_3\beta_2^2 \ . }}
The parameters~$\ga_i$ also satisfy the cubic torus equation, just
like the parameters~$\alpha_i$ in \Cubicalpha.  Indeed, the $\gamma_i$
become much simpler when written in terms of the uniformizing variables and,
as we will discuss in later sections,  \Defzeta\ merely represents
the addition formula for theta functions.  

If we now drop the two trivial D-branes in the matrix factorization \MtwotwoCgauge{}
(each corresponding to $J=\oneone_{1\times 1}$ and $E=W \oneone_{1\times 1}$, 
or vice versa), we finally obtain
for the condensate~$R$ the following $2\times 2$ factorization:
\eqn\MtwotwoCgaugeB{
                   J_R(\ga)\,=\,\pmatrix{-L_1(\ga) & -Q_2(\ga) \cr
                                    \hphantom{-}L_2(\ga) & -Q_1(\ga)} \ , \qquad
                   E_R(\ga)\,=\,\pmatrix{-Q_1(\ga) &\hphantom{-}Q_2(\ga) \cr
                                    -L_2(\ga) &-L_1(\ga) } \ . }
We can readily identify it with $\bar S(\ga)=(-E(\ga),-J(\ga))$, which corresponds 
to an anti-brane ({\it c.f.} \Mtwo{}) depending on the open-string modulus, $\ga$.

%%%%%%%%%%%%%%%%%%%%%%%%%%%%%%%%%%
\newsec{Holomorphic vector bundles and matrix factorizations }
%%%%%%%%%%%%%%%%%%%%%%%%%%%%%%%%%%

As we discussed earlier,  the classification of $B$-type branes is
given by that of holomorphic vector bundles\foot{More generally,
$B$-type branes are described in terms of coherent sheaves, which is a notion
more general than vector bundles. As we will explain later and as
pointed out in \AshokZB, in order to describe sheaves that are not
vector bundles, the construction detailed below needs to be refined
by using equivariant $R$-symmetry.} and this is equivalent to finding
matrix factorizations.  Since the primary focus of this paper is
upon matrix factorization, and how to generate new matrix factorizations
from old ones via tachyon condensation, we will discuss in some
detail  how one can extract the bundle data from the matrices.  The
converse  construction is also possible \Laza, but  we will not address
it directly here.  

More concretely, we will see in the following how one can associate a bundle with each
of the matrix factors, $J$ and $E$. We will denote these bundles
by $\cE_J$ and $\cE_E$, respectively. The charges of the branes will then be determined by 
$\cE_J$ and the charges of the anti-branes by $\cE_E$. The impatient reader, who would
like to skip the technical details, is invited to move on to Section~3.4, where we summarize our findings with regard to the brane charges.

Before starting with the discussion, one should also note that the bundle data that
we seek has a natural ambiguity coming from the fact that the
coordinates, $x_j$, are themselves sections of the line bundle,
$\cL^3$, and since we allow ourselves to multiply and divide vectors
by the $x_j$, any bundle data that we get will be ambiguous up to
tensoring by powers of this line bundle.  Therefore, $\cE(r,d)$,
will be indistinguishable from  $\cE(r,d \pm 3 r)$; note that this
is compatible with the ambiguity induced by the large radius
monodromy in the K\"ahler moduli space.

There is a closely related issue with the definitions of $J(x)$ and $E(x)$:  The 
fact that  the $x_j$ are sections of a line bundle means that the matrices
will generically have non-trivial transformations between patches on $\Sigma$.
That is, suppose that between two patches one has $x_j \to g\, x_j$ for
some transition function, $g$.   Then one has $W(x) \to g^3 W(x)$ and
\eqn\mattrans{J(g x) ~=~ g^k \, G_1(g) \, J(x) \, G_2(g)^{-1} \,, \qquad 
E(g x) ~=~ g^{3-k} \, G_2(g)\, E(x) \, G_1 (g)^{-1} \,,}
where $k\in \IZ$ is arbitrary and $G_1$ and $G_2$ are, in fact, matrices that 
describe the action of the $R$-symmetry.  We will discuss $R$-symmetry in
more detail in Section 5, but here we note that if the matrix elements of $J$ and $E$  
each have a well-defined (but possibly different) degree, then $G_1$
and $G_2$ will be diagonal with integer powers of $g$.

%%%%%%%%%%%%%%%%%
\subsec{The direct approach}
%%%%%%%%%%%%%%%%%

We start by taking the most naive approach to the problem and seeing
how far we can get.   Indeed, the most elementary way to exhibit the holomorphic 
vector bundles associated with a matrix factorization is to look at the kernels of the matrix
factors.  Given an $n \times n$-matrix factorization of the form \matfac,  it follow that 
\eqn\rankdefn{ {\rm det}(J) ~=~ W^{n-r} \,, \qquad {\rm det}(E) ~=~ W^r \,,}
for some $r \in \IZ_+$.   On the surface, $\Sigma$, defined by $W=0$
in a complex projective space, the kernel  of $E$ is thus a rank-$r$ vector 
bundle, $\cE_J$, on $\Sigma$ and it is spanned by the
columns of $J$.  Conversely, the columns of $E$ span the rank-$(n-r)$
vector  bundle,  $\cE_E$,  that is the kernel of $J$.  
In terms of the objects, $P$, in \DbP, the idea is that
we are extracting the data about the condensation that leads to $P$ by 
passing to the kernels of the maps.   The two bundles extracted in this way correspond to
the ``constituent'' $D$-brane and its anti-$D$-brane, $P_0$ and
$P_1$ in \DbP.  In more physical terms, the same, naive argument
\GreeneUT\ that leads from the bulk Landau-Ginzburg model to the
surface $W=0$ in projective space, when extended to the boundary
suggests that one should look at zeroes of $J$ and $E$, for branes
and anti-branes respectively, on the boundary.  A proper justification
of this argument may, however, require the boundary linear sigma-model
\HeHo.

There is a technical problem with the naive construction above:  
The matrices may have non-trivial transition properties \mattrans\ and so taking the
linear span of columns may not be well-defined upon $\Sigma$.
The simplest remedy is to consider the kernels of: 
\eqn\goodJE{\hat J(x) ~\equiv~   \, G_1 (x_p^{-1}) \, J(x) \, G_2(x_p^{-1})^{-1} \,, 
\qquad  \hat E(x) ~\equiv~  G_2(x_p^{-1})\, E(x) \, G_1 (x_p^{-1})^{-1} \,,}
for some $p=1,2,3$ and where $G_1$ and $G_2$ are defined in
\mattrans.  This is similar to the gauge transformation in the sense
of \gauge\ but it is not invertible in the polynomial ring.  
I does, however, make the entries of $\hat J$ and $\hat E$ into 
rational functions of the $x_j$.  However, $\hat J$ and $\hat E$ have
well-defined kernels and images on $\Sigma$ 
(because $\hat G_1 = \hat G_2 = \oneone$) and thus 
$\cE_{\hat  J}$ and $\cE_{\hat  E}$ are well defined in $\Sigma$.
This is what we really mean by ``multiplying and dividing by powers
of $x_j$,'' and this is the reason for the ambiguity, $\cE(r,d+ 3 j r)$
for $j \in \IZ$, in the bundle data.  We will proceed with the mild abuse in 
terminology by thinking of   $\cE_J$ and $\cE_E$ as vector bundles, 
with the understanding
that they can be turned into vector bundles using the foregoing construction.
 
Since the matrices are polynomials in the $x_j$, the spanning sets (the
matrix columns) of these vector bundles are holomorphic.   
These vector bundles will clearly have transition functions  induced from
those of the $x_j$ and it is tempting to assume that 
these will be sufficient to define the vector bundles $\cE_J$ and $\cE_E$.
However, it is not that simple: The columns of $J$ (or $E$) are not linearly 
independent and  therefore do not constitute a basis.   The obvious remedy
is to choose a linearly independent subset of columns and use this as a basis, but
the problem is that such a choice of basis cannot be done globally on $\Sigma$:
One needs to introduce further patches on $\Sigma$ and use different sets of columns 
in each of these new patches.  There will also be   non-trivial
transition functions between such  patches.  Mathematically, this amounts
to constructing an explicit  local trivialization of the vector bundle.
These transition functions can be written
as rational functions of the matrix elements and the patches can be arranged so that
these transition functions are holomorphic on the intersections.  Thus one
can easily see that $\cE_J$ and $\cE_E$ are holomorphic vector bundles, and
one can, in principle, compute their properties in this manner.
 
For example, consider the $3 \times 3$ factorization given by \JEthree.
Every matrix element of $J$ has degree one and every matrix element of $E$ 
has degree two and so $G_1 = G_2 = \oneone$.  
Since ${\rm det}(J) = W$, $\cE_E$ has rank $1$, and so the columns of $E$ must
all be multiples of one another, which can easily be verified (mod $W=0$).  
Thus $\cE_E$ must be a line bundle
on $\Sigma$, and the matrices $G_1$ and $G_2$ are trivial.
 Now observe that the $j^{\rm th}$ column of $E$ vanishes identically
if $x_j = \alpha_1$, $x_{j+1} = \alpha_3$ and $x_{j+2} = \alpha_2$, where the
subscripts are taken mod $3$.  One therefore needs to use at least two of the columns
in two different patches, $U_1$ and $U_2$,  if one is to define the line bundle globally.  
The transition functions between these two patches will be a ratio of quadratic 
functions in the $x_j$, and thus meromorphic on $\Sigma$. Indeed, the patches can be chosen
so that the transition function on $U_1 \cap U_2$ is biholomorphic.    

From this simple example, we
see that to characterize the line bundle we not only need $G_1$, $G_2$ and the 
properties of  the $x_j$, but also the  non-trivial transition functions between patches in which
different sets of columns are linearly independent.  Thus the naive construction
of the holomorphic vector bundles $\cE_J$ and $\cE_E$ works nicely, but
it may not result in the simplest, canonical description of the bundle. 

To simplify and generalize this discussion, we start with a brief review of
holomorphic vector bundles on a torus.

%%%%%%%%%%%%%%%%%
\subsec{Holomorphic vector bundles on an elliptic curve}
%%%%%%%%%%%%%%%%%

To define the elliptic curve, $\Sigma_q$, 
corresponding to the surface $\Sigma$ with complex structure modulus, $q$, it is 
most convenient to think of it as an annulus in the complex plane with the interior
and exterior edges identified.  That is, one considers 
$\IC^* \equiv \IC \backslash\{0\} $ with the identification $z \sim q z$
and where $q \in \IC^*$ is a parameter.  It is also useful to use the
``additive parametrization'' where $q= e^{2\pi i \tau}$, $z= e^{2\pi i \xi}$ and
$\xi \sim \xi+1 \sim \xi+\tau$.   There is a natural
projection map, $\pi:  \IC^* \rightarrow \Sigma_q$,  and given a holomorphic vector 
bundle on $\Sigma_q$, one can pull it back to $\IC^*$.   One can then show
(see, for example, \refs{\PolishchukDB,\PolishchukAppell}) that the pull-back 
must be a trivial bundle on $\IC^*$, and so the bundle on $\Sigma_q$ is determined 
entirely by the ``gluing matrix'' under $z \sim q z$.   The trivial, rank-$r$ bundle on
$\IC^*$ is simply $\IC^* \times \IC^r$, and to obtain a rank $r$ bundle on $\Sigma_q$ one
must specify an invertible $r \times r$ matrix, $A(z)$, of  holomorphic  functions on 
$\IC^*$ and then one has:
\eqn\bundleequiv{\cV_r(A) \equiv \{(z,v) \in \IC^* \times \IC^r :  \ \ 
(z,v) ~\sim~ (q\, z\,, A(z)\,v) \} \,.}
Conversely, the set of non-trivial bundles is determined by the choices
of $A(z)$ up to gauge equivalence: $A(z) \to B(q z) A(z) B^{-1}(z)$
for some invertible $r \times r$  holomorphic matrix, $B(z)$.  Thus we
have a complete characterization of the vector bundles on $\Sigma_q$.

There is also a well-known result of Atiyah \Atiyah\ that states that
every indecomposable holomorphic vector bundle on $\Sigma_q$ is 
characterized by three parameters: (i)  the rank, $r$, (ii) the first
Chern number, $c_1$, and (iii) a point, $y$ on $\Sigma_q$.  Indeed, one can always
write $\cV_r(A)$ as $\tilde \cL \otimes \cV_r(e^N)$ where $\tilde  \cL$ is an 
appropriately chosen line bundle and $N$ is a {\it constant}, indecomposable, nilpotent
$r \times r$ matrix  \PolishchukDB.  We will discuss and illustrate this result
extensively in subsequent sections, but here we will merely note that the 
classical theta function:
\eqn\thetadefn{\hat \theta(z, y) ~\equiv~ \sum_{n \in \IZ} q^{{1 \over 2} n (n-1)} \,
(y \, z)^n \,,}
is a global holomorphic section of the line bundle  $\cV_1( y^{-1} z^{-1})$ on 
$\Sigma_q$,  where $A_y(z) =  y^{-1} z^{-1}$.
Note that this transition function involves a factor of $ z^{-1}$, 
which has winding number $-1$, and thus $c_1 =1$. 
Also observe that the transition matrix for $\hat\theta(z,q\,y)$ 
becomes $A_{qy}(z)=q^{-1}A_y(z)$, which is gauge equivalent to 
$A_y(z)$ with $B(z)=z^{-1}$. Hence both $\hat\theta(z,y)$ and 
$\hat\theta(z,qy)$  are global sections of the same line bundle 
$\cV_1( y^{-1} z^{-1})$ and therefore the parameter~$y$ may be thought of as 
living in $\Sigma_q$.

Returning to the line bundle spanned by the columns of
the $3 \times 3$ matrix, $E$, of \JEthree, an  elementary theta-function identity 
reveals that the  matrix elements of $E$ may be rewritten in terms of theta functions:
\eqn\Eij{E_{ij} ~=~ {\eta^2(\tau) \over \alpha_1\alpha_2\alpha_3}\,  
\mu_i(\xi - \zeta) \, \mu_j(\xi + \zeta) \,,}
where $x_\ell = \mu_\ell(\xi)$, $\alpha_\ell = \mu_\ell(\zeta)$ and 
$\eta(\tau)$ is the Dedekind $\eta$-function. 
Thus the columns of $E$ are all {\it holomorphic} multiples of the
single (nowhere-vanishing) basis  vector:
\eqn\Ethreebas{v_1 ~\equiv~ e^{3 \pi i (\xi - \zeta)} \, 
\left(\matrix{\mu_1(\xi - \zeta) \cr 
 \mu_2(\xi - \zeta) \cr \mu_3(\xi - \zeta) }\right) \,.}

The phase pre-factor has been included so as to render $v_1$ periodic
under $\xi \to \xi+1$.  Under $\xi \to \xi + \tau$ one has 
$v_1 \to - y^3 z^{-3} v_1 $, where $z = e^{2 \pi i \xi}$ and $y = e^{2 \pi i \zeta}$, 
and therefore the holomorphic transition function $A(z)$ reads
\eqn\uetrans{A(z)~=~-y^3\,z^{-3} \ . }
Thus the  kernel of $J$ leads to the line bundle, $\cV_1( y^{-3} z^{-3}) =
(\cV_1( y^{-1} z^{-1}))^3$, or $\cE(1,3)$,
 with $c_1 =3$  and a  parameter, $y$.   This bundle is, modulo the
 ambiguity outlined at the beginning of the section, equivalent to
 $\cE(1,3k), k \in \IZ$ and both  $\cE(1,0)$ and  $\cE(1,-3)$ are the ``anti-bundles''
 of two of the bundles whose charges were listed in \shiftedcharges.  
 We will discuss the relation
 with all the charges listed in \shiftedcharges\ in more detail after we have identified all the matrix
 bundles $\cE_J$, $\cE_E$, associated with the $\twobytwo$ and  $3\times 3$ factorizations.

%%%%%%%%%%%%%%%%%
\subsec{MCM modules}
%%%%%%%%%%%%%%%%%

We started this section by specializing to the surface, $W=0$, and
then identifying the holomorphic vector bundles associated with the matrix factorization.
There is another, somewhat more abstract, approach that realizes these
vector bundles in terms of Maximal Cohen-Macaulay (MCM)  modules.
Since this description is the standard approach to matrix factorizations in the 
mathematics literature, it is useful to relate it to our discussion here.  Moreover,
the formulation in terms of MCM modules can be used to show that 
holomorphic vector bundles are sufficient to reconstruct the  matrix factorization 
(see, for example, ref.~\Laza).   

The first step is to introduce a local ring, $\cR$, defined by the 
superpotential, $W$.  That is, let   $\cP =\IC[x_1,\dots, x_m]$ be the ring 
of complex polynomials in the $x_j$, and let $[W]$ be the ideal of $\cP$
 generated by the superpotential, $W$.  One then defines a new ring, $\cR$, by:
\eqn\Rdefn{ \cR ~\equiv~   \IC[x_1,\dots, x_m]/ [W] \,,}
which is simply the ring of polynomials taken modulo $W$.  Once again one
considers the kernels and images of the matrices  $E$ and $J$, but this time
one thinks of these kernels and images as  {\it modules} over the ring $\cR$.  
Specifically, $J$ and $E$
are now thought of as maps from $\cR^n$ to $\cR^n$, and the columns of
the matrices, of course, generate the images of these maps.  
For instance, the columns of $J$ generate a module, $\cM$, however, this module  
is not necessarily freely generated.   That is,   elements of $\cM$ can generically 
be written only as a non-trivial linear combination (over the ring~$\cR$) of
{\it all} the columns, and yet the columns are not linearly independent. 
There are thus relations between the columns of $J$, and the set of such 
relations are called the first syzygy, $\Omega_1(\cM)$, of $\cM$, which is also 
an $\cR$-module.   Indeed, the columns of the matrix $E$ exactly generates this 
set of relations.  Again,  the module, $\Omega_1(\cM)$, is generically not 
freely generated and so there are relations between
the relations and this defines the second syzygy,  $\Omega_2(\cM)$, of $\cM$.
In a matrix factorization the set of such relations between the columns 
of $E$ is of course generated by the columns of
$J$ again.  In other words, we have $\Omega_2(\cM)= \cM$.
This identity, in fact, defines an MCM module.

One can summarize the foregoing by stating that the following is
an exact sequence:
\eqn\MCMdefn{\xymatrix{ \dots  \ar@{->}[r]^E & \cR^n  \ar@{->}[r]^J & \cR^n
\ar@{->}[r]^E & \cR^n  \ar@{->}[r]^J & \cR^n 
\ar@{->}[r]^E  & \cR^n \ar@{->}[r]  &\cM   \ar@{->}[r]   &0 \,,}}
with a similar sequence with $E$ and $J$ interchanged.  

The ideas and techniques of MCM modules are very useful if one is
going to generate matrix factorizations from bundle data \Laza,
but since this is not our focus here, we will not need to discuss these ideas
any further.

%%%%%%%%%%%%%%%%%%%%%%%%%%%%%%%%%%
\newsec{Some examples of matrix factorizations and their vector bundles}
%%%%%%%%%%%%%%%%%%%%%%%%%%%%%%%%%%

Here we will describe, in detail,  the vector bundles associated with
the $2 \times 2$ and $3 \times 3$ factorizations.  Before we proceed to the
examples, we need to summarize some of the properties of the classical theta 
and Appell functions that arise in the description of holomorphic vector bundles on
a torus.  More details may be found in Appendix A.

%%%%%%%%%%%%%%%%%
\subsec{Classical elliptic functions}
%%%%%%%%%%%%%%%%%

In the present paper we mainly deal with line bundles  and rank two vector 
bundles. We therefore  briefly discuss their canonical sections, which are given by
Riemann theta functions and Appell functions, respectively.
The standard definitions are:
\eqn\basictheta{\vartheta(\xi)  ~=~ \vartheta(\xi |\tau) ~\equiv~ \sum_{n \in \IZ}  
q^{{1\over 2}\, n^2 } \, z^n \,, }
\eqn\basicappell{\kappa(\rho, \xi) ~=~ \kappa(\rho, \xi |\tau) ~\equiv~ \sum_{n \in \IZ}  
{q^{{1\over 2}\, n^2 } \, z^n \over  q^n - y}\,, }
where $y \ne q^m, m \in \IZ$ and 
\eqn\qazdefns{q~\equiv~ e^{2\pi i\,   \tau  }\,, \quad  z~\equiv~ e^{2\pi i \, \xi }\,, 
\quad  y~\equiv~ e^{2\pi i\, \rho } \,.}

These functions satisfy the following periodicity relations:
\eqn\perioda{\vartheta(\xi+1)  ~=~ \vartheta(\xi) \,, \qquad 
\vartheta(\xi+\tau)  ~=~q^{-{1\over 2}}\, z^{-1}\, \vartheta(\xi)  \,,}
\eqn\periodb{\eqalign{\kappa(\rho, \xi+1) ~=~ & \kappa(\rho+1, \xi)~=~  \kappa(\rho, \xi)\,, 
\qquad  \kappa(\rho, \xi+\tau) ~=~  y \, \kappa(\rho, \xi)~+~ \vartheta(\xi) \,,\cr 
\kappa(\rho+\tau, \xi) ~=~  & q^{-{1 \over 2}}\, z \, (y \, \kappa(\rho, \xi)~+~ \vartheta(\xi)) \,.}}

Consider these functions on the torus whose fundamental cell in 
$\IC$ is defined by  $\xi \sim \xi + 1\sim \xi + \tau $.  It is a fairly familiar fact 
that the theta functions provide global holomorphic  sections of line bundles 
on this torus:  They are periodic under $\xi \to \xi + 1$ and have a transition
function of  $ q^{-{1\over 2}} e^{-2\pi i \, \xi }$ under $\xi \to \xi + \tau$.  The fact 
that this function has winding number $-1$ on the circle defined by $\xi \in [0,1]$
means $\vartheta(\xi)$ is a holomorphic section of the line bundle, 
$\cV_1(q^{-{1\over 2}} z^{-1})$, with $c_1 =+1$ on the torus.

The Appell function provides a global holomorphic section of a non-trivial
rank two   vector bundle, $\cF$. In particular, the vector:
\eqn\vecsec{\left({\kappa(\rho,\xi) \atop \vartheta(\xi) } \right) }
is globally holomorphic,   is periodic under $\xi \to \xi + 1$, but 
has  the non-trivial transition matrix under $\xi \to \xi + \tau$:
\eqn\transmat{A(z) ~=~ \left(\matrix{y &  1 \cr 0 & q^{-{1 \over 2}}\, z^{-1}}\right) \,.}
This vector bundle, $\cF$, is a non-trivial extension of $\cV_1(q^{-{1\over 2}} z^{-1})$
by $\cV_1(y )$ :
\eqn\test{\xymatrix{ 0 \ar@{->}[r]  & \cV_1(y )  \ar@{->}[r]  & \cF 
\ar@{->}[r]  & \cV_1(q^{-{1\over 2}} z^{-1})  \ar@{->}[r]  &0 \,.}}

On the cubic curve, $\Sigma$, we need the theta functions appearing in 
\mudefn\ (see Appendix~A for more details).   The functions, $\mu_\ell$, 
have the following periodicity properties:
\eqn\muperiods{ \mu_\ell(\xi + 1) ~=~ - \mu_\ell(\xi) \,, \qquad
\mu_\ell(\xi + \tau) ~=~ - q^{-{3 \over 2}} \, z^{-3} \, \mu_\ell(\xi) \,,}
but it is also useful to note that:
\eqn\musubperiods{ \mu_\ell(\xi + \coeff{1}{3} ) ~=~ - \omega^{-(\ell-1)}\, 
\mu_\ell(\xi) \,, \qquad  \mu_\ell(\xi -  \coeff{1}{3} \, \tau) ~=~ - q^{-{1 \over 6}} \, 
z \, \mu_{\ell+1}(\xi) \,,}
where the subscript on $\mu_{\ell+1}$ is taken mod 3.  Holomorphy and
the periodicity properties on the torus defined by  $\xi \sim \xi + {1 \over 3}
\sim \xi + \tau $ uniquely define the functions $\mu_{\ell}(\xi)$.   Note that 
\muperiods\ implies that $\mu_{\ell}(\xi)$ is not a function on $\Sigma$, but
is a section of the line bundle, $\cL^3$.  To be more precise, 
$e^{3 \pi i \xi} \mu_\ell(\xi)$ is periodic under $\xi \to \xi +1$, and is
a section of $\cL^3 =\cV_1(z^{-3})$, where $\cL \equiv \cV_1(z^{-1})$.
 
Since the functions $\mu_\ell(\xi)$ are  global holomorphic 
sections of $\cL^3$, each of 
the $\mu_\ell(\xi)$ has three zeroes.  In particular, one has:
\eqn\muvals{
\mu_1(0) ~=~ \mu_1(\coeff{1}{3}) ~=~ \mu_1(\coeff{2}{3}) ~=~ 0\,,
\qquad \mu_2(0) ~=~ -\mu_3(0) ~=~ i\, \eta(\tau) \,.}
The values of the $\mu_\ell$ at other third-periods can then be deduced
from this using \musubperiods.
 
One can define Appell functions, $\Lambda_\ell$, associated with the 
$\mu_\ell$ (see Appendix A for details).  These Appell functions are
defined by their periods:
\eqn\Lambdaperiods{ \Lambda_\ell (\rho, \xi+ \coeff{1}{3}) ~=~ - \omega^{-(\ell-1)}\, 
\Lambda_\ell (\rho, \xi) \,, \qquad \Lambda_\ell (\rho, \xi+ \tau) ~=~y^3\, 
\Lambda_\ell (\rho, \xi) ~+~ \mu_\ell(\xi) \,.}
%

%%%%%%%%%%%%%%%%%
\subsec{The $\thrbythr$ factorization revisited}
%%%%%%%%%%%%%%%%%

Consider, once again, the matrix factorization defined by \JEthree.  
We have seen that $\cE_E$ is a line bundle and that a
global, non-vanishing  holomorphic section can be taken to 
be\foot{For simplicity, we have dropped the pre-factor in \Ethreebas.}:
\eqn\Ethreebas{v_1 ~\equiv~ \left(\matrix{\mu_1(\xi - \zeta) \cr 
 \mu_2(\xi - \zeta) \cr \mu_3(\xi - \zeta) }\right) \,.}

The bundle $\cE_J$ has rank two, and may be defined, via
the kernel of $E$,  as the set of vectors orthogonal to:
\eqn\horvec{(\mu_1(\xi + \zeta)\,,\mu_2(\xi + \zeta)\,,\mu_3(\xi + \zeta)) \,.}
Given that the columns of $J$ are expressed as sections 
of a line bundle, one might, at first, expect that the rank two, holomorphic 
vector bundle, $\cE_J$,  is itself a trivial sum of line bundles.  For example,
one might try taking the last two columns of $J$ as a basis.  However,
for $\xi = \zeta + {n \over 3}$ one has:
\eqn\xalrelns{ x_\ell ~=~ - \omega^{-n\, (\ell-1)}\, \alpha_\ell \,,}
and thus the last two columns of $J$ are multiples of one another, and
so they do not represent a good global basis for the vector bundle.
In the neighborhood of such points one must use a different pair of
columns as a basis, and the transition function is:
\eqn\colrelns{ \left(\matrix{\alpha_3 \, x_2 \cr 
\alpha_2 \, x_1  \cr \alpha_1 \, x_3 }\right)    ~=~ - {\mu_1 (\xi + \zeta) \over
\mu_3(\xi + \zeta)}\,  \left(\matrix{\alpha_1 \, x_1 \cr 
\alpha_3 \, x_3  \cr \alpha_2 \, x_2 }\right)  ~-~  {\mu_2(\xi + \zeta) \over
\mu_3(\xi + \zeta)}\,   \left(\matrix{\alpha_2 \, x_3 \cr 
\alpha_1 \, x_2  \cr \alpha_3 \, x_1 }\right)   \,.}
This change of basis is singular at the three zeroes
of $\mu_3(\xi + \zeta)$ ({\it i.e.} at $\xi = \zeta + {2 \tau \over 3}+ {n \over 3}$).
One therefore has to break the torus into patches if one is to use
the columns of $J$ as a basis.

On the other hand, one can use Appell
functions to obtain a basis of holomorphic 
sections for the  two-dimensional vector bundle, $\cE_J$.  Consider the 
vectors:
\eqn\goodbas{  v_1 ~\equiv~  \left(\matrix{\alpha_1 \, x_1 \cr 
\alpha_3 \, x_3  \cr \alpha_2 \, x_2 }\right)  \,, \qquad 
v_2 ~\equiv~  \left(\matrix{\alpha_1 \, \Lambda_1(\rho,\xi)  \cr 
\alpha_3 \, \Lambda_3(\rho,\xi) \cr \alpha_2 \, \Lambda_2(\rho,\xi) }\right)   \,.}
One can show that 
\eqn\vtwocond{\alpha_1 \, \mu_1(\xi + \zeta) \, \Lambda_1(\rho,\xi) ~+~ 
\alpha_3 \, \mu_2(\xi + \zeta) \, \Lambda_3(\rho,\xi) ~+~ 
\alpha_2 \, \mu_3(\xi + \zeta) \, \Lambda_2(\rho,\xi) ~=~ 0 \,,}
provided that:
\eqn\yval{ \rho ~=~ \coeff{1}{2}\, \tau ~-~ \zeta \,.}
This means that $v_2$ is in the kernel of $E$.  One can prove this identity
by considering the periodicity properties of the function, $F(\xi,\zeta)$, 
defined to be the left-hand side of \vtwocond.  
One first notes that  the ``anomalous'' shift term in the 
Appell functions amounts to shifting $v_2$ by $v_1$ and 
since $v_1$ is in the kernel of $E$, the shift term does not
contribute to $F(\xi,\zeta)$ under $\xi \to \xi+ \tau$.  This means 
that, considered  either as a function of $\xi$ or as a function 
of $\zeta$, $F(\xi,\zeta)$  represents a global
section of a line bundle.   One then uses holomorphy and standard theta-function
methods to write it in terms of theta functions or, in this instance, conclude that it is zero
provided that one chooses the Appell function parameter according to \yval.

Using the same kind of argument one can establish the following identities:
\eqn\othercolsa{  \mu_2(\xi - \zeta) \, v_1 ~-~ e^{6\pi i \zeta}\, 
\Lambda_2(\coeff{\tau}{2}\,, \xi - \zeta) \, v_2 ~=~   -i q^{-{9 \over 8}} \,
{\eta^3(3\tau) \over \eta(\tau)} \, e^{3\pi i (\xi+ \zeta)}\, 
 \left(\matrix{\alpha_3 \, x_2 \cr 
\alpha_2 \, x_1  \cr \alpha_1 \, x_3 }\right) \,,  } 
\eqn\othercolsb{  \mu_3(\xi - \zeta) \, v_1 ~-~ e^{6\pi i \zeta}\, 
\Lambda_3(\coeff{\tau}{2}\,, \xi - \zeta) \, v_2 ~=~    i q^{-{9 \over 8}} \,
{\eta^3(3\tau) \over \eta(\tau)} \, e^{3\pi i (\xi+ \zeta)}\, 
 \left(\matrix{\alpha_2 \, x_3 \cr 
\alpha_1 \, x_2  \cr \alpha_3 \, x_1 }\right) \,.} 
In other words, the set of columns of $J$ are given by holomorphic
linear combinations of the holomorphic vectors, $v_1$ and $v_2$.

Finally, observe that the vectors $u_j \equiv e^{3 \pi i \xi} v_j$ are periodic
under $\xi \to \xi+1$ and under $\xi \to \xi+\tau$ they have the transition matrix:
\eqn\ujtrans{A(z) ~=~ 
\left(\matrix{q^3  \, e^{-6 \pi i \zeta}  &  1 \cr  0 &  z^{-3}} \right) \,.}
Summarizing, the bundle, $\cE_J$, is the non-trivial, non-split bundle 
with $(r(\cE_J),c_1(\cE_J) )= (2,3)$.  Up to the ambiguity of tensoring with
$\cL^{3n}$, this is
equivalent to $\cE(2,-3)$, which indeed belongs to the charges listed in
\shiftedcharges.

%%%%%%%%%%%%%%%%%
\subsec{Vector bundles of the $2 \times 2$ factorization}
%%%%%%%%%%%%%%%%%

Consider the  $2 \times 2$ factorization given by \Mtwo.  
The determinants of $E$ and $J$ are both $W$ and so the kernels
of both are one-dimensional.  Because  the matrix elements have different 
degrees, the matrices in \mattrans\ are now non-trivial, that is, taking
$g = -q^{-3/2} e^{-6\pi i \xi}$ one has: 
\eqn\mattrans{\eqalign{J(x(\xi + \tau) ) ~=~ & g^k \, G_1(g) \, J(x(\xi)) \, G_2(g)^{-1} \,, \cr 
E( x(\xi + \tau) ) ~=~ & g^{3-k} \, G_2(g)\, E(x(\xi) ) \, G_1 (g)^{-1} \,,}}
where
\eqn\GLGR{G_1 ~=~ \pmatrix{ g^2  & 0 \cr  0 &  g}  \,, \qquad
G_2 ~=~ \oneone_\twobytwo \,.}
Since $G_2 = \oneone$ there are no technical issues in defining $\cE_E$. 
The columns of $E$ must  be
proportional to each other, and by looking at the common zeroes of 
the $L_j$ and $Q_j$ one can argue that the kernel should be spanned
by theta functions of characteristic $2$.  However there is a simpler way
to obtain the kernels of the matrices using the functions $\mu_\ell$.

By taking constant ({\it i.e.} independent of $\xi$) linear combinations
of the rows of $J$ in \JEthree, one can show that 
\eqn\twoa{\widetilde L_1 \,  \mu_1(\xi - \zeta)  ~+~ \widetilde L_2 \,  
(\mu_2 (\xi - \zeta)  + \mu_3(\xi - \zeta) ) ~=~ 0 \,, }
where 
\eqn\twoa{\widetilde L_1 ~\equiv~ (\alpha_1^2\, x_1 + \alpha_2^2\, x_2+ 
\alpha_3^2\, x_3 ) \,, \qquad \widetilde L_2 ~\equiv~ (\alpha_2 \,\alpha_3 \,  x_1
+ \alpha_1 \,\alpha_3 \,  x_2  +\alpha_1 \,\alpha_2 \,  x_3)  \,.}
One can similarly derive relations between $\mu_1$ and $\mu_2 + \omega^k \mu_3$.
Taking the latter relationships for $k=1,2$ one can multiply by each by factors
that are linear in the $x_j$ so as to obtain a relationship of the form:
\eqn\twob{ \widetilde Q_1 \,  (\mu_2(\xi - \zeta)  + \mu_3(\xi - \zeta) )~-~
\widetilde Q_2 \, \mu_1(\xi - \zeta)  ~=~ 0 \,,}
where the $\widetilde Q_j$ are quadratic in the $x_\ell$.  Thus we get a $2 \times 2$
matrix with a null vector of the form:
\eqn\nullvec{  \left(\matrix{\tilde s_1(\xi) \cr  \tilde s_2(\xi)}\right)~\equiv~   
\left(\matrix{   \mu_2(\xi - \zeta ) + \mu_3(\xi - \zeta )  \cr  
 \mu_1(\xi - \zeta ) }\right)    \,,}
This matrix is gauge equivalent to $J_\twobytwo$ in \Mtwo, except one must 
replace $\alpha_j$ in \Mtwo\ according to:
\eqn\alpharep{ \alpha_1 \to    \alpha_2 \, (\alpha_3^3 - \alpha_1^3) \,,
\quad \alpha_2 \to    \alpha_1 \, (\alpha_2^3 - \alpha_3^3) \,,
\quad  \alpha_3 \to    \alpha_3 \, (\alpha_1^3 - \alpha_2^3) \,.}
Using \Eij\ at $\xi = \zeta$, along with \muvals\ and \musubperiods,
one can see that the replacement  \alpharep\ is equivalent to
\eqn\zetarep{ \zeta ~\to~  2\, \zeta + \coeff{1}{3}\, \tau \,.}

To understand the underlying vector bundle, one should note that
this derivation of  the matrix factorization implicitly relies on the fact that 
$\tilde s_1(\xi)$ and  $\tilde  s_2(\xi)$ have a common zero at $\xi =\zeta$ and so 
\nullvec\ vanishes identically at one point.  To cover this vanishing point we
could find another section and the transition function, but it is simpler to note that  there is  
a global,  {\it nowhere-vanishing} section of  this bundle given by:
\eqn\svec{ 
\left(\matrix{s_1(\xi) \cr  s_2(\xi)}\right) ~\equiv~  
{e^{3 \pi i \xi}  \over   \vartheta( (\xi - \zeta) - \coeff{1}{2} ( 1 + \tau)) \, ) } \,
\left(\matrix{   \mu_2(\xi - \zeta ) + \mu_3(\xi - \zeta)  \cr   \mu_1(\xi - \zeta) }\right)    \,.}
The $s_j$  are thus theta functions of characteristic $2$ and satisfy:
\eqn\speriods{ s_j(\xi+1) ~=~ s_j(\xi) \,, \qquad  s_j(\xi+\tau) ~=~ z^{-2}\, s_j(\xi)  \,,}
and so the bundle $\cE_E$ is simply $\cE(1,2)$.

Conversely, given the $s_j$, the ratio $s_2/s_1$ is an elliptic function
with two poles and two zeroes on the torus, and so there is a unique 
way to write it in the form:
\eqn\linratio{{s_2 \over s_1}  ~=~ 
{a_1 x_1 + a_2 x_2 +a_3 x_3 \over b_1 x_1 + b_2 x_2 + b_3 x_3} \,,}
and this defines $\widetilde L_1$ and $\widetilde L_2$.  Similarly, one can write 
$s_2/s_1$ as a ratio of quadratics in the $x_j$.   Each quadratic, $\widetilde Q_j$,
has six parameters, one of which is a scale and the other five
can be used to adjust the locations of the six zeroes on the torus.\foot{The
sum of the zeroes of $Q_j$ must be zero modulo the lattice $\xi \sim \xi+1$, 
$\xi \sim \xi+\tau$.}
The locations of two zeroes in each $\widetilde Q_j$ are fixed by the $s_j$, 
while the other zeroes of $\widetilde Q_1$ must coincide with those
of $\widetilde Q_2$, but are otherwise free.  Thus there is a three-parameter
family of ways of writing $s_2/s_1$ as ratio of quadratics.  On the other hand,
one can trivially generate such a quadratic ratio by multiplying the numerator and 
denominator of  \linratio\ by the same arbitrary  linear function of the $x_j$. This
choice of linear function has three parameters, and so the quadratic ratio
is unique up to gauge transformations.  Thus one can reconstruct the
$2 \times 2$-matrix factorization uniquely from  \svec.  

By keeping track of the locations of the zeroes in the foregoing argument,
it is easy to check that one can write the columns of $E$ as:
\eqn\Jholrels{\left(\matrix{ \widetilde L_1  \cr  -\widetilde L_2}\right) ~=~
 \chi_1(\xi) \,  \left(\matrix{s_1(\xi) \cr  s_2(\xi)}\right) \,, \qquad  
\left(\matrix{ \widetilde Q_2  \cr 
 \widetilde Q_1}\right) ~=~ \chi_2(\xi) \, \left(\matrix{s_1(\xi) \cr  s_2(\xi)}\right) \,,}
for some global holomorphic sections, $\chi_1(\xi)$ and $\chi_2(\xi)$, of
$\cE(1,1)$ and $\cE(1,4)$ respectively.  From this one can write the columns
of $J$ as:
\eqn\Eholrels{\left(\matrix{  \widetilde Q_1  \cr   \widetilde  L_2}\right) ~=~
s_2(\xi) \,  \left(\matrix{\chi_2(\xi) \cr  - \chi_1(\xi)}\right) \,, \qquad  
\left(\matrix{- \widetilde Q_2  \cr 
 \widetilde L_1}\right) ~=~ - s_1 (\xi) \, \left(\matrix{\chi_2(\xi) \cr  - \chi_1(\xi)}\right)  \,,}
and hence the columns of $\cE_J$ correspond to the bundle 
$\cE(1,1)$ (or  $\cE(1,4)$).

%%%%%%%%%%%%%%%%%
\subsec{Summary}
%%%%%%%%%%%%%%%%%

In this section, we considered both the $2\times 2$ and $3\times
3$ factorizations and determined the data of the bundles $\cE_J$ and
$\cE_E$, associated with the matrix factors, $J$ and $E$.  We now like to match
these data to the $RR$ charges of the
$B$-branes. Specifically, as mentioned before, the charges of the
branes will be associated with $\cE_J$, while the anti-branes will
be associated with $\cE_E$.

To proceed, recall that for extracting the bundle data from the
matrix factorization, we had to impose $W\equiv 0$, and therefore
we implicitly characterized the branes in the large radius limit.
Thus we should refer to the $RR$ charges listed in \shiftedcharges,
which apply to the large radius limit. Indeed we find a perfect
match between the data of $\cE_J$, and the charges of {\it one} of the
members of $\{S_a\}$ or $\{L_a\}$, respectively, provided we tensor
uniformly with the line bundle $\cL^{-3}$ (recall that brane charges
are defined only up to such shifts, reflecting the monodromy around
the large radius limit). There is also an analogous match between
the data of $\cE_E$, and the charge of one of the anti-branes in
$\{\bar S_a\}$ or $\{\bar L_a\}$, respectively. 
We have summarized these findings in \ltab\MFProp.

\bigskip
%%%%%%%%%%%%%%%%%%%%%%%%%%%%%%%%%%
\tabinsert\MFProp{Here we summarize some properties of the short- and long-diagonal branes, 
associated to the $2\times 2$ and the $3\times 3$-matrix factorizations.
In particular the relationship between the bundle data 
of the matrix factors $\cE_J$ and $\cE_E$,
and the large radius $RR$~charges is shown.}
{
\begintable
\| \multispan{2}\ctr{$2\times 2$ factorization }
\| \multispan{2}\ctr{$3\times 3$ factorization }\crthick
Bundle \hfill
\| \ctr{$\cE_J$} \vb \ctr{$\cE_E$}
\| \ctr{$\cE_J$} \vb \ctr{$\cE_E$} \cr
Matrix bundle $\cE$ \hfill
\| \ctr{$\cE(1,1)$} \vb \ctr{$\cE(1,2)$} \| \ctr{$\cE(2,3)$} \vb \ctr{$\cE(1,3)$} \cr
$\cE\otimes\cL^{-3}\simeq (r,c_1)^{LR}$ \hfill
\| \ctr{$(1,-2)^{LR}$} \vb \ctr{$(1,-1)^{LR}$} 
\| \ctr{$(2,-3)^{LR}$} \vb \ctr{$(1,0)^{LR}$}  \cr
Brane \& anti-brane in the\hfill
\| \ctr{\vctr{$S_1$}} \vb \ctr{\vctr{$\bar S_3$}}
\| \ctr{\vctr{$L_1$}} \vb \ctr{\vctr{$\bar L_3$}} \crnorule
positive cone\hfill
\| \vb \| \vb
\endtable }
%%%%%%%%%%%%%%%%%%%%%%%%%%%%%%%%%%

Note that if we view the charge lattice in \lfig\branes\ as the
$SU(3)$ root lattice, the specific charges of the vector bundles
at large radius lie within what one may call the positive cone, or
fundamental region, spanned by the ``simple roots'' $S_1$ and $S_2$;
this is indeed natural from the point of view of quiver representations
(see the appendix of ref.~\DouglasQW).

What then about the other charges in the lists \shiftedcharges,
some of which are related to more general objects than vector bundles
(like $D0$-branes described by point-like sheaves)? 
The point is, of course, that in order to obtain the TFT on the elliptic
curve from the Landau-Ginzburg model, one
must implement a $\IZ_3$ orbifold projection, and it is only then that the
full (quantum) $\IZ_3$ orbits of the branes will emerge. This is what we will discuss
in Section~5, however before doing so, we will first complete the discussion
of $P_\thrbythr$ by analyzing its open-string moduli space.

%%%%%%%%%%%%%%%%%%%%%%%%%%%%%%%%%%
\newsec{Compactifying the moduli space of the $3\times 3$ factorization.}
%%%%%%%%%%%%%%%%%%%%%%%%%%%%%%%%%%

We now return to the $3\times 3$-matrix factorization, $P_\thrbythr$, in order
to examine in greater detail its open-string moduli space. This will
eventually lead us to an exceptional $4\times 4$ matrix
factorization, which appears only at a certain point in that
moduli space, and which compactifies the moduli space of the $\thrbythr$ 
factorization $P_\thrbythr$.  It is this  $4\times 4$ matrix
factorization which naturally appears in the boundary fermion construction.

First of all one observes that the $3\times 3$ matrices, $J=J_\thrbythr$ 
and $E=E_\thrbythr$,
given in \JEthree\ are quasi-(double-)periodic for $\zeta\rightarrow
\zeta+1$ and $\zeta\rightarrow \zeta+\tau$ due to the
quasi-periodicity~\muperiods\ of the parameters~$\alpha_\ell$ in
terms of the uniformizing open-string modulus $\zeta$ of eq.~\alphauni.
In other words, such a shift in $\zeta$ results in a multiplication
of both $J$ and $E$ by an overall factor. However, the factors of
$J$ and $E$ are inverse to each other, and hence they are easily
compensated by a gauge transformation~\gauge. Therefore, modulo
gauge transformations the $3\times 3$-matrix factorization is indeed
periodic for $\zeta\rightarrow \zeta+1$ and $\zeta\rightarrow
\zeta+\tau$.

However, it is conceivable that there are further identifications in the
open-string moduli space of $\zeta$ due to additional gauge equivalences: 
According to eq.~\musubperiods\ the uniformizing functions 
$\mu_\ell(\zeta)$ of $\alpha_\ell$ have also nice periodicity properties
 for one-third of a period. In particular, for the $3\times 3$ matrices $J$ 
 and $E$, we readily find:
\eqn\transthird{J(\zeta+\coeff{1}{3})\,=\,
   C_1\,\pmatrix{\omega^2\,\alpha_1 \, x_1 & \omega^{\phantom 1}\,
   \alpha_2 \, x_3 & \phantom{\omega^2}\,\alpha_3 \, x_2   \cr 
                 \phantom{\omega^2}\,\alpha_3 \, x_3 & \omega^2\,
                 \alpha_1 \, x_2 & \omega^{\phantom 1}\,\alpha_2 \, x_1  \cr
                 \omega^{\phantom 1}\,\alpha_2 \, x_2 & \phantom{\omega^2}\,
                 \alpha_3 \, x_1 & \omega^2\,\alpha_1 \, x_3}\ , }
and
\eqn\transtauthird{J(\zeta+\coeff{\tau}{3})\,=\,
    C_2\,\pmatrix{\alpha_3 \, x_1 & \alpha_1 \, x_3 & \alpha_2 \, x_2   \cr 
                  \alpha_2 \, x_3 & \alpha_3 \, x_2 & \alpha_1 \, x_1  \cr
                  \alpha_1 \, x_2 & \alpha_2 \, x_1 & \alpha_3 \, x_3}\ , }
with overall factors~$C_1$ and $C_2$. Hence, apart from these factors,
for $\zeta\to\zeta+\coeff{1}{3}$ the entries of $J$ are multiplied by cube roots of 
unity, whereas for $\zeta\to\zeta+\coeff{\tau}{3}$ the coefficients of $x_\ell$ in $J$ 
are permuted.  However,   both transformations can be compensated by suitable gauge 
transformations, \ie\  $J(\zeta)=U_{L,1}\,J(\zeta+\coeff{1}{3})\,U_{R,1}$ and 
$J(\zeta)=U_{L,2}\,J(\zeta+\coeff{\tau}{3})\,U_{R,2}$ 
with\foot{One also has   $E(\zeta)=U_{R,1}^{-1}\,E(\zeta+\coeff{1}{3})\,U_{L,1}^{-1}$ 
and $E(\zeta)=U_{R,2}^{-1}\,E(\zeta+\coeff{\tau}{3})\,U_{L,2}^{-1}$.}
\eqn\transone{U_{L,1}\,=\,\Diag{1,\,\omega^2,\,\omega} \ , 
               \qquad U_{R,1}\,=\,\coeff{1}{C_1}\,\Diag{\omega,\,\omega^2,\,1} \ , }
and
\eqn\transtwo{U_{L,2}\,=\,
   \pmatrix{0 & 1 & 0 \cr 0 & 0 & 1 \cr 1 & 0 & 0} \ , \qquad
   U_{R,2}\,=\,\coeff{1}{C_2}\,\pmatrix{0 & 1 & 0 \cr 0 & 0 & 1 \cr 1 & 0 & 0} \ . }
We conclude that, modulo gauge transformations, the $3\times 3$ 
matrix factorization is double-periodic in the open-string modulus~$\zeta$ 
with $\zeta\sim\zeta+\coeff{1}{3}$ and $\zeta\sim\zeta+\coeff{\tau}{3}$, 
and thus obtain for the $3\times 3$-matrix factorization the following
toroidal open-string moduli space:
\eqn\mspacethree{{\frak M}^{3\times 3}_\zeta = \{\zeta \in \IC\,|\,\zeta \sim \zeta +
\coeff{1}{3}\,,\, \zeta \sim \zeta +\coeff{\tau}{3}\,\,\} \ . }
One might suspect that there are other identifications in the open-string 
moduli space~\mspacethree\ of the $3\times 3$-matrix factorization and 
that the space ${\frak M}^{3\times 3}_\zeta$ is yet again just the covering 
space of the true moduli space. However,  the construction of the vector 
bundles encoded in the columns of the matrix $E$ and in the columns of 
the matrix $J$ shows that the corresponding vector bundle transition matrices 
\uetrans\ and \ujtrans\ contain both the factor $e^{2\pi i\,\cdot 3\zeta}$, 
which independently confirms the stated periodicity of $J$ and $E$. 
Thus ${\frak M}^{3\times 3}_\zeta$ is indeed the open-string moduli space 
for the $3\times 3$-matrix factorization.

The reduction of the periodicity by one third will be an important ingredient in 
the construction of the $3\times 3$-matrix factorization via tachyon 
condensation from $2\times 2$ matrix building blocks as discussed 
in Section~5.

Our next task is to examine the moduli-space~\mspacethree\ of the 
factorization $P_\thrbythr$ in greater detail. For $\zeta\to 0$ the uniformizing function, 
$\alpha_1$, approaches zero, {\it c.f.} eq.~\muvals, for which the $3\times 3$ matrix 
factorization becomes singular.\foot{At first sight there seem to be nine 
singularities, namely $\alpha_\ell=0$ for the three choices of $\ell$, 
which, according to \musubperiods\ and \muvals, have each three zeroes. 
However,  eqs.~\transthird\ and \transtauthird\ show that all nine choices 
are gauge-equivalent.} Note that, in contrast to the singularity in the 
matrix factorization $P_\twobytwo$, this singularity is not a mere gauge 
artifact,  because it cannot be removed by a gauge transformation~\gauge. 
This observation, however, is at first puzzling since there is no obvious physical 
reason for a singularity in the toroidal open-string moduli space.

However, we should keep in mind that the topological $B$-type $D$-branes 
really are objects in a category with certain equivalence relations, and a 
given matrix-factorization is just a particular realization of a topological 
$B$-type $D$-brane. In other words, singularities of matrix factorizations 
can also be an artifact of using the wrong representative for a $D$-brane 
in a given patch in the open-string moduli space. As has been discussed 
in ref.~\HoriJA, the apparent singularity encountered in the $3\times 3$-matrix 
factorization is indeed of this type, and this is what we want to 
make manifest in the following. 

Following ref.~\HoriJA,  we replace as a first step the $3\times 3$ matrix 
factorization by a $4\times 4$-matrix factorization that is obtained by 
adding a trivial brane-anti-brane pair to \JEthree, \ie:
\eqn\threeequiv{J_{4\times 4}\,=\,\pmatrix{ {W\over\alpha_1} \cr & J(\alpha) } \ , \qquad
                E_{4\times 4}\,=\,\pmatrix{ \alpha_1 \cr & E(\alpha) } \ . }                
The next task is to perform an appropriate gauge transformation, \gauge,
so that the singularity disappears for $\zeta\to 0$. This is achieved by first 
rewriting the open-string modulus, $\zeta$, by $\zeta=\lambda-\rho$.\foot{This step is a little 
{\it ad hoc} but it is motivated by viewing the brane $L(\zeta)$ as a composite
of the two branes $\bar L(\lambda)$ and $\bar L(-\rho)$.}
Using eq.~\Eij\ with $x_\ell$, $\alpha_\ell$ replaced by 
$\beta_\ell$ and $\gamma_\ell$ respectively, we may re-express 
$\alpha_\ell$ by\foot{Strictly speaking the parameters, 
$\alpha_\ell$, must be rescaled by a homogeneous factor, \ie\ 
$\alpha_\ell\to\eta^2(\tau)\mu_3(\lambda+\rho)\,\alpha_\ell$.}
\eqn\alphanew{\eqalign{\alpha_1 \,&=\, \beta _2^2 \gamma _1 
\gamma _3-\beta _1 \beta _3 \gamma _2^2 \ , \cr
\alpha_2 \,&=\, \beta _1^2 \gamma _1 \gamma _2-\beta _2 
\beta _3 \gamma _3^2 \ , \cr
\alpha_3 \,&=\, \beta _3^2 \gamma _2 \gamma _3-\beta _1 \beta _2 
\gamma _1^2 \ , }}
where $\alpha_\ell$ and $\beta_\ell$ are uniformized by $\beta_\ell=\mu_\ell(\lambda)$ 
and $\gamma_\ell=\mu_\ell(\rho)$. Now the limit $\alpha_1\to 0$ becomes 
$\gamma_\ell \to \beta_\ell$, which in terms of the uniformizing parameters 
translates into $\rho\to \lambda$. In terms of these auxiliary variables we perform the 
gauge transformation
\eqn\threegaugetrans{U_L\,=\,
\pmatrix{1 & {\beta_1\over\alpha_1\gamma_1}x_1 & {\beta_2\over\alpha_1\gamma _2}x_2 & 
                       {\beta_3\over\alpha_1\gamma _3}x_3 \cr
         0 & 1 & 0 & 0 \cr
         0 & 0 & 1 & 0 \cr
         0 & 0 & 0 & 1 } \ , \qquad 
U_R\,=\,\pmatrix{1 & 0 & 0 & 0 \cr
         -{\gamma_1\over\alpha_1\beta_1}x_1 & 1 & 0 & 0 \cr
         -{\gamma_2\over\alpha_1\beta_2}x_2 & 0 & 1 & 0 \cr
         -{\gamma_3\over\alpha_1\beta_3}x_3 & 0 & 0 & 1 } \ , }
which generates the following $4\times 4$-matrix factorization:
\eqn\threeequivtrans{\eqalign{
\widetilde J_{4\times 4}\,&=\,
\pmatrix{ 0 & {\beta_1\over\gamma_1}x_1^2-{\beta_1\gamma_1\over\gamma_2\gamma_3}x_2x_3 & 
              {\beta_2\over\gamma_2}x_2^2-{\beta_2\gamma_2\over\gamma_1\gamma_3}x_1x_3 & 
              {\beta_3\over\gamma_3}x_3^2-{\beta_3\gamma_3\over\gamma_1\gamma_2}x_1x_2 \cr
        {\beta_1\gamma_1\over\beta_2\beta_3}x_2x_3-{\gamma_1\over\beta_1}x_1^2 & 
              \alpha_1\,x_1 & \alpha_2\,x_3 & \alpha_3\,x_2 \cr
        {\beta_2\gamma_2\over\beta_1\beta_3}x_1x_3-{\gamma_2\over\beta_2}x_2^2 & 
              \alpha_3\,x_3 & \alpha_1\,x_2 & \alpha_2\,x_1 \cr
        {\beta_3\gamma_3\over\beta_1\beta_2}x_1x_2-{\gamma_3\over\beta_3}x_3^2 & 
              \alpha_2\,x_2 & \alpha_3\,x_1 & \alpha_1\,x_3 } \ , \cr
\widetilde E_{4\times 4}\,&=\,
\pmatrix{ \alpha_1 & -{\beta_1\over\gamma_1}x_1 & -{\beta_2\over\gamma_2}x_2 & -{\beta_3\over\gamma_3}x_3 \cr
          {\gamma_1\over\beta_1}x_1 & -{\alpha_1\over\alpha_2\alpha_3}x_2x_3 & 
                   {1\over\alpha_3}x_3^2+{\beta_3\gamma_3\over\alpha_2\beta_1\gamma_2}x_1x_2 & 
                   {1\over\alpha_2}x_2^2+{\beta_2\gamma_2\over\alpha_3\beta_1\gamma_3}x_1x_3 \cr
          {\gamma_2\over\beta_2}x_2& 
                   {1\over\alpha_2}x_3^2+{\beta_3\gamma_3\over\alpha_3\beta_2\gamma_1}x_1x_2 & 
          -{\alpha_1\over\alpha_2\alpha_3}x_1x_3 & 
                   {1\over\alpha_3}x_1^2+{\beta_1\gamma_1\over\alpha_2\beta_2\gamma_3}x_2x_3 \cr
          {\gamma_3\over\beta_3}x_3 & 
                   {1\over\alpha_3}x_2^2+{\beta_2\gamma_2\over\alpha_2\beta_3\gamma_1}x_1x_3 & 
          {1\over\alpha_2}x_1^2+{\beta_1\gamma_1\over\alpha_3\beta_3\gamma_2}x_2x_3 & 
          -{\alpha_1\over\alpha_2\alpha_3}x_1x_2 } \ , }}
which we will denote by $\widetilde P_\fobyfo$.
We want to emphasize again that $\widetilde P_\fobyfo$ describes the same 
physical $D$-brane as the original factorization $P_\thrbythr$. 
As we will see in a moment, this factorization is indeed most canonical 
from the Landau-Ginzburg point of view: while $P_\thrbythr$ is odd-dimensional 
so that it cannot be given a standard representation in terms of boundary fermions, 
it is the equivalent factorization $\widetilde P_\fobyfo$ that can be given a concise
expression in terms of Landau-Ginzburg fields.

In addition, $\widetilde P_\fobyfo$ has, by construction, the virtue that it is also well-defined for 
$\zeta=0$, and thus we may proceed and take the limit $\alpha_1\to 0, 
\gamma_\ell\to\beta_\ell$. Upon doing so, we prefer to perform yet another 
gauge transformation:
\eqn\fourgaugetrans{\eqalign{
U_L\,&=\,\pmatrix{
   \beta_2\left(\beta_1^3-\beta_3^3\right) & 
      -\left({\beta_3^2\over\beta_1\beta_2}+a\right)x_1 &
      -2 a\, x_2 & 
      \left({\beta_1^2\over\beta_2\beta_3}-3 a\right)x_3 \cr
   0 & 1 & 0 & 0 \cr
   0 & 0 & 1 & 0 \cr
   0 & 0 & 0 & 1 }  \ , \cr
U_R\,&=\,{1\over \beta_2(\beta_1^3-\beta_3^3)}\ \Transpose{U_L} \ ,}}
in order to obtain the following form of $\widetilde P_\fobyfo$ at $\zeta=0$:
\eqn\Mtwofour{\eqalign{
\widetilde J_{4\times 4}^0\,&=\,
\pmatrix{ 0 & x_1^2-a\,x_2x_3 & x_2^2-a\,x_1 x_3 & x_3^2-a\,x_1 x_2 \cr
          -x_1^2+a\,x_2 x_3 & 0 & \phantom{-}x_3 & -x_2 \cr
          -x_2^2+a\,x_1 x_3 & -x_3 & 0 & \phantom{-}x_1 \cr
          -x_3^2+a\,x_1 x_2 & \phantom{-}x_2 & -x_1 & 0 } \ , \cr
\widetilde E_{4\times 4}^0\,&=\,
\pmatrix{ 0 & -x_1 & -x_2 & -x_3 \cr
          x_1 & 0 & -x_3^2+a\,x_1 x_2 & \phantom{-}x_2^2-a\,x_1 x_3 \cr
          x_2 & \phantom{-}x_3^2-a\,x_1 x_2 & 0 & -x_1^2+a\,x_2 x_3 \cr
          x_3 & -x_2^2+a\,x_1 x_3 & \phantom{-}x_1^2-a\,x_2 x_3 & 0 } \ . }}
We will denote this limit by $\widetilde P_\fobyfo^0$. In terms of the boundary BRST
operator, $\cQ$, we can rewrite this factorization in a very simple form
\eqn\canonform{\widetilde P_\fobyfo^0:\ \ \ 
\cQ=\sum_{\ell=1}^3 \left( x_\ell\,\pi_\ell + 
\coeff{1}{3}\del_{x_\ell}W(x,a)\,
\bar\pi_\ell \right) \ , }
where $W(x,a)$ is the Landau-Ginzburg potential in \cubic\ and $\pi_\ell,\bar\pi_\ell$,
$\ell=1,2,3$, are boundary fermions obeying
$\{\pi_i,\pi_j\}=\{\bar\pi_i,\bar\pi_j\}=0$,
$\{\pi_i,\bar\pi_j\}=\delta_{ij}$.\foot{That this represents a valid
boundary BRST operator that obeys $\cQ^2=W(x,a)\id$, follows from the fact 
that, due to homogeneity of the Landau-Ginzburg potential, $W(x,a)$, we can write
$W(x,a)\oneone={1\over3}\sum_\ell x_\ell\,\del_{x_\ell}W(x,a)\oneone
\equiv{1\over3}\sum_{i,j}x_i\pi_i\,\del_jW(x,a)\bar\pi_j$.} Indeed,
for an appropriately chosen basis of the Clifford algebra, for which
the chirality gamma matrix, $\gamma_5$, takes the form 
$\gamma_5=\Diag{1,\ldots,1,-1,\ldots,-1}$, the boundary BRST operator 
\canonform\ 
is precisely given by the matrix factorization \Mtwofour, \ie,
\eqn\canonformA{
\cQ=\pmatrix{ 0 & \widetilde J_{4\times 4}^0 \cr
              \widetilde E_{4\times 4}^0 & 0 } \ . }
For non-vanishing open-string modulus, $\zeta$, \canonform\ gets deformed in leading
order to $\cQ\to\cQ+\zeta \delta\cQ$, where
\eqn\deltaQ{
\delta\cQ\sim \pi_1\pi_2\pi_3 + 
              \left(
(a^3-1)x_1x_2x_3\right) \bar\pi_1\bar\pi_2\bar\pi_3 + \ldots \ . }
The ellipsis indicates terms that are linear and quadratic in $x$ and are 
multiplied by products of boundary fermions of the type $\bar\pi_i\pi_j\pi_k$ 
and $\pi_i\bar\pi_j\bar\pi_k$. Note that the deformation, $\delta\cQ$, 
contains a constant, $x$-independent term,
which generates the constant entry, $\alpha_1$, of the matrix
$\tilde E_{4\times 4}$ 
in eq.~\threeequivtrans.

Let us summarize and analyze the foregoing results. Starting
from the matrix factorization, $P_\thrbythr$, we encountered a
singularity. This singularity was removed by taking an equivalent
description in terms of the $4\times 4$-matrix factorization
$\widetilde P_\fobyfo$ in~\threeequivtrans, which, however, is {\it
reducible} due to an extra constant entry in the  matrix, $E$. This
constant entry disappears as $\zeta\to0$, which implies that the
corresponding factorization $\widetilde P_\fobyfo^0$ becomes {\it
irreducible}. This irreducible factorization is rigid in the sense
that it exists only at one point in the open-string moduli space,
and specifically is not part of a parametric family of {\it irreducible} matrix
factorizations.\foot{There does, however, exist another family of irreducible
$4\times 4$-matrix factorizations (denoted by $P_\fobyfo$,
without the tilde) which describes $D$-branes
with different charges ({\it c.f.}~Section~7).}
For a simple depiction of the situation, see \lfig\exceptional.
 
%%%%%%%%%%%%%%%%%%%%%%%%%%%%
 \vskip1mm
\figinsert\exceptional{
This shows the open-string moduli space of the factorization
$\widetilde P_\fobyfo$.  It is compactified at $\zeta=0$ by an
exceptional indecomposable object, which is rigid in the sense that
any deformation of it produces a decomposable object. It is associated
with the most canonical factorization, \canonform, of the Landau-Ginzburg
superpotential. In physical terms, it can be interpreted as a single
rigid anti-$D2$ brane, and the deformation away from it corresponds
to adding and pulling apart an extra $D0$-$\bar {D0}$-brane
pair.}{1.8in}{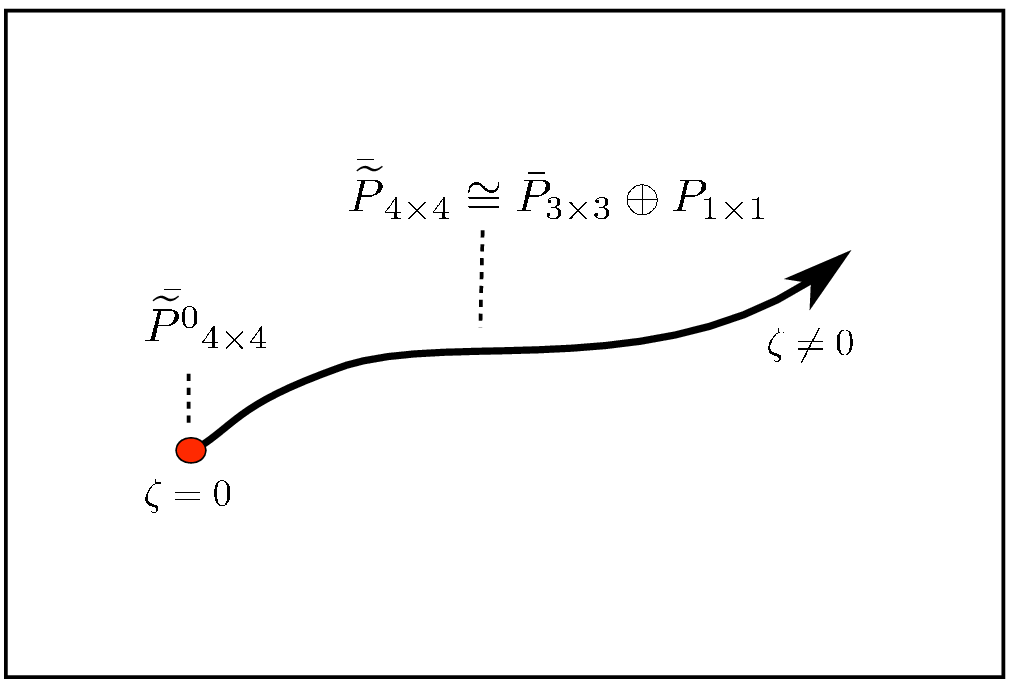}
%%%%%%%%%%%%%%%%%%%%%%%%%%%%

In order to give this mathematical description a physical interpretation,
we observe that the long-diagonal brane, $\bar L_3$, is associated
to the line bundle, $\cE(1,0)$, and located in the positive
cone of the charge lattice in \lfig\branes.
Furthermore, according to \shiftedcharges, the brane $\bar L_3$ has
the (large radius) $RR$~charges of a pure $\bar{D2}$-brane. This,
however, does not necessarily imply that the brane $\bar L_3$
corresponds to a pure $\bar{D2}$-brane, rather more generally it
describes a $\bar{D2}$-brane with a $D0$-$\bar{D0}$-brane pair
resolved on its world-volume. We will now argue that the
rigid exceptional matrix factorization, $\bar{\widetilde P}_\fobyfo^0$,
describes a pure $\bar{D2}$-brane, which also on physical grounds
does not depend on an open-string modulus, whereas the matrix
factorization, $\bar P_{3\times 3}$, captures the situation of a
$\bar{D2}$-brane with an additional resolved $D0$-$\bar{D0}$-brane
pair.\foot{Note again that such $D$-brane configurations cannot be
distinguished at the level of K-theory.} This
$D$-brane configuration is indeed expected to depend on
a (relative) open-string modulus.

This line of arguments can be substantiated by using the relationship
between line bundles and divisors. Each line bundle in the class,
$\cE(1,0)$, can be specified by a divisor $\cO(\zeta_1-\zeta_2)$,
where $\zeta_1$ and $\zeta_2$ denote points on the torus, $\Sigma$.
The distance between them corresponds to the open-string modulus
$\zeta$ that appears in the transition function, $A(z)$, of the
line bundle $\cE(1,0)$. Physically, the point $\zeta_1$ describes
the position of a $D0$-brane, whereas the point $\zeta_2$ refers
to the position of a $\bar{D0}$-brane.  If, on the other hand,
$\zeta_1$ and $\zeta_2$ coincide, or in other words, if the $D0$-brane
is on top of the $\bar{D0}$-brane, we arrive at the divisor of a
trivial line bundle, which precisely corresponds to the exceptional
$4\times 4$-matrix factorization of a pure, rigid $\bar{D2}$-brane.

The appearance of distinguished matrix factorizations at special
points in the moduli space has been observed before in ref.~\Laza.
According to the classification of holomorphic vector bundles on
the torus~$\Sigma$ \Atiyah, each vector bundle of rank~$r$ and degree
zero is classified by a degree zero line bundle on $\Sigma$, namely
by the determinant line bundle of the vector bundle in question.
Moreover, each line bundle of degree zero has vanishing first Chern
class and according to eq.~\bundleequiv\ is given by the transition
function $A(z)=y^{-1}$ on~$\IC^*$~\refs{\PolishchukAppell,\PolishchukDB}.
Note that for $y=1$, or for $\zeta=0$, the transition
function becomes the identity and hence the determinant bundle
becomes the trivial line bundle.\foot{Those
``exceptional'' bundles of rank $r$ and degree zero are also distinguished
by the fact that they have a global non-trivial section.} In
ref.~\Laza{} it is shown that such ``exceptional'' vector bundles in
${\cal E}(r,0)$ with trivial determinant bundle, correspond to
exceptional MCM modules ${\cal M}_r$,\foot{The MCM modules
${\cal M}_r$ are self-dual $\cal R$-modules, i.e. ${\cal M}_r \simeq
\Hom({{\cal M}_r,\cal R})$ \Laza.} and furthermore give rise to
exceptional matrix-factorizations, which are isolated in the sense
that there exist no  (open string) deformations that would stay  within the class
of {\it irreducible} factorizations.

To conclude this section, we consider how the bundle data encoded
in $J_\thrbythr$ and $E_\thrbythr$ evolves over the moduli space, 
${\frak M}^{3\times 3}_\zeta$.  Consider first the anti-D-brane, $\bar L$, for which 
(up to an unimportant sign) the r\^oles of $J_\thrbythr$ and $E_\thrbythr$ 
are exchanged, \ie\ the vector bundle data is encoded in the matrix 
$E_\thrbythr$ given in \JEthree. At a generic point in the open-string 
moduli space ${\frak M}^{3\times 3}_\zeta$, the vector bundle spanned 
by the columns of $E_\thrbythr$ is associated to a line bundle in 
$\cE(1,0)$ (\cf\   eq.~\Ethreebas).  Therefore one might naively expect that 
the exceptional $4\times 4$-matrix factorization at 
$\zeta=0$, corresponds to the exceptional rank one MCM module~${\cal M}_1$
 (it is actually the anti-version, $\bar{\widetilde P}_\fobyfo^0$, that we talk about here, as we started out with the anti-brane, $\bar L$).
This, however, is not true, because we should keep in mind that this $4\times 4$ 
matrix factorization has been constructed by adding a trivial brane-anti-brane pair. 
In particular the new column of $E_{4\times 4}$ in \threeequiv\ has increased 
the number of linearly independent columns of $E_{3\times 3}$ by one, and thus 
the $4\times 4$-matrix factorization is correctly identified with the rank two 
exceptional module,  ${\cal M}_2$.\foot{The self-duality of $\cM_2$ is reflected 
in the fact that the matrices $J_{4\times 4}$ and $E_{4\times 4}$ in \Mtwofour\ are 
antisymmetric.}  Comparison with ref.~\Laza\ does indeed confirm that the 
exceptional $4\times 4$-matrix factorization
is associated to the exceptional MCM module ${\cal M}_2$.

On the other hand, if we now consider the the $D$-brane, $L$, the
 bundle data are encoded in the matrix~$J$ of \JEthree. 
As explained in section~3.2, the $D$-brane $L$ gives rise to the rank two vector 
bundle ${\cal E}(2,3)$. As before, in order to analyze the limit $\zeta\to 0$,
we must not neglect the added trivial brane-anti-brane pair. However, 
 the matrix~$J$ is only enhanced by the block-diagonal entry, $W$, 
 (\cf\ eq.~\threeequiv), which does not increase the number of linear independent 
 columns of $J$ because $W\equiv 0$ on the torus, $\Sigma$. Therefore 
 the exceptional $4\times 4$-matrix factorization assigned to the 
 brane, $L$, at $\zeta=0$ must still be associated to a vector 
 bundle in ${\cal E}(2,3)$. Once again comparing with ref.~\Laza\ reveals 
 that the $4\times 4$-matrix factorization~\Mtwofour\ is really 
 mapped to the exceptional $4\times 4$-matrix factorization in $\cE(2,3)$, 
which, furthermore, is associated to the first syzygy module, $\Omega_1(\cM_2)$, 
of $\cM_2$. Note that this is precisely in accord with the physical picture 
because passing over to the first syzygy module, $\Omega_1(\cM_2)$, of $\cM_2$
corresponds to switching from the anti-$D$-brane, $\bar L$, to the $D$-brane,~$L$.

%%%%%%%%%%%%%%%%%%%%%%%%%%%%%%%%%%
\newsec{Tachyon condensation}
%%%%%%%%%%%%%%%%%%%%%%%%%%%%%%%%%%

The aim of this section is to generate new matrix factorizations via 
tachyon condensation.  In particular, we will show that any matrix 
factorization on the elliptic curve may, in principle, be obtained by 
iteratively condensing building blocks of $2\times 2$-matrix factorizations. 
So as to generate new matrices in a controlled fashion, we introduce 
the notion of equivariant matrix factorizations.   This enables us to
explain how different choices of tachyons for a given pair of matrix
factorizations will lead to different condensates.  Conversely, if
we seek a certain condensate, equivariance provides a very
useful set of constraints upon the form of the requisite tachyon.   

In the present and the next two sections we will use these ideas to generate
explicitly all rank two matrix factorizations.   

%%%%%%%%%%%%%%%%%
\subsec{Equivariant matrix factorizations}
%%%%%%%%%%%%%%%%%

In order to systematically generate, via tachyon condensation, a particular 
matrix factorization for a given set of $RR$~charges, it is necessary to 
have control over the $RR$~charges of the constituents of the condensate. 
For instance, we need to distinguish among the three $D$-branes 
comprised in the $2\times 2$-matrix factorization, $S$, and in the 
$3\times 3$-matrix factorization, $L$, in order to realize the different 
composites illustrated in \lfig\condensation. This is achieved by 
refining the description of $B$-type $D$-branes in terms of equivariant 
matrix factorizations along the lines of  \AshokZB\ (see also \WalcherTX). 

Since our analysis takes place at the Gepner point
of the K\"ahler moduli space, the appropriate Landau-Ginzburg 
model is specified in terms of the superpotential~\cubic\ {\it together} 
with the natural $\IZ_3$ orbifold action $\rho(k)$, $k\in \IZ_3$
\eqn\Orbact{\rho:\rho(k)\,x_\ell\,\mapsto\,\omega^k\,x_\ell \quad {\rm with} 
\quad \omega\,=\,e^{2\pi i\over 3} \ . }
Obviously, this orbifold action must also be taken into account in the 
characterization of $B$-type $D$-branes, for which the equivariant 
formulation of matrix factorizations becomes the appropriate framework. 
In practice this means that a $D$-brane, $P$, represented by a 
$n\times n$-matrix factorization \DbP\ is supplemented by two 
$\IZ_3$ representations $R_0$ and $R_1$ of dimension~$n$:
\eqn\DbPorb{P\equiv\left[\xym{(P_1,R^P_1) \ar@/^/[d]^{E_P} \cr 
(P_0, R^P_0) \ar@/^/[u]^{J_P}}\right] \ , }
such that, in addition to the factorization condition \matfac, one also 
requires the equivariance condition  \AshokZB:
\eqn\equivcon{\eqalign{
J_P(x)\,&=\,R^P_1(k^{-1})\,J_P(\rho(k)x)\,R^P_0(k) \ , \cr
E_P(x)\,&=\,R^P_0(k^{-1})\,E_P(\rho(k)x)\,R^P_1(k) \ . }}
Note that these relations also require an adjustment of the 
notion of gauge transformations. Namely, it is easy to infer 
that a gauge transformation \gauge\ induces also a conjugation 
transformation acting on the representations~$R^P_0$ and $R^P_1$ by
\eqn\gaugerep{R^P_0(k)\rightarrow  U_R^{-1}\,R^P_0(k)\,U_R \ , 
        \qquad R^P_1(k)\rightarrow U_L\,R^P_1(k)\,U_L^{-1} \ . }

We demonstrate this idea by going through the 
matrix factorizations that we have encountered so far. 
The easiest example is the trivial $D$-brane configuration, $V$, given 
by the $1\times 1$-matrix factorization, $J_V=1$, $E_V=W$, 
which corresponds to the vacuum.  For this configuration the equivariance 
conditions \equivcon\ are fulfilled as long as $R^V_0(k)=R^V_1(k)=
\omega^{a\,k}$ for any $a=1,2,3$. The triviality of the D-brane configuration 
manifests itself in the fact that the two representations 
$R^V_0(k)$ and $R^V_1(k)$ must be the same.

For the $2\times 2$-matrix factorization, $S$,  the equivariance yields 
three  possible representations~$R^S_0$ and $R^S_1$, which read 
\eqn\equivreptwo{R^S_0(k)\,=\,\Diag{\omega^{(1-a)\,k},\,\omega^{(1-a)\,k}} \,,
\qquad  R^S_1(k)\,=\,\Diag{\omega^{-a\,k},\,\omega^{-(a+1)\,k}} \ . }
Taking into account the orbifold $\IZ_3$ action therefore
results in three different $2\times 2$-matrix factorizations, 
which are distinguished by $a=1,2,3$. Thus in the orbifolded Landau-Ginzburg 
model the $2\times 2$-matrix factorization, $S_a$, gains an 
additional label, $a$, which specifies the choice of representations 
$R^S_0$ and $R^S_1$. 

Analogously, we find for the $3\times 3$-matrix factorization three possible
representations
\eqn\equivrepthree{R^L_0(k)\,=\,\omega^{(1-a)\,k}\,\id_{3\times 3} \ , \qquad 
                   R^L_1(k)\,=\,\omega^{-(a+1)\,k}\,\id_{3\times 3} \ ,}
which give rise to the three distinct branes,
$L_a$, labeled by $a=1,2,3$.

Note that the equivariant labels allow us to unambiguously 
distinguish among all the short- and long-diagonal branes illustrated in \lfig\branes. 
This use of equivariance  means that we can go beyond the vector bundles
discussed in ref.~\Laza\ and enables us to deal with objects that
are more general than vector bundles in the large radius limit.
In particular, recall that one of the short branes, $S_2$, is the $D0$-brane and 
this is not associated to a vector bundle but rather to a point-like sheaf.  

Tachyon condensation provides the opportunity to make extensive direct 
tests of this equivariant formulation of the $\IZ_3$ orbits of branes.  
We will now  examine this in detail  but first we need the appropriate
equivariant modification of the open-string spectrum \AshokZB. 
Just as in \FermCoh, a boundary changing operator $\Psi_{(P,Q)}$ of 
an open string stretching between the brane~$P$ and $Q$ can 
be pictured by the diagram:
\eqn\FermCohequ{\xym{(P_1,R_1^P) \ar@/^/[d]^{E_P} \ar[rrd] && 
		(Q_1,R_1^Q) \ar@/^/[d]^{E_Q}  \cr
                  (P_0,R_1^P) \ar@/^/[u]^{J_P} \ar[rru] && (Q_0,R_0^Q) \ar@/^/[u]^{J_Q} }
                  \setbox0=\hbox{$\scriptstyle \psi_1$}
                  \setbox1=\hbox{\kern-\wd0 $\scriptstyle\psi_0$}
                  \hbox{\kern-21.5ex \raise 13pt \box0 \lower 10pt \box1}
                  \hbox{\kern+20ex} \ .} 
In the equivariant context, the physical state conditions, \PsiPhys, 
of the maps, $\psi_0$ and $\psi_1$, of (the fermionic) operator, $\Psi_{(P,Q)}$,
are now supplemented by the equivariance condition\foot{A similar 
equivariance condition applies for the bosonic operators.}
\eqn\equivconferm{\eqalign{\psi_0(x)\,&=\,R_1^Q(k^{-1})\,
\psi_0(\rho(k)x)\,R_0^P(k) \ , \cr
\psi_1(x)\,&=\,R_0^Q(k^{-1})\,\psi_1(\rho(k)x)\,R_1^P(k) \ . }}
This condition amounts to a selection rule upon the original
(non-equivariant) tachyon spectrum.  The selection rule tells
us which, if any, tachyon appears between different members
of the $\IZ_3$ families.  It also tells us the $\IZ_3$ label of
the condensed state.  We now illustrate this with some 
examples.

%%%%%%%%%%%%%%%%%
\subsec{A reprise of tachyon condensations of $P_\twobytwo$}
%%%%%%%%%%%%%%%%%

In calculating tachyon condensations, our {\it modus operandi} consists
of two basic steps:
First we analyze the boundary changing operators of the constituents  in 
terms of equivariant matrix factorizations. Then we build up the composites  via 
the cone construction, \tachcondiagram, and simplify the result (if necessary) 
using gauge transformations, \gauge\ and \gaugerep.
We begin by returning to the examples of tachyon condensation already 
introduced in section 1.2. 

The simplest example is brane/anti-brane annihilation between 
 $S_a(\alpha)$ and $\bar S_b(\beta)$. Recall that the fermionic identity operator, 
 \TachB, appears  in the open-string spectrum of the boundary changing 
sector provided that $\alpha=\beta$.  Moreover, equivariance imposes an 
additional constraint on the existence of this operator, \ie\ due to \equivconferm\ 
the representation $R_0^S$ and $R_1^{\bar S}$ acting on $\psi_0=\id_{2\times 2}$ 
and the representations $R_1^S$ and $R_0^{\bar S}$ acting on $\psi_1=\id_{2\times 2}$ 
must pairwise coincide. In other words the fermionic identity operator, \TachB,  
can only form a condensate of $S_a(\alpha)$ with $\bar S_a(\alpha)$.\foot{In 
equivariant matrix factorizations going from the brane~$P$ to the anti-brane, 
$\bar P$, does not only exchange the matrices $(J_P,E_P)$ with $(-E_P,-J_P)$ 
but is also accompanied by a flip of the representations $(R_0^P,R_1^P)$ to 
$(R_1^P,R_0^P)$.} 

Applying the cone construction  to the constituents $S_a(\alpha)$ and $\bar 
S_a(\alpha)$ using the boundary changing operator, \TachB, one obtains  
the following composite:
\eqn\vaccond{J_V\,=\,\pmatrix{J_{2\times 2}(\alpha) & \id_{2\times 2} \cr 0 & 
-E_{2\times 2}(\alpha)} \ , \qquad  E_V\,=\,\pmatrix{E_{2\times 2}(\alpha) &
 \id_{2\times 2} \cr 0 & -J_{2\times 2}(\alpha)} \ . }
Note that the constant entries in the matrices, $J_V$ and $E_V$,  mean that
one may make elementary simplifications by ``row and column elimination''. 
 Here we simply observe that because there are  two independent constant entries 
in $J_V$ and in $E_V$,  the method of  row and column elimination 
shows that the composite, \vaccond,
is gauge equivalent to four trivial brane-anti-brane pairs. Thus, within the 
$D$-brane category of $B$-type $D$-branes, the tachyon 
condensation of $S_a$ and $\bar S_a$ describes the annihilation
to the vacuum:
\eqn\Twoanni{S_a(\zeta)\ \condense_\Psi\ \bar{S}_a(\zeta) \Longrightarrow V \ , }
where $\zeta$ is given by $\alpha_\ell=\mu_\ell(\zeta)$.
 
The other example considered in Section 1.2 was  the 
condensation of  $S_a(\alpha)$ with $S_b(\beta)$.  
The relevant boundary-changing operator, $\Psi_{(S_a,S_b)}(\alpha,\beta)$, 
is given in eq.~\TachC\  and the condensation process has already been described
in section 1.3.    Here,  we want to re-examine the discussion
so as to elucidate the selection rules arising from equivariance. 
First, the form of the tachyon, $\Psi_{(S_a,S_b)}(\alpha,\beta)$, 
constraints the equivariance labels, $a$ and $b$. That is to say, that evaluating condition 
\equivconferm\ together with the representations \equivreptwo\ of $S_a$ and $S_b$ 
yields the relation $b=a+1$.  Hence, the boundary changing operator  
$\Psi_{(S_a,S_b)}(\alpha,\beta)$ only appears in the open-string cohomology 
for the ``equivariant pair'' $S_a(\alpha)$ and $S_{a+1}(\beta)$, and only for such 
a pair can this fermionic boundary operator be used to construct the composite,
$\bar S_c(\gamma)$, determined in eq.~\MtwotwoCgaugeB.

Furthermore, we can also determine the equivariance label, $c$, of the resulting 
composite, $\bar S_c(\gamma)$.   The representations, \equivreptwo, 
of the constituents,  $S_a(\alpha)$ and $S_{a+1}(\beta)$, yield, for the 
untransformed composite, \MtwotwoC\ the representations
\eqn\MtwotwoCreps{\eqalign{R_0(k)\,&=\,\Diag{ \omega^{-a\,k},
\omega^{-a\,k},\omega^{(1-a)\,k},\omega^{(1-a)\,k}} \ , \cr
R_1(k)\,&=\,\Diag{ \omega^{(2-a)\,k},\omega^{(1-a)\,k},
\omega^{-a\,k},\omega^{-(a+1)\,k} } \ ,}}
As discussed in section 1.3,  by acting with a gauge transformation, $U_L$, $U_R$, 
the composite, \MtwotwoC, can be cast into the $4\times 4$-matrix factorization 
\MtwotwoCgauge. However, this gauge transformation also induces a 
corresponding conjugation action on the representations, \MtwotwoCreps, 
according to \equivconferm, and hence, after dropping all the trivial brane-anti-brane 
pairs, we readily read off the resulting equivariant $2\times 2$-matrix factorization, 
which turns out to be $\bar S_{a+2}(\gamma)$.

Before we move on to the next example, we can gain some further insight into 
the open-string parameters, $\gamma_\ell$,  given in \Defzeta, by rewriting
them in terms of uniformizing parameter along the lines of eq.~\alphauni. Namely, 
with $\alpha_\ell=\mu_\ell(\zeta)$ and $\beta_\ell=\mu_\ell(\lambda)$ the 
parameter $\gamma_\ell$ can be uniformized by $\gamma_\ell=\mu_\ell(\zeta-\lambda)$, 
where we used eq.~\Eij\ after substituting $x_\ell$ by $\beta_\ell$. 
Thus schematically the tachyon condensation process is summarized by
\eqn\TwoTwotobarTwo{
S_a(\zeta) \ \condense_{\Psi_{(S_a,S_{a+1})}}\ S_{a+1}(\lambda) 
\Longrightarrow \bar S_{a+2}(\zeta-\lambda) \ . }

The next task is to construct the long-diagonal $D$-branes, $L$,  by tachyon 
condensation of two short-diagonal $D$-branes~$S$. From \lfig\condensation\ we 
readily infer that one is required to condense the brane $S_a(\alpha)$ 
with $\bar S_{a+2}(\beta)$ in order to reach $L$ as a condensate. 
Thus we seek a fermionic boundary changing operator, 
$\Psi_{(S_a,\bar S_{a+2})}(\alpha,\beta)$,  in the spectrum of open-strings 
stretching from $S_a$ to $\bar S_{a+2}$.\foot{The precise argument is actually a bit 
more involved: From the $A$-model mirror picture we know that $S_a$ intersects 
with $\bar S_{a+2}$ once, and hence we expect (at least) either one 
fermionic operator or one bosonic operator in the spectrum. The orientation 
of the intersection tells us whether the operator is bosonic or fermionic. 
Since the intersection of $S_a$ and $S_{a+1}$ has the same orientation as 
$S_a$ with $\bar S_{a+2}$ the fermionic operator, $\Psi_{(S_a,S_{a+1})}$, 
implies the presence of a fermionic operator $\Psi_{(S_a,\bar S_{a+2})}$.} 
By once again taking advantage of the equivariance condition, \equivconferm,  
we readily deduce the degrees of the entries of the fermionic boundary 
changing operators:
\eqn\TachForLdegree{\psi_0\,=\,\pmatrix{ l_1(x) & l_2(x) \cr l_3(x) & l_4(x) } \ ,
\qquad  \psi_1\,=\,\pmatrix{ l_5(x) & q(x) \cr c & l_6(x) } \ .  }
Here $c$ has degree zero, $l_1$, \dots, $l_6$ are linear and $q$ is quadratic in 
$x_\ell$. A detailed analysis of the corresponding fermionic cohomology 
element along the lines of eq.~\PsiPhys\ and \PsiExact\ eventually reveals:
\eqn\TachForLdegree{
\psi_0\,=\,\pmatrix{ G_1(\alpha,\beta) & G_2(\alpha,\beta) \cr 
G_3(\alpha,\beta) & G_1(\beta,\alpha) } \ , \qquad
\psi_1\,=\,\pmatrix{ G_4(\alpha,\beta) & H(\alpha,\beta) \cr 
C(\alpha,\beta) & G_4(\beta,\alpha) } \ ,}
with
\eqn\TachForLentriesC{C(\alpha,\beta)\,=\,-\alpha_3\beta_3
\left(\alpha_3\beta_2-\alpha_2\beta_3\right) \ , }
and
\eqn\TachForLentriesL{\eqalign{
   G_1(\alpha,\beta)\,&=\,-\left(\alpha_3\beta_1+\alpha_1\beta _3\right)x_2
                          +{\alpha_3\left(\beta_2^3+\beta_3^3\right)
                            \over\beta_1\beta_3}x_3  \ , \cr
   G_2(\alpha,\beta)\,&=\,\left(\alpha_2\beta_3-\alpha_3\beta_2\right) \, x_2 \ , \cr
   G_3(\alpha,\beta)\,&=\,\left(\alpha_3\beta_2-\alpha_2\beta_3\right)x_1
                         +\left({\alpha_3\beta_2^2\over\beta_3}-{\alpha_2^2\beta_3\over\alpha_3}\right)x_3 \ , \cr 
   G_4(\alpha,\beta)\,&=\,{\alpha_3\beta_2\left(\alpha_3\beta_2-\alpha_2\beta_3\right)\over\beta_1\beta_3}x_1
                         +{\alpha_3\left(\alpha_2\beta_1+\alpha_1\beta_2\right)\over\beta_2}x_2
                         -{\alpha_2\alpha_3\left(\beta_2^3+\beta_3^3\right)\over\beta_1\beta_2\beta_3}x_3 \ , }}
and
\eqn\TachForLentriesQ{\eqalign{
   H(\alpha,\beta)\,&=\,{\alpha_2\beta_2\left(\alpha_3\beta_2-\alpha_2\beta_3\right)\over\alpha_1\alpha_3\beta_1\beta_3}x_1^2
                       +{\left(\alpha_2\alpha_3\beta_1^2-\alpha_1^2\beta_2\beta_3\right)
                              \over\alpha_2\alpha_3\beta_2\beta_3}x_2^2 \cr
                &\quad +{\left(\alpha_2^3\beta_1^2\beta_3-\alpha_1^2\alpha_3\beta_2^3\right)
                              \over\alpha_1\alpha_2\alpha_3\beta_1\beta_2\beta_3}x_1x_2
                              +{\left(\alpha_3^3\beta_2^3-\alpha_2^3\beta_3^3\right)
                              \over\alpha_1\alpha_2\alpha_3\beta_1\beta_2 \beta _3}x_1x_3
                       +{\alpha_3\beta_2-\alpha_2\beta_3\over\alpha_2\beta_2}x_2x_3 \ . }}
Next we use this fermionic boundary-changing operator, 
$\Psi_{(S_a,\bar S_{a+2})}(\alpha,\beta)$, to form the composite 
\eqn\TachConL{\hat J_L(\alpha,\beta)\,=\,\pmatrix{-E_{2\times 2}(\beta) & 
\psi_0(\alpha,\beta) \cr 
 0 & J_{2\times 2}(\beta) } \ , \qquad \hat E_L(\alpha,\beta)\,=
 \,\pmatrix{-J_{2\times 2}(\beta) & \psi_1(\alpha,\beta) \cr 
0 & E_{2\times 2}(\beta) } \ . }
Note that the matrix $\hat E_L$ contains a constant entry, which allows us to perform  
row and column eliminations so as to obtain a $3\times 3$-matrix factorization. 
Moreover, from the degrees of the entries of $\hat E_L$, it is apparent that 
the row and column operations precisely remove the linear entries in $\hat E_L$. 
Therefore the matrix $\hat E_L$ becomes, after removing the trivial brane-anti-brane 
pair, precisely the matrix $E_{3\times 3}$ with only quadratic entries. 
Conversely,  after row and column elimination,  $\hat J_L$ contains
only linear entries and takes the form the matrix~$J_{3\times 3}$. 

The final task is to relate the result of this  row and column elimination 
to the standard form of the $3\times 3$-matrix factorization stated in \JEthree, and
thereby  determine the precise form of the brane, $L$.  This is achieved
by an appropriate gauge transformation that allows us to determine the 
open-string parameter of the composite in terms of $\alpha_\ell$ and $\beta_\ell$. 
The result of this, straightforward but tedious, analysis yields 
(in terms of the parameters $\zeta$ and $\lambda$ of 
the uniformized functions $\alpha_\ell=\mu_\ell(\zeta), \beta_\ell=\mu_\ell(\lambda)$):
\eqn\TwoTwoToThree{
S_a(\zeta) \ \condense_{\Psi_{(S_a,\bar S_{a+2})}}\ \bar S_{a+2}(\lambda) 
\Longrightarrow L_a(\coeff{1}{3}(\zeta-\lambda))  \ . }

Before we conclude this section,  a few comments are in order: First,
note that the factor $\coeff{1}{3}$ in the uniformizing function~$\mu_\ell$
correctly reproduces the periodicity of the $3\times 3$-matrix
factorization observed in eq.~\mspacethree. Furthermore, the
enhancement of the $3\times 3$-matrix factorization to the $4\times
4$-matrix factorization observed in section~3.4 also becomes
manifest. Namely, the indecomposable $4\times 4$-matrix factorization
is obtained for $\lambda\to\zeta$, or $\beta_\ell\to\alpha_\ell$. In
this limit, the constant entry \TachForLentriesC\ vanishes and the
composite, \TachForLdegree,  becomes gauge equivalent to the
exceptional factorization $\widetilde P_\fobyfo^0$ in \Mtwofour.
 
%%%%%%%%%%%%%%%%%%%%%%%%%%%%%%%%%%
\newsec{Constructing more general matrix factorizations}
%%%%%%%%%%%%%%%%%%%%%%%%%%%%%%%%%%

So far we have analyzed some examples of tachyon condensation among the known
factorizations $P_\twobytwo$ and $P_\thrbythr$, and shown how the
latter  can be obtained as a condensate of the $\twobytwo$
factorization. We have also seen how the $4 \times 4$ factorization
$\widetilde P_\fobyfo$, which is more naturally associated with
boundary fermions, can be produced by condensation and how it
connects to the moduli space of the $3 \times 3$ factorization.

We now wish to go beyond this and obtain new factorizations.  There
are several techniques that we can use to do this, and in this
section we will describe them and determine the  tachyons
that we will need to create a number of new factorizations.  In the
subsequent sections, we will present a list of the resulting
factorizations and discuss some of their properties.

%%%%%%%%%%%%%%%%%
\subsec{New matrix factorizations and tachyons from transpositions}
\subseclab\Transposes
%:transposes
%%%%%%%%%%%%%%%%%
Given a matrix factorization $P=(J,E)$, one can obtain another matrix 
factorization by transposing the matrices $J$ and $E$. We denote the
resulting factorization by
\eqn\MFtranspose{\Transpose{P}=(\Transpose{J},\Transpose{E}) \ .
}
According to the equivariance condition \equivcon,  transposition 
also acts upon the equivariant representations, $R_{0,1}^P$, by:
\eqn\MFTranposeRep{R_0^{\Transpose{P}}(k)\,=\,R_1^{\Transpose{P}}(k^{-1}) \ , \qquad
                   R_1^{\Transpose{P}}(k)\,=\,R_0^{\Transpose{P}}(k^{-1}) \ .}

Applying the transposition operation to the branes $S_a$ and $L_a$ we find
(modulo simple gauge transformations) the corresponding transposed
matrix factorizations to be:
\eqn\MFtransposecases{\Transpose{S_a}(\zeta) \,\sim\, \bar S_{-a-1}(\zeta) \ , \qquad
                      \Transpose{L_a}(\zeta) \,\sim\, L_{-a}(-\zeta) \ . }
In general the transposed matrix factorization can give rise to new matrix factorizations,
which are not related to the original factorization by changing the open-string modulus or
by passing over to the anti-brane. We will encounter  such an example in  with 
the two $5\times 5$ factorizations, $P_{5\times 5}$ and $P_{5\times 5}^{\rm T}$, that
describe two distinct classes of $D$-branes.

In the language of MCM modules, the transposed matrix factorization is associated to the dual
MCM module. In particular, eq.~\MFtransposecases\ illustrates that the brane, $L$, at $\zeta=0$
is invariant under transposition, which reflects the self-duality of the corresponding
exceptional MCM module at $\zeta=0$ in the open-string moduli space as discussed in Section~4.

Suppose we have a known tachyon, $\Psi_{(P,Q)}$,
satisfying the physical state condition \PsiPhys. Then the transposed physical state condition reads
\eqn\PsiPhysT{\eqalign{
0\,&=\,\psi_0^{\rm T}\,E_Q^{\rm T}
+J_P^{\rm T}\, \psi_1^{\rm T}\ , \cr
                      0\,&=\,\psi_1^{\rm T}\,J_Q^{\rm T}
+E_P^{\rm T}\,\psi_0^{\rm T} \ ,}}
which we readily identify with the physical state conditions for a fermionic boundary changing
operator, $\Psi_{(\Transpose{Q},\Transpose{P})}\sim(\Transpose{\psi_0},\Transpose{\psi_1})$,
of an open string stretching between $\Transpose{Q}$ and $\Transpose{P}$.

To illustrate this procedure we choose as an example the fermionic boundary
changing operator, $\Psi_{(S_a(\lambda),L_{a+1}(\zeta))}$ of an open string 
stretching between $S_a(\lambda)$ and $L_{a+1}(\zeta)$. This immediately gives
rise to the boundary changing operator, $\Psi_{(\Transpose{L_{a+1}}(\zeta),\Transpose{S_a}(\lambda))}$,
of an open string stretching between $\Transpose{L_{a+1}}(\zeta)$ and $\Transpose{S_a}(\lambda)$, which
we can finally convert according to \MFtransposecases\ to the fermionic boundary changing operator
\eqn\Tmap{\Psi_{(L_{-a-1}(-\zeta),\bar S_{-a-1}(\lambda))}\sim
          \Psi_{(\Transpose{L_{a+1}}(\zeta),\Transpose{S_a}(\lambda))} \ , }
where those two boundary changing operators are related by the same 
gauge transformation that gives rise
to the corresponding identification in eq.~\MFtransposecases.

%%%%%%%%%%%%%%%%%
\subsec{Creating bound states at threshold}
%%%%%%%%%%%%%%%%%

In obtaining  matrix factorizations via condensation, there is a
significant difference between the situation where the D-brane
charges, $(r,c_1)$, of the end product are co-prime or have common
factors.  In particular, when they have a common factor, it turns
out that there is a simple canonical procedure for obtaining the
matrix factorization as a ``bound state at threshold''.

To illustrate the basic physical idea it is easiest to use the A-brane 
mirror picture where
the charges $(r,c_1)$ correspond to  winding numbers $(p,q)$.  The
tension $m$ of a BPS-saturated $D1$-brane is then simply proportional
to its length on the covering space of $\Sigma$, as shown in
\lfig\branes. That is, 
\eqn\BPSbound{
m\ =\ |q+\rho\, p|\ ,
}
where $\rho$ is the complex structure parameter of the mirror torus.
It coincides with the K\"ahler parameter of the original $B$-model,
which decouples in the topologically twisted theory so that we have no 
control over it.\foot{In the Landau-Ginzburg
model, which corresponds a $\IZ_3$ orbifold of the torus CFT, $\rho$
is fixed to be a third root of unity, but in the orbifolded Landau-Ginzburg
model, the K\"ahler modulus $\rho$ becomes a free parameter.}  However,
things are simple on the torus; in particular, there are no lines
of marginal stability on the K\"ahler moduli space, so all what matters is that
the tension is linear in the charges. This implies
that if we only know the charges, $(p,q)$, we cannot distinguish a
single string with winding numbers $(np,nq)$ from $n$ strings, each
with winding number $(p,q)$.  Such single-string configurations are
commonly referred to as bound states at threshold \SenYI.\foot{Of
course, when $(p,q)$ are co-prime, all multiple string configurations
with the same net winding number  cannot satisfy the BPS bound
because the triangle inequality implies that their total length
will not be minimal.} In general, it is often unclear whether a
combination of states with equal charges will actually form a true
bound state at threshold, or whether the state is simply a
multi-particle superposition of the constituents.

In matrix factorizations, where we have
additional information in form of the open-string moduli,  we can,
in fact, easily see the difference between a bound state at threshold 
and a mere direct sum of its constituents. In the following,
we consider only the condensation of a pair of identical branes
but more general configurations can be treated in a similar way.

Consider first a pair of identical branes, $P$, but with 
distinct position moduli, $\alpha$ and $\beta$. It is clear that 
as long as $\alpha\not=\beta$, the
branes  do not intersect and thus there isn't a  tachyon that
 could possibly lead to bound state formation. The combined
system is characterized by a block-diagonal matrix factor
\eqn\directsum{J_{PP}(\alpha,\beta)\,=\,\pmatrix{J_P(\beta) & 0 \cr 
0 & J_P(\alpha) } \ , 
}
and similarly for $E_{PP}$.  The vector bundles corresponding to
$J_{PP}$ and $E_{PP}$  are thus trivial direct sums.  
What is the moduli space of such a configuration?\foot{We thank Robert 
Helling for asking this
question, which then lead to the following discussion.} The naive answer is 
that  it is simply given by the direct product of the individual
moduli spaces. However, there exists a gauge transformation that
switches the r\^oles of the block matrices. That is, $J_{PP}(\alpha,\beta)$
is gauge equivalent to $J(\beta,\alpha)_{PP}$, and since gauge equivalences
are equivalences in the full category, it follows that the open-string
moduli space is given by the symmetrized product of the
individual moduli spaces, \ie,
\eqn\symprod{{\frak M}_{PP}\,=\, {\rm Sym}^2(\Sigma) \ , 
}
which is entirely to be expected because the $D0$-branes are 
indistinguishable.

The identifying gauge transformations have a fixed point when the branes 
coincide and this leads to an orbifold singularity in the moduli
space when $\beta=\alpha$.   This is reflected in the physics in that, when
the branes move on top of one another, the
cohomology jumps and a tachyon appears (in a similar manner to the
results discussed in Section~1).\foot{Again, both a bosonic
and a fermionic tachyon appear at $\beta = \alpha$, so that there is no
net contribution to the intersection index.} Since the physical state
 condition at $\beta=\alpha$
is identical to the equation that determines the boundary preserving
endomorphism,  $\Omega_P$, the new operator
$\Omega_{(P,P)}\equiv\Psi_{(P,P)}$ has the exactly the same form
as $\Omega_P$, however, due to the implicit Chan-Paton labels it is really
a boundary changing operator because it acts between different
$D$-branes.

Because the ``tachyon'' $\Omega_{(P,P)}=(\Omega_{(P,P)}^{(J)},\Omega_{(P,P)}^{(E)})$ has charge one, it is a
marginal operator of the theory that couples to a dimensionless modulus,
$\zeta_\Omega$:
\eqn\notdirectsum{J_{PP}(\alpha,\zeta_\Omega)\,=\,\pmatrix{J_P(\alpha) & 
\zeta_{\Omega}\, \Omega_{PP}^{(J)}(\alpha) \cr 
0 & J_P(\alpha) } \ . 
}
Switching on
$\zeta_\Omega$ condenses the two-brane system
into one indecomposable object, which is a genuine bound state at
threshold.  Indeed, the off-diagonal terms in \notdirectsum\ 
generically transform non-trivially between coordinate patches on the 
torus and thereby make the vector bundle defined by \notdirectsum\ 
into a non-split extension of the original vector bundles.
This clearly shows the usefulness of the extra moduli
information carried by the factorization, as the existence of such
an object cannot be inferred from the $K$-theory charges alone.
Mathematically, the interpretation of this extra degree of freedom
is that it resolves the singularity of ${\frak M}_{PP}$.  We have depicted the
situation in \lfig\threshold.

\vskip 3mm
%%%%%%%%%%%%%%%%%%%%%%%%%%%%
\figinsert\threshold{The upper part of this diagram shows the moduli space,
${\frak M}_{PP}\simeq {\rm Sym}^2(\Sigma)$, of two identical branes but
with independent moduli $\alpha$ and $\beta$.   A new branch emerges at 
the singularity at $\alpha=\beta$.   Switching on $\zeta_\Omega$ moves
the factorization onto this new branch and the
reducible two-brane system condenses into an indecomposable bound 
state at threshold.}{1.5in}{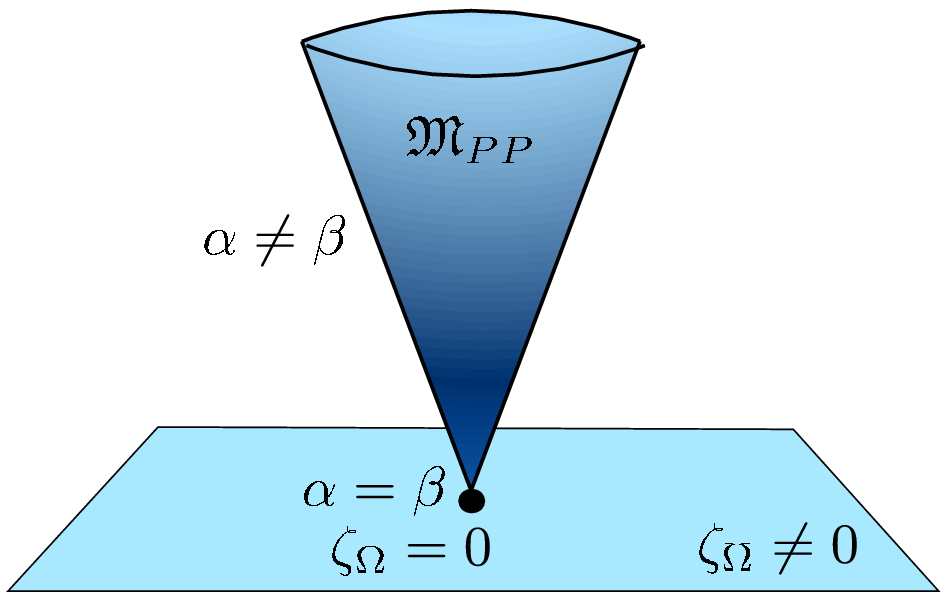}
%%%%%%%%%%%%%%%%%%%%%%%%%%%%

The tachyon, $\Omega_{(P,P)}$, responsible for 
creating bound states at threshold, can be explicitly 
obtained by following an observation in ref.~\BrunnerMT.    
Let $P=(J_P(\alpha), E_P(\alpha))$ be a matrix factorization, 
where $\alpha_\ell$ are the usual functions of the  brane moduli:  
$\alpha_\ell=\mu_\ell(\zeta)$.  One has
\eqn\MFcondn{
E_P(\alpha) \cdot J_P(\alpha) = W\ \id\ .
}
Taking the derivative of the above equation with respect to $\zeta$, one obtains
\eqn\MFcondna{
E_P(\alpha) \cdot \partial_\zeta J_P(\alpha) + 
\partial_\zeta E_P(\alpha) \cdot J_P(\alpha) =
0\ ,
}
which becomes the physical state condition, \PsiPhys, for the following 
tachyon between identical branes:
%
%\eqn\boundpsi{
%\Omega_ {(P,P)}: \ \ \cases{&$\psi_0 = \partial_\zeta J_P(\alpha)$ \cr
%&$\psi_1 = \partial_\zeta E_P(\alpha)$\ .  }
%}
\eqn\boundpsi{
\Omega_{(P,P)}^{(J)}\ =\ \partial_\zeta J_P(\alpha),\qquad
\Omega_{(P,P)}^{(E)}\ =\ \partial_\zeta E_P(\alpha).
}
That this operator is not exact follows from the fact that its boundary preserving
version describes the non-trivial marginal operator associated with the open-string modulus, 
$\zeta$.

For our present purposes, it is simpler to work with derivatives  with respect 
to the $\alpha_\ell$, rather than with respect to the flat coordinate, $\zeta$. 
However, the $\alpha_\ell$ are not independent since they must lie on the cubic 
curve. This is easily taken into account by taking 
a particular linear combinations of derivatives. We find that the
\eqn\alphader{
\cD(\alpha)\equiv
\sum_\ell \dalpha_\ell {{\partial}\over{\partial \alpha_\ell}}  \ ,
}
where $\dalpha_\ell \equiv \mu_\ell(-2 \zeta)$, does the job {\sl in lieu} of 
the $\zeta$-derivative in \boundpsi.\foot{The $\dalpha_\ell$ satisfy the following two identities,
enabling us to show that $\cD\propto \partial/\partial\zeta$: 
$\hphantom{a}\qquad\qquad\sum_\ell \dalpha_\ell \alpha^2_\ell=0\quad  
{\rm and}\quad  \dalpha_1 \alpha_2 \alpha_3 
+\dalpha_2 \alpha_3 \alpha_1+\dalpha_3 \alpha_1 \alpha_2 =0\ .$} 
To summarize, we can write
the following concise cohomology representative for the ``tachyonic modulus'':
\eqn\omegatrick{
\Omega_{(P,P)}(\al)\ =\ \cD(\alpha)\cQ_P(\alpha)\ .
}
%

%%%%%%%%%%%%%%%%%
\subsec{Completing the links in the quiver diagram:  Tachyons between $S$ and $L$}
%%%%%%%%%%%%%%%%%

The spectrum of open strings connecting long branes  to short 
branes/anti-branes, falls into four distinct classes (ignoring equivariance
labels):
\eqn\longshorttachyons{
\Psi_{(S,L)}\ ;\ 
\Psi_{(L,S)}\ ;\quad 
\Psi_{(L,\bar{S})}\ {\rm and}\ 
\Psi_{(\bar{S},L)}\ .
}
The tachyons for the last two classes may be obtained from the first two
using transposition as  outlined in Section~6.1.
Thus, we shall focus here on the first
two classes alone. The prediction based on the quiver diagram shown
in \lfig\quiver\ is that there should be one tachyon  of type 
$\Psi_{(S_a,L_{a+1})}$,
one tachyon of type $\Psi_{(S_a,L_{a})}$ and two tachyons of type
$\Psi_{(L_{a},S_{a+1})}$. The 
degrees of the terms in the relevant tachyon operators 
are obtained by using the 
equivariance condition \equivconferm. We find that the number
of solutions to  \PsiPhys\ is indeed consistent with the quiver 
diagram.

We now present the explicit expressions that we find for the tachyons.  
We denote the moduli associated with the $L$ and $S$
branes by $\alpha_\ell\equiv\mu_\ell(\zeta)$  and $\beta_\ell
\equiv\mu_\ell(\lambda)$, respectively, and let 
$\gamma_\ell \equiv \mu_\ell(-\zeta-\lambda)$.  We then find the following
rectangular matrices for the first kind of tachyons:
\eqn\twothreea{
\Psi_{(S_{a},L_{a+1})}:\ \cases{ &$\psi_0= \pmatrix{
\alpha_1 \gamma_3 & \alpha_3 \gamma_2 \cr
 \alpha_2 \gamma_2 & \alpha_1 \gamma_1 \cr
 \alpha_3 \gamma_1 & \alpha_2 \gamma_3
}$ \ ,\cr
&$\psi_1=\pmatrix{
-\beta_3 \gamma_3 & {{x_1 \alpha_2 
\gamma_1}\over{\alpha_1 \beta_1}}+{{x_3 \alpha_1 
\gamma_1}\over{\alpha_3 \beta_2}}-{{x_2 \alpha_1 
\gamma_2}\over{\alpha_2 \beta_2}} \cr
 -\beta_3 \gamma_1 & {{x_1 \alpha _1 
\gamma_2}\over{\alpha_3 \beta_1}}+{{x_3 \alpha_3 
\gamma_2}\over{\alpha_2 \beta_2}}-{{x_2 \alpha_3 
\gamma_3}\over{\alpha_1 \beta_2}} \cr
 -\beta_3 \gamma _2 & -\frac{x_2 \alpha_2 
\gamma_1}{\alpha_3 \beta_2}+\frac{x_1 \alpha_3 
\gamma_3}{\alpha_2 \beta_1}+\frac{x_3 \alpha_2 
\gamma_3}{\alpha_1 \beta_2}}$ \ ,
}}
where $a=1,2,3$. This tachyon gives rise to the composite as given below:
\eqn\compone{
S_a(\lambda)\ \condense_{\Psi_{(S_a,L_{a+1})}} \ L_{a+1}(\zeta)
\Longrightarrow\, S_{a+1}(3\zeta+\lambda)\ .
}
In arriving at \compone, 
we used the constant entries in the tachyon to carry out row and column
reductions and discarded three `trivial' 
$P_{1\times 1}$ and $\bar{P}_{1\times 1}$ vacuum pieces.

The tachyon $\Psi_{(S_{a},L_{a})}$ can be shown to be gauge equivalent to
$x_1$ times the tachyon $\Psi_{(S_{a},L_{a+1})}$   in \twothreea:
\eqn\twothreeax{
 \Psi_{(S_{a},L_{a})} ~=~ x_1 \ \Psi_{(S_{a},L_{a+1})} \ .
}
It is easy to see that any function linear in the $x_i$ multiplying
$\Psi_{(S_{a},L_{a+1})}$ will satisfy the physical state condition.
However, it turns out that there is only one non-trivial cohomology
element that can be obtained in this way.  This is the tachyon that
leads to a new $5\times 5$ factorization that will be discussed
later.

The last kind of tachyons comes with a multiplicity of two, and we 
distinguish them by adding a superscript:
\eqn\twothreeb{
\Psi^{(1)}_{(L_{a},S_{a+1})}:\ \cases{
&$\psi_0=\pmatrix{
\frac{x_2 \gamma_3}{\beta_1 \beta_3}-\frac{x_1 \alpha_3 \gamma_1}{\alpha_2 
\beta_1 \beta_3} & \frac{x_2 \alpha_2 \gamma_1}{\alpha_3 \beta_2 \beta_3}
+\frac{x_3 \gamma_2}{\beta_1
\beta_2} & -\frac{x_3 \gamma_1}{\beta_1 \beta_2}+\frac{x_2 \alpha_3 \gamma_2}{\alpha_2
\beta_1 \beta_2}+\frac{x_1 \alpha_1 \gamma_2}{\alpha_3 \beta_1 \beta_3} \cr
 0 & -\gamma_2 & \gamma _1
}$\ ,\cr
&$\psi_1=\pmatrix{
\frac{x_1 \alpha _3 \gamma _1}{\alpha _1 \alpha _2 \beta _1}+\frac{x_2 \gamma _1}{\alpha _3 \beta_3}
+\frac{x_3 \gamma_2}{\alpha_2 \beta_1} & \frac{x_1 \gamma_3}{\alpha_3 
\beta_3}-\frac{x_2 \gamma_2}{\alpha_1 \beta_3} & 
\frac{x_1 \beta_2 \gamma_2}{\alpha_3 \beta_1
   \beta_3}-\frac{x_3 \gamma_1}{\alpha_1 \beta _1} \cr
 \frac{x_2 \left(\alpha_3^2 \beta_3 \gamma_1-\alpha _2^2 \beta_1 \gamma_2\right)}{\alpha _1
   \alpha _2 \alpha_3 \beta_1 \beta_3} & \frac{x_2 \gamma_3}{\alpha_3 \beta _3}-\frac{x_1
   \gamma_1}{\alpha_2 \beta_3} & \frac{x_1 \gamma_2}{\alpha_3 \beta_3}-\frac{x_2 \gamma_3}{\alpha_2 \beta_1}
}$,}
}
and
\eqn\twothreec{
\Psi^{(2)}_{(L_{a},S_{a+1})}:\ \cases{
&$\psi_0=\pmatrix{
\frac{x_3 \gamma_2}{\beta_1 \beta_2}-\frac{x_2 \alpha_2 \gamma_3}{\alpha_1 
\beta_1 \beta_2}-\frac{x_1 \alpha_3 \gamma_3}{\alpha_2 \beta_1 \beta_3}
 & \frac{x_1 \alpha _2 \gamma_2}{\alpha_1 \beta_1 \beta_3}
-\frac{x_2 \gamma_1}{\beta_1 \beta_3} & -\frac{x_2 \alpha_1
   \gamma_2}{\alpha_2 \beta_2 \beta_3}-\frac{x_3 \gamma_3}{\beta_1 \beta_2} \cr
 -\gamma_2 & 0 & \gamma_3
}$\ ,
\cr
&$\psi_1=\pmatrix{
\frac{x_2 \gamma_3}{\alpha_3 \beta_3}-\frac{x_1 \gamma_1}{\alpha_2 \beta_3} & \frac{x_3
\gamma_2}{\alpha_3 \beta_1}-\frac{x_1 \beta_2 \gamma_3}{\alpha_2 \beta_1 \beta_3} &
 -\frac{x_1 \alpha_2 \gamma_2}{\alpha_1 \alpha_3 \beta_1}-\frac{x_2 \gamma_2}{\alpha_2
\beta_3}-\frac{x_3 \gamma_3}{\alpha _1 \beta _1} \cr
\frac{x_1 \gamma_2}{\alpha_1 \beta _3}-\frac{x_2 \gamma_1}{\alpha_2 \beta_3} & \frac{x_2
\gamma_2}{\alpha_1 \beta_1}-\frac{x_1 \gamma _3}{\alpha_2 \beta_3} 
& \frac{x_2 \left(\alpha_2 \beta_3^2 \gamma_3-\alpha_1 
\beta_1^2 \gamma_2\right)}{\alpha_1 \alpha_2 \beta_1 \beta_2 \beta_3} }$.
}
}
We can form the same composite with either one of the tachyons, using
as before the constant entries in $\psi_0$ to carry out row and column reductions and 
thereby discarding one trivial, $P_{1\times 1}$, vacuum piece.  The 
result is the ${4\times 4}$-matrix factorization, $\bar P_{4 \times 4}$, as shown 
in \lfig\condensation\ and discussed in Section~7.

%%%%%%%%%%%%%%%%%%%%%%%%%%%%%%%%%%
\newsec{All rank two factorizations obtained via tachyon condensation}
%%%%%%%%%%%%%%%%%%%%%%%%%%%%%%%%%%

We will now use the tachyons found in the previous to write down 
explicitly all rank-two  matrix factorizations.  Note that these
do not describe all rank-two vector bundles, but only those for
which the charge vector, $(r,c_1)$, lies in the positive cone. 
By equivariance, a general rank-two vector bundle for which $(r,c_1)$ is outside 
of the positive cone, can be mapped to a bundle within the positive cone but
with $r>2$. These correspond to higher rank matrix factorizations.

From \lfig\branes\ and \lfig\condensation\ one can see that there are
six different factorizations to be considered. These are distinguished
by their Chern numbers, $c_1$, and we will take $-2 < c_1 \le +3$.
These factorizations are shown on the right edge of the diagram in
\lfig\condensation.

There is one minor subtlety which arises from exceptional matrix factorizations.
Recall from Section~4 that the moduli space of some bundles may have singularities 
and that their smooth resolution can involve an extension
of the matrix factorization and thus, to a corresponding extension of the underlying
bundle to one of higher rank.  We encountered this with the ``exceptional'' bundle
that appears in the special matrix factorization,  
$\bar{\widetilde P}_\fobyfo^0$.  Such factorizations are rigid and do
not exist at generic moduli.  Thus our statement about there being
only six  different factorizations is meant to be true only for generic moduli.  

We have already  seen that the vector bundle $\cE(2,3)$ is obtained
from $P_{3\times 3}$, and thus it remains to construct the factorizations
with $c_1 =0, \pm 1, \pm 2$.    Below we will explain how to do this explicitly,
and obtain a new $4\times 4$-matrix factorization, 
denoted by $P_{4\times 4}$ and $\bar P_{4\times 4}$, which describes the vector 
bundles $\cE(2,2)$ and $\cE(2,-2)$ respectively; 
a $6 \times 6$-matrix factorization,  $P_{6\times 6}$,  and  corresponding
to  $\cE(2,0)$; and   a $5 \times 5$-matrix factorization,  $P_{5\times 5}$, which
gives $\cE(2,1)$. The bundle  $\cE(2,-1)$ is associated with the 
transpose, $P^{\rm T}_{5\times 5}$.

%%%%%%%%%%%%%%%%%
\subsec{Chern number $\pm 2$}
%%%%%%%%%%%%%%%%%

Since we already know that the $2\times2$ factorizations correspond to
line bundles with $c_1=\pm 1$, the construction of bound states at threshold 
will lead us to rank two bundles with $c_1=\pm2$ respectively.
Using \omegatrick, we consider the $4\times 4$-matrix factorization given by
\eqn\MFfourb{
P_{4\times 4}:\ \cases{
&$J=\pmatrix{J_{2\times2}(\alpha) & \psi_0 \cr
0 & J_{2\times2}(\alpha)
}$ \ ,\cr
&$E=\pmatrix{E_{2\times2}(\alpha) & \psi_1 \cr
0 & E_{2\times2}(\alpha)
}$\ .
}
}
The explicit form of the tachyon that we obtain is given below:
\eqn\MFtwoOmega{\eqalign{
\psi_0 &=\cD(\alpha) J_{2\times2}(\alpha) \cr 
&= \pmatrix{ \big[
\frac{2 x_1x_2 \dalpha_2 }{\alpha_1 \alpha_3}
-\frac{\dalpha_3 x_1^2}{\alpha_3^2}
- \frac{2 x_2^2 \dalpha_1}{\alpha_2 \alpha_3}
+\frac{x_3^2 \dalpha_2}{\alpha_2^2}\big] & 
\big[
(\frac{2 x_2 \dalpha_1}{\alpha_2 \alpha_3}
-\frac{2 x_3 \dalpha_3}{\alpha_1 \alpha_2}) x_1
-\frac{2 \dalpha_2 x_1^2}{\alpha_1 \alpha_3}
-\frac{x_2^2 \dalpha_3}{\alpha_3^2}\big] \cr
 x_3 \dalpha_1-x_2 \dalpha_3 & x_1 \dalpha_3-x_3 \dalpha_2
}\ ,\cr
\psi_1 &= \cD(\alpha) E_{2\times2}(\alpha) =\pmatrix{
 x_1 \dalpha_3-x_3 \dalpha_2 & 
\big[\frac{2 \dalpha_2 x_1^2}{\alpha_1 \alpha_3}
+(\frac{2 x_3 \dalpha_3}{\alpha_1 \alpha_2}
-\frac{2 x_2 \dalpha_1}{\alpha_2 \alpha_3}) x_1
+\frac{x_2^2 \dalpha_3}{\alpha_3^2}\big] \cr
 x_2 \alpha\alpha_3-x_3 \dalpha_1 & 
[-\frac{\dalpha_3 x_1^2}{\alpha_3^2}
+\frac{2 x_2 \dalpha_2 x_1}{\alpha_1 \alpha_3}
-\frac{2 x_2^2 \dalpha_1}{\alpha_2 \alpha_3}+\frac{x_3^2
   \dalpha_2}{\alpha_2^2}]
}\ ,
} }
where $\dalpha_\ell$ denotes $\mu_\ell(-2\zeta)$. The matrix bundles, $\cE_J$
and $\cE_E$, for the $4\times4$ factorization are
associated with the vector bundles $\cE(2,2)$ and  $\cE(2,-2)$ respectively.   
Because of the behavior of   $P_{2\times 2}$ under
transposition, it follows that   $P^{\rm T}_{4\times 4}$ is gauge equivalent to 
$\bar{P}_{4\times 4}$.

%%%%%%%%%%%%%%%%%
\subsec{Chern number  $0$}
%%%%%%%%%%%%%%%%%

The $3\times 3$ factorization gives us two matrix bundles, $\cE_J$
and $\cE_E$, one having rank one and degree zero and the other
having rank two and degree $3$. Therefore, the composite factorization
will consist of one matrix bundle of rank two and degree zero, plus a
bundle of rank four and degree $6$, of the following form:
\eqn\MFfourb{
P_{6\times 6}:\ \cases{
&$J=\pmatrix{-E_{3\times3}(\alpha) & \psi_0 \cr
0 & -E_{3\times3}(\alpha)
}$\ \cr  
&$E=\pmatrix{-J_{3\times3}(\alpha) & \psi_1 \cr
0 & -J_{3\times3}(\alpha)
}$ \ .
}
}
The explicit form of the tachyon is then given by (with $\dalpha_l$ denoting
$\mu_l(-2\zeta)$)
\eqn\MFthreeOmega{\eqalign{ 
\psi_0 &= \cD(\alpha) E_{3\times3}(\alpha) = \pmatrix{
-\frac{\dalpha_1 x_1^2}{\alpha_1^2}-\frac{2 x_2 x_3 \dalpha_1}{\alpha_2 \alpha_3} &
   -\frac{\dalpha_3 x_3^2}{\alpha_3^2}-\frac{2 x_1 x_2 \dalpha_3}{\alpha_1 \alpha_2} &
-\frac{\dalpha_2 x_2^2}{\alpha_2^2}-\frac{2 x_1 x_3 \dalpha_2}{\alpha_1 \alpha_3} \cr
-\frac{\dalpha_2 x_3^2}{\alpha_2^2}-\frac{2 x_1 x_2 \dalpha_2}{\alpha_1 \alpha_3} &
   -\frac{\dalpha_1 x_2^2}{\alpha_1^2}-\frac{2 x_1 x_3 \dalpha_1}{\alpha_2 \alpha_3} &
-\frac{\dalpha_3 x_1^2}{\alpha_3^2}-\frac{2 x_2 x_3 \dalpha_3}{\alpha_1 \alpha_2} \cr
 -\frac{\dalpha_3 x_2^2}{\alpha_3^2}-\frac{2 x_1 x_3 \dalpha_3}{\alpha_1 \alpha_2} &
   -\frac{\dalpha_2 x_1^2}{\alpha_2^2}-\frac{2 x_2 x_3 \dalpha_2}{\alpha_1 \alpha_3} &
   -\frac{\dalpha_1 x_3^2}{\alpha_1^2}-\frac{2 x_1 x_2 \dalpha_1}{\alpha_2 \alpha_3}
}\,, \cr
\psi_1 &=  \cD(\alpha) J_{3\times3}(\alpha)=\pmatrix{
 x_1 \dalpha_1 & x_3 \dalpha_2 & x_2 \dalpha_3 \cr
 x_3 \dalpha_3 & x_2 \dalpha_1 & x_1 \dalpha_2 \cr
 x_2 \dalpha_2 & x_1 \dalpha_3 & x_3 \dalpha_1
} 
 \,.
}}
Recalling the results  for $P_{3 \times 3}$ from  \ltab\MFProp, we see that 
the matrix bundles, $\cE_J$ and $\cE_E$, are
associated with the vector bundles $\cE(2,0)$ and $\cE(4,6)$ respectively.
Note that $P^{\rm T}_{6\times 6}(\zeta)$ is gauge equivalent to 
$P_{6\times 6}(-\zeta)$, which
follows from the known behavior~\MFtransposecases\ of $P_{3\times 3}$ under transposition.

%%%%%%%%%%%%%%%%%
\subsec{Chern number  $\pm 1$}
%%%%%%%%%%%%%%%%%

These two composites arise as $5\times 5$-matrix factorizations that are
transposes of each other. The composite obtained from the tachyon
$\Psi_{(\bar{S}_{a},\bar{L}_{a})}$ is given by
\eqn\MFfivea{
P_{5\times5}:\ \cases{
&$J=\pmatrix{-E_{3\times3}(\alpha) & \psi_0(\alpha,\beta) \cr
0 & -E_{2\times2}(\beta)
}$ \ ,\cr
&$E=\pmatrix{-J_{3\times3}(\alpha)& \psi_1(\alpha,\beta) \cr
0 & -J_{2\times2}(\beta)
}$\ ,}}
where the tachyon is explicitly given by:
\eqn\fivebyfivetach{
\Psi_{(\bar{S}_{a},\bar{L}_{a})}:\ \cases{ 
&$\psi_0=\pmatrix{
-\beta_3 \gamma_3x_1  & {{ \alpha_2 
\gamma_1x_1^2}\over{\alpha_1 \beta_1}}+{{ \alpha_1 
\gamma_1x_1 x_3}\over{\alpha_3 \beta_2}}-{{ \alpha_1 \gamma_2x_1 x_2}
\over{\alpha_2 \beta_2}} \cr
 -\beta_3 \gamma_1x_1  & {{ \alpha_1 \gamma_2x_1^2}
\over{\alpha_3 \beta_1}}+{{ \alpha_3 \gamma_2x_1 x_3}
\over{\alpha_2 \beta_2}}-{{ \alpha_3 \gamma_3x_1 x_2}
\over{\alpha_1 \beta_2}} \cr
 -\beta_3 \gamma _2x_1  & -\frac{ \alpha_2 \gamma_1x_1 x_2}{\alpha_3 \beta_2}
+\frac{ \alpha_3 \gamma_3x_1^2}{\alpha_2 \beta_1}+\frac{ \alpha_2 
\gamma_3x_1 x_3}{\alpha_1 \beta_2}}$ \ , \cr
&$\psi_1= \pmatrix{
\alpha_1 \gamma_3 x_1& \alpha_3 \gamma_2 x_1\cr
 \alpha_2 \gamma_2x_1 & \alpha_1 \gamma_1 x_1\cr
 \alpha_3 \gamma_1x_1 & \alpha_2 \gamma_3 x_1
}$ \ ,
}}
where $\gamma_\ell$ denotes $\mu_\ell(-\zeta-\lambda)$. 
For $P_{5\times 5}$, the matrix bundle $\cE_J$ 
is  associated with
the vector bundle $\cE(2,5) \equiv \cE(2,-1)$. The factorization
as given in \MFfivea\ seemingly depends on  two open-string
moduli, $\zeta$ and $\lambda$, of the $3\times3$ and $2\times2$ 
matrix factorizations that form the composite. However, 
one can carry out a gauge transformation such that
the factorization  depends on only the combination
$(3\zeta+\lambda)$. This combination is to be identified with the physical
open-string
modulus of the composite. The form of the factorization after such a gauge transformation
has been given in Appendix C.

As we already mentioned in Section~6, the remaining $5\times 5$ matrix
factorization, corresponding to $\cE(2, 1)$,  can be obtained
 by transposition and we will therefore denote it by
$P^{\rm T}_{5\times 5}$.
This then completes the list of all matrix factorizations related 
to rank two bundles.

%%%%%%%%%%%%%%%%%%%%%%%%%%%%%%%%%%
\newsec{Summary and conclusions}
%%%%%%%%%%%%%%%%%%%%%%%%%%%%%%%%%%

Our purpose has been to show how matrix factorizations,
physically realized in terms of boundary Landau-Ginzburg models,
can be used as an effective computational tool for dealing with
topological $B$-type branes.  Once again we stress that to precisely
understand these $B$-branes, one must understand the derived category
and not merely the $K$-theory of the branes.  That is, it is not
sufficient to classify the branes in terms of their charges, but one must
also understand the equivariant maps between the branes (as well
as the dependence of all these quantities on the complex structure
moduli).  In more physical terms, this means classifying the tachyons
between branes and understanding the condensates they produce.  We
have seen that matrix factorizations provide a very explicit way
of realizing the branes and of performing such computations.  Moreover
the description of $B$-type branes via matrix factorization appears
to be complete in that the results of refs.~\refs{\OrlovB,\HeHo} show
that the  derived  category of topological $B$-type  $D$-branes is
isomorphic to the category of matrix factorizations
\refs{\kontsevichU,\OrlovA}\ (at least for a large class of
geometries).  Thus we have an explicit and relatively simple way
of representing the category of topological $B$-branes, as well as performing
computations that capture the dependence on moduli, in a manner
that is easily accessible for physicists.

To flesh-out  this picture, we have shown how such computations
work in practice, and while we have focused on the cubic torus,
the ideas involved appear universal.  We showed how the bundle data
for the branes and anti-branes can be extracted directly from the
matrices.  These bundle data are ambiguous in a way that is familiar
from the bulk linear sigma model:  Generically, there is a monodromy around the
``large-radius'' limit in the K\"ahler moduli space that makes the
bundle data ambiguous up to an overall tensoring with the line
bundle $\cL^{N}$.  We have seen that this ambiguity is mirrored in
the $B$-model by a corresponding ambiguity in extracting the bundle
data from the matrix factors. Moreover, the $\ZZ_N$ monodromy at
the Gepner point in the  K\"ahler moduli space is matched by the
$\ZZ_N$ orbit of brane charges that one can associate with a given
matrix factorization.\foot{Note that these two monodromies
are very different: tensoring with a line bundle merely shifts
the Chern number $c_1$,  while the $\ZZ_N$ monodromy at the Gepner point changes all
the Chern numbers, including the rank.  Indeed, these two monodromies
taken together generate a (sub-)group of the full duality group of
the theory, and thus a given matrix factorization will describe a whole
orbit under this group.} We have seen that these charges can be
resolved by refining the matrix factorization in terms of equivariant
$R$-symmetry.  In this way, the Landau-Ginzburg model is able to
describe general sheaves and not merely vector bundles.  A
more complete treatment, for instance based on the forthcoming
description of the boundary linear sigma model \HeHo, may be needed
in order to precisely track states between Landau-Ginzburg and large radius
phases, and determine how the bundle data evolve over the K\"ahler
moduli space.

Having explicitly shown how the matrix factorizations encode the
details of the underlying vector bundles, we showed, in some detail,
how tachyon condensation works within this framework. In particular,
we demonstrated how repeated tachyon condensation of the $2 \times 2$
factorization can be used to construct, at least in  principle, the
matrix factorizations corresponding to general vector bundles on the
torus.  Indeed, we showed how to obtain  all previously known
factorizations, including their dependence on moduli, from the $2
\times 2$ factorization (prior to this paper, the connection
with the $3 \times 3$ factorization was only known implicitly, based
upon the $RR$ charges).  We went on to construct all the matrix
factorizations corresponding to rank two bundles on the torus.

One of the crucial ingredients in understanding and organizing the
process of tachyon condensation was the application of equivariant
$R$-symmetry \refs{\AshokZB,\WalcherTX}.  The resulting orbits of
branes are labeled by representations of $\IZ_N$, and when translated
to the language of  matrix factorization, these
 representations impose selection rules upon tachyons,
and so effectively organize the tachyon spectrum between any given
pair of branes. This enabled us to determine the correct
tachyon to generate each and every possible condensate of the members
of different $\IZ_N$ multiplets.

One of the other interesting features we uncovered in this study
of tachyon condensation, and the bundles associated with matrix
factorization, is how the bundle structure can jump, apparently
discontinuously, on the open-string moduli space of the branes.  We saw how such
behavior is easily, and smoothly, described by matrix factorizations.
For example,  the $3 \times 3$ factorization is apparently singular
at one point in its moduli space, however, we showed 
(following ref.~\HoriZD) that this singularity could be resolved smoothly by passing
to $4 \times 4$ factorization that, at generic points in the moduli
space, is equivalent to the $3 \times 3$ factorization.  We found
that the  $4 \times 4$ factorization  corresponds to a pair of
rank two vector bundles, and, at generic  points  in the moduli
space, one of these rank two bundles is split, \ie, it is a trivial
sum of line bundles, but at the special point, where the $3 \times
3$ factorization is singular, both the rank two vector bundles are
indecomposable. We interpreted this configuration physically in terms of a
pure anti-$D2$-brane, which is rigid; deforming away from the special point
corresponds to adding and pulling apart an extra $D0$-$\bar{D0}$-brane pair, which
amounts to a reducible bundle configuration depending on an open-string modulus.

A general class of examples that exhibits similar behavior was
encountered in studying bound states at threshold.  In this instance
one combines two parallel branes and, if their moduli do not match,
then there is no tachyon and the overall vector bundle is simply
the sum of the component bundle  for each brane.  However when the
moduli coincide then there is a new tachyon, formally given by
a boundary preserving operator, and this condenses the branes
into a true bound state at threshold that corresponds to a non-split
extension of the component vector bundles.

Another, related feature that we found interesting is that the gauge
symmetry, inherent to any matrix factorization, sometimes leads to
identifications in the open-string moduli space; this can impose a
non-trivial global structure on it,  like an unexpected shortening of
periodicities, or orbifold points that connect to new branches.

We believe that many of the ideas and techniques we have used here
will be applicable to more general Calabi-Yau manifolds.  Once one
finds a set of matrix factorizations and suitable tachyonic, boundary
changing cohomology elements, one should be able to build more by
condensation.\foot{Of course, a major new ingredient and complication
is the problem of stability \DouglasGI, but since this is tied to
the K\"ahler parameters it is not clear to what extent this can be
addressed within the topological $B$-model; for some ideas in this
direction, see ref.~\WalcherTX.  It is also known that some of the
permutation branes cross lines of marginal stability, \ie, they can
decay, when going to large radius. It is thus natural to carry out
Seiberg dualities even in the topological B-model, where they appear
as a change of basis.} However, the obvious approach of trying to
find matrix factorizations pertaining to something like the quintic
threefold does not seem to be straightforward.  One can easily
convince oneself by elementary counting arguments that matrix
factorization of a generic quintic may be difficult:  One simply
totals up the degrees of freedom in the matrix elements, taking
into account the gauge invariances, and subtracts the constraints
imposed by \matfac. The result is generically negative, and thus
one should expect interesting new matrix factorizations of the
complete quintic only for special branes, and/or for special points
in its complex moduli space.   

For example, at the Fermat point one
has all the matrix factorizations arising from tensoring the matrices
corresponding to  Recknagel-Schomerus states  of the underlying
minimal models; the task would be to try to extend these
away from the Fermat point. However, while for the torus the sections
relevant for the simplest branes are explicitly known to be given
by theta and Appell functions, much less is known about sections
on threefolds. One the other hand, deformations of branes on
threefolds will often be obstructed, so that there are no open
string moduli to begin with; this may facilitate the construction
of the corresponding matrix factorizations.

Moreover, it might also be rewarding to focus on a kind
of ``hybrid'' description of $B$-type branes.  The same counting
arguments that show that matrix factorizations for the complete
quintic are rare, also show that if there are only three variables
of any degree, then there will generically be matrix factorizations
with possibly several moduli.  This  suggests that it might be
productive to look at $B$-type branes defined via vector bundles
on sub-manifolds.  For example, by adding two further equations to
the function(s) that define the Calabi-Yau manifold, one can recast
the Calabi-Yau, at least locally, as a Riemann surface fibered over
a base, which may be amenable to a treatment similar to the one
described in this paper.  Specifically, if one focuses on elliptically
fibered Calabi-Yau manifolds, one may make direct use of many
of the results presented here; this is presently under investigation.

\goodbreak
\bigskip
\noindent {\bf Acknowledgments:} \medskip \noindent
We would like to thank Robert Helling and Manfred Herbst for helpful
comments, and especially Ilka Brunner, Dennis Nemeschansky and
Johannes Walcher for participating in various stages of this work.
The work of NW is supported in part by the DOE grant DE-FG03-84ER-40168.

%%%%%%%%%%%%%%%%%%%%%%%%%%%%%%%%%%
\appendix{A}{Theta functions and Appell functions}
%%%%%%%%%%%%%%%%%%%%%%%%%%%%%%%%%%

The theta functions and Appell functions are defined by:
\eqn\basictheta{\vartheta(\xi)  ~=~ \vartheta(\xi |\tau) ~\equiv~ \sum_{n \in \IZ}  
q^{{1\over 2}\, n^2 } \, z^n \,, }
\eqn\basicappell{\kappa(\rho, \xi) ~=~ \kappa(\rho, \xi |\tau) ~\equiv~ \sum_{n \in \IZ}  
{q^{{1\over 2}\, n^2 } \, z^n \over  q^n - y}\,, }
where $y \ne q^m, m \in \IZ$ and
\eqn\qazdefns{q~\equiv~ e^{2\pi i\,   \tau  }\,, \quad  z~\equiv~ e^{2\pi i \, \xi }\,, 
\quad  y~\equiv~ e^{2\pi i\,\rho } \,.}

These functions also satisfy the periodicity relations:
\eqn\perioda{\vartheta(\xi+1)  ~=~ \vartheta(\xi) \,, \qquad 
\vartheta(\xi+\tau)  ~=~q^{-{1\over 2}}\, z^{-1}\, \vartheta(\xi)  \,,}
\eqn\periodb{\eqalign{\kappa(\rho, \xi+1) ~=~ & \kappa(\rho+1, \xi)~=~ 
\kappa(\rho, \xi)\,,  \qquad  \kappa(\rho, \xi+\tau) ~=~  y \, \kappa(\rho, \xi)~+~ 
\vartheta(\xi) \,,\cr 
\kappa(\rho+\tau, \xi) ~=~  & q^{-{1 \over 2}}\, z \, (y \, \kappa(\rho, \xi)~+~ \vartheta(\xi)) \,.}}
In addition, the Appell functions satisfy an interchange identity:
\eqn\intident{   \vartheta\big(\rho - \half(1+\tau)\big) \,  \kappa(\rho, \xi)~=~  
-q^{{1\over 2}}\, z \, y^{-1}  \vartheta (\xi)  \, 
\kappa\big(\xi +  \half(1+\tau), \rho - \half(1+\tau)\big)  \,.}
By taking the limit as  $\rho \to 0$ on both sides of this identity one
can then show that:
\eqn\specvalone{  \kappa(\rho,   - \half(1+\tau) ) ~=~  
 {q^{-{1 \over 8}} \, \eta^3(\tau) \over \vartheta (\rho -  \half(1+\tau)) }  \,,  }
where 
\eqn\etadefn{\eta(\tau) ~\equiv~ q^{1 \over 24}\, \prod_{n=1}^\infty
(1 - q^n) }
is the Dedekind $\eta$-function.

To uniformize the curve \cubic, we need the theta functions with characteristics:
\eqn\thetadefn{ \Theta \bigg[{ {c_1 \atop c_2}} \bigg  |\,\xi \,, \tau\bigg]\ =\
\sum_{m \in \IZ}  q^{(m+c_1)^2/2} e^{2\pi i(\xi +c_2)(m+c_1)}\,,}
and then define:
\eqn\mudefn{\eqalign{
\mu_\ell(\xi) &~\equiv~ \mu_\ell(\xi|\tau)  ~=~   \omega^{(\ell -1)}  \, 
\Theta \bigg[{ {{1 \over 3} (1-\ell) -{1 \over 2} \atop -{1 \over 2}} }\, \bigg |\,3\,\xi ,3
\,\tau \bigg]\,, \cr
& ~=~    i\, (-1)^{(\ell-1)}\, q^{{3 \over 2} ( {1 \over 2} +
{1 \over 3}\, (\ell-1) )^2 } \, z^{-   ( {3 \over 2}  +   (\ell-1) )  }\,   \vartheta( 3\, \xi - 
( \coeff{3}{ 2}  +   (\ell-1) )\tau -\coeff{1}{ 2}   | 3\,\tau) \,, }}
where $\omega \equiv e^{2 \pi i /3}$ and $\ell=1,2,3$.  One can then show that 
\cubic\ is uniformized by taking $x_\ell = \mu_\ell(\xi)$ with the modulus,
$a$, related to $\tau$ via:
\eqn\ataureln{\Big({3\, a\, (a^3 +8) \over  a^3 - 1} \Big)^3~=~  j(\tau) \,.}

The associated Appell functions are then defined by:
\eqn\Lambdadefn{\eqalign{
\Lambda_\ell (\rho, \xi) &~\equiv~  i\, (-1)^{(\ell-1)}\, q^{{3 \over 2}( {1 \over 2} +
{1 \over 3}\, (\ell-1) )^2 } \, z^{-   ( {3 \over 2}  +   (\ell-1) )  }\, \sum_{n \in \IZ}
{\big(q^3\big)^{{1 \over 2} \, n^2} \, \big( -q^{-( {3 \over 2}  +   (\ell-1) )}\, z^3 
\big)^n \over \big(q^3\big)^{n}  - y^3\,q^{( {3 \over 2}  +   (\ell-1) )} } \cr
& ~=~   i\, (-1)^{(\ell-1)}\, q^{{3 \over 2}( {1 \over 2} +
{1 \over 3}\, (\ell-1) )^2 } \, z^{-   ( {3 \over 2}  +   (\ell-1) )  }  
\cr & \qquad \qquad  \qquad  \qquad
\kappa \big(3\, \rho +  (\coeff{3}{ 2}  +   (\ell-1) )\tau    \,,
 3\, \xi - ( \coeff{3}{ 2}  +   (\ell-1) )\tau -\coeff{1}{ 2}  \, |\,3\,\tau \big) \,.}}

We now sketch some of the manipulations involving the $\mu_l$ that we
used in the main text of the paper. A change
in the sign of the argument of $\mu_l$ is the permutation (up to an overall
sign):
\eqn\signperm{
\mu_1(-\zeta)=-\mu_1(\zeta)\ ,\quad \mu_2(-\zeta)=-\mu_3(\zeta)\ ,\quad
\mu_3(-\zeta)=-\mu_2(\zeta)\ .
}
In the following, let us denote $\mu_\ell(\zeta)$, $\mu_\ell(\lambda)$ and $\mu_\ell(-\zeta-\lambda)$ by $\alpha_\ell$, $\beta_\ell$ and $\gamma_\ell$,
 respectively. 
The addition formulae for the $\mu_\ell(\zeta)$ may be deduced from \Eij. For
example, by setting $\xi=-\zeta$, one obtains formulae such as the one given
below  with the argument of the $\mu_\ell$ doubled:
\eqn\doubled{
\dalpha_1 \equiv \mu_1(-2\zeta)= {{\alpha_1 (\alpha_2^3 - \alpha_3^3)}\over{ i\eta(\tau)^3}}\ .
}
Similar manipulations also lead to \zetarep\ and \alphanew.
The following identities also follow from \Eij:
\eqn\gammarels{\eqalign{\alpha_1 \beta_1 \gamma_1+\alpha_2 \beta_2
\gamma_2+\alpha_3 \beta_3 \gamma_3=0 \ , \cr
\alpha_1 \beta_2 \gamma_3+\alpha_2 \beta_3 \gamma_1+\alpha_3 \beta_1 \gamma_2
=0 \ , \cr
\alpha_1 \beta_3 \gamma_2 +\alpha_2 \beta_1 \gamma_3+ \alpha_3 \beta_2 \gamma_1
=0 \ . }}
These identities are useful in solving the physical state condition 
and in identifying the open-string modulus of condensates.

%%%%%%%%%%%%%%%%%%%%%%%%%%%%%%%%%%%%%%%%%%%%%%%%%%e
\appendix{B}{Row and column elimination of matrix factorizations}
%%%%%%%%%%%%%%%%%%%%%%%%%%%%%%%%%%%%%%%%%%%%%%%%%%%

Specifying a matrix factorization, given by $J$ and $E$ satisfying
\matfac, corresponds to choosing a representation for the
$\IC[x_1,x_2,x_3]$-module homomorphism $J$ and $E$. In general,
however, there are different representatives for the same module
homomorphisms. In physical terms, the choice of representative means
that we can act upon a given matrix factorization with a gauge
transformation to obtain another matrix factorization that satisfies
the defining condition \matfac\ and thus describes the same D-brane
configuration. From the point of view of the category, such gauge
transformations act trivially on the objects and thus lead to an
equivalent description.

For a $n\times n$-matrix factorization, the gauge transformations
are given by eq.~\gauge{} with the transformation matrices $U_L$
and $U_R$ as elements of $GL(n,\IC[x_1,x_2,x_3])$. That is to say
$U_L$ and $U_R$ are invertible matrices in the polynomial ring
$\IC[x_1,x_2,x_3]$.\foot{A matrix is invertible in $\IC[x_1,x_2,x_3]$
if and only if its determinant is a non-vanishing constant.}

In the following we often make use of the gauge freedom \gauge{}
in order to rewrite a given D-brane configuration in a convenient
gauge. There are three basic gauge transformations which allow us
to simplify matrix factorizations step by step. The idea is to
simplify in each step one of the two matrices of a matrix factorization,
{\it e.g.}, $J$. A gauge transformation acting on $J$ via $U_L$ and
$U_R$ then also defines according to \gauge{} the corresponding
gauge transformation acting on $E$.

The first basic gauge transformations are simply given by multiplying a matrix row~$r$ and a matrix column~$t$ of $J$ by non-zero constants $a$ and $b$ respectively. For a $n\times n$-matrix factorization this gives rise to the diagonal transformations matrices
\eqn\rowmulti{\eqalign{U_L^\times(a,b)&= 
                          \Diag{1,\ldots,1,a,1,\ldots,1}\, , \,
                       U_R^\times(a,b)= 
                          \Diag{1,\ldots,1,b,1,\ldots,1} \, , \cr
                       U_L^{\times}(a,b)^{-1}&=
                          \Diag{1,\ldots,1,{1\over a},1,\ldots,1}\, , \,
                       U_R^{\times}(a,b)^{-1}= 
                          \Diag{1,\ldots,1,{1\over b},1,\ldots,1}\, .}}
The second kind of gauge transformations are given by either adding the row $r$ of $J$, multiplied by a polynomial, $p_r(x)$, to the row $s$ of $J$, or analogously by adding the column $t$ of $J$, multiplied by a polynomial, $p_c(x)$, to the column $u$ of $J$. The appropriate transformation matrices are respectively given by
\eqn\rowred{\eqalign{U_L^-(r,s,p_r(x))\, & =  \, \id_{n\times n}\,
                                               +\,p_r(x)\,\Gamma_{s,r} \ , \qquad
                       U_R^-(r,s,p_r(x))\, = \, \id_{n\times n} \ , \cr
                       U_L^-(r,s,p_r(x))^{-1}\, & = \, \id_{n\times n}\,
                                               -\,p_r(x)\,\Gamma_{s,r} \ , \qquad
                       U_L^-(r,s,p_r(x))^{-1}\, = \, \id_{n\times n} \ , }}
and
\eqn\columnred{\eqalign{U_L^|(t,u,p_c(x))\, & = \, \id_{n\times n} \ , \qquad
                       U_R^|(t,u,p_c(x))\,  =  \, \id_{n\times n}\,
                                               +\,p_c(x)\,\Gamma_{t,u} \ , \cr
                       U_L^|(t,u,p_c(x))^{-1}\, & = \, \id_{n\times n} \ , \qquad 
                       U_L^|(t,u,p_c(x))^{-1}\,  = \, \id_{n\times n}\,
                                               -\,p_c(x)\,\Gamma_{t,u} \ , \cr
                       }}
with
\eqn\GammaMat{\left(\Gamma_{r,s}\right)_{kl}\, 
                = \,\cases{$1$ & \hbox{for}\ $r=k, s=l$ \ , \cr\mathstrut\cr $0$ & else \ . }}

With these simple transformation rules we can already deduce an
important property of a given matrix factorizations~$P$: If either
the matrix $J$ or the matrix $E$ contains a constant, that is to
say a non-vanishing entry of degree $0$, the matrix factorization~$P$
simplifies as follows: For concreteness let us assume that the
matrix $J$ of an $n\times n$-matrix factorization has a non-vanishing
constant in the top left corner. First, we act on the matrix
factorization with a transformation of the type \rowmulti{} in order
to normalize this constant to $1$. Then we apply $(n-1)$ times the
gauge transformations  \rowred\ with $r=1$, $s=2, \ldots, n$ and
appropriate polynomials~$p_r(x)$ such that the first row of $J_P$
becomes $(1, 0, \ldots, 0)$. In a third step we apply the gauge
transformation \columnred\ $n-1$ times with $t=1$, $u=2, \ldots,
n$ and with suitable polynomials~$p_c(x)$ so as to also reduce the
first column to $(1, 0, \ldots, 0)$. After this chain of gauge
transformations we obtain a gauge equivalent matrix factorization.
As a consequence of \matfac\ both the first row and the first column
of $E$ have automatically been transformed into $(W, 0, \ldots,
0)$! Hence the original $n\times n$-matrix factorization is really
a $(n-1)\times(n-1)$-matrix factorization with a trivial irrelevant
summand, $P_{1\times1}$. This technique, which reduces the dimension of a matrix
factorization by applying gauge transformations, is in the main
text referred to ``row and column elimination''.

\appendix{C}{An explicit form of the $5\times5$ matrix factorization}

We present a form of the $5\times5$ matrix factorization with the
open-string modulus is given by $\nu_l$.
This is gauge equivalent to the form given in \MFfivea\ when
$\nu_l\sim \mu_l(3\zeta+\lambda)$.
\eqn\newfivebyfive{
J_{5\times5} =\pmatrix{
\frac{x_1^2}{\nu _3} & \frac{x_2^2}{\nu _2} & \frac{x_3^2-3 a x_1 x_2}{\nu _1} & \frac{x_1
   x_2}{\nu _3} & -\frac{x_1 x_2 \nu _2}{\nu _1 \nu _3} \cr
 0 & x_3 \nu _1 & -x_2 \nu _2 & x_1 \nu _2 & 0 \cr
 -x_3 \nu _1 & 0 & x_1 \nu _3 & 0 & x_2 \nu _3 \cr
 x_2 \nu _2 & -x_1 \nu _3 & 0 & -\frac{x_2 \nu _1^2-x_1 \nu _2^2+x_3 \nu _3^2}{\nu _1} & \frac{x_2 \nu _1^2
   \nu _2-x_1 \left(\nu _2^3+\nu _3^3\right)}{\nu _1^2} \cr
 0 & 0 & 0 & x_3 \nu _1-x_2 \nu _3 & x_1 \nu _3-x_3 \nu _2
}}
\eqn\fivebyfiveE{
E_{5\times 5} =\pmatrix{
x_1 \nu _3 & -\frac{x_1 x_2}{\nu _2} & 
\widehat{Q}_2
& \frac{x_2 \left(x_2 \nu _3-x_3 \nu _1\right)}{\nu _2 \nu _3} & -\frac{x_2
   \left(x_2 \nu_1^2-x_1 \nu _2^2+x_3 \nu _3^2\right)}{\nu _1 \nu _2 \nu _3} \cr
 x_2 \nu _2 & 
\widehat{Q}_1
& \frac{x_1 x_2
   \nu _2^2}{\nu _1 \nu _3^2} & \frac{x_1 \left(x_3 \nu _2-x_1 \nu _3\right)}{\nu _3^2} & -\frac{x_1
   \left(x_1 \left(\nu _2^3+\nu _3^3\right)-x_2 \nu _1^2 \nu _2\right)}{\nu _1^2 \nu _3^2} \cr
 x_3 \nu _1 & -\frac{x_2^2}{\nu _2} & \frac{x_1^2}{\nu _3} & 0 & -\frac{x_1 x_2}{\nu _3} \cr
 0 & \left(\frac{x_1^2}{\nu _2}-\frac{x_1x_3}{\nu _3}\right) & \frac{x_2 \left(x_1 \nu _3-x_3 \nu
   _2\right)}{\nu _3^2} & \frac{x_3 \nu _1 \left(x_1 \nu _3-x_3 \nu _2\right)}{\nu _2 \nu _3^2} & 
\widehat{Q}_3
\cr
 0 & \frac{x_1 \left(x_2 \nu _3-x_3 \nu _1\right)}{\nu _2 \nu _3} & \frac{x_2 \left(x_2 \nu _3-x_3 \nu
   _1\right)}{\nu _3^2} & \frac{x_3 \nu _1 \left(x_2 \nu _3-x_3 \nu _1\right)}{\nu _2 \nu _3^2} & 
\widehat{Q}_4
}}
where
$$\eqalign{
\widehat{Q}_1 &= \frac{\nu _2 x_1^2-3 a x_2 \nu _3 x_1+x_3^2 \nu _3}{\nu _1 \nu _3}  \ ,
\cr
\widehat{Q}_2&=
-\frac{x_2^2}{\nu_3}+\frac{3 a x_1 x_2}{\nu_1}-\frac{x_3^2}{\nu_1} \ ,\cr
\widehat{Q}_3&=
\frac{-\nu_1 \nu_2 \nu_3 x_2^2-x_3 \nu_1^2 \nu_2 x_2+x_1 x_3 \left(\nu_2^3+\nu_3^3\right)}{\nu_1 \nu_2 \nu_3^2}\ , \cr
\widehat{Q}_4&=
\frac{\nu_2 \nu_3 x_1^2+x_3 \nu_2^2 x_1-x_2 x_3 \nu_1^2-x_3^2 \nu_3^2}{\nu_2 \nu_3^2}\ .\cr
}
$$

\listrefs
\vfill
\eject
\end